\documentclass[numberedappendix]{emulateapj}  
\usepackage{subfigure} 
\usepackage[hyperfootnotes=false,naturalnames=true,pdfstartview=FitH,pdfpagemode=UseNone]{hyperref}
\hypersetup{pdfauthor={Dan Coe}, pdftitle={LensPerfect A1689}}
\let\chapter=\section  
\usepackage{placeins}  

\newcommand{\Nimages}{135\ }
\newcommand{\Ngals}{42\ }
\newcommand{\Nknotimages}{168\ }
\newcommand{\Nknotss}{55\ }
\newcommand{\Nensemble}{54\ }  

\newcommand{\NFWfitM}{$r_s = 338^{+105}_{-71}$ kpc, $c_{200} = 7.6\pm1.3$ ($c_{vir} = 9.6^{+1.7}_{-1.6}$)}
\newcommand{\NFWfitMW}{$r_s = 258^{+54}_{-40}$ kpc, $c_{200} = 9.2\pm1.2$ ($c_{vir} = 11.5^{+1.5}_{-1.4}$)}

\newcommand{\NFWfitMbestp}{($r_s = 338$ kpc, $c_{200} = 7.6$ [$c_{vir} = 9.6$])}

\newcommand{\NFWfitMc}{($c_{200} = 7.6\pm1.3$)}

\newcommand{\NFWfitMWct}{$c_{200} = 9.2\pm1.2$}  
\newcommand{\NFWfitMWcvirt}{$c_{vir} = 11.5^{+1.5}_{-1.4}$}  
\newcommand{\hunits}{h_{70}^{-1}}
\newcommand{\NFWfitMWMvir}{$M_{vir} = 2.0^{+0.5}_{-0.3} \times 10^{15} M_\odot \hunits$}
\newcommand{\NFWfitMWM}{$M_{200} = 1.8^{+0.4}_{-0.3} \times 10^{15} M_\odot \hunits$}
\newcommand{\NFWfitMWrvir}{$r_{vir} = 3.0\pm0.2 ~ {\rm Mpc} ~ \hunits$}
\newcommand{\NFWfitMWr}{$r_{200} = 2.4^{+0.1}_{-0.2} ~ {\rm Mpc} ~ \hunits$}
\newcommand{\NFWfitMcval}{$7.6\pm1.3$}
\newcommand{\NFWfitMWcval}{$9.2\pm1.2$}

\newcommand{\NFWMRE}{$R_E = 47.0\arcsec \pm 1.2\arcsec$ ($143^{+3}_{-4}$ kpc)}

\newcommand{\SersicfitMW}{$R_e = 273.6$ kpc, $\kappa_e = 0.2939$, $n = 1.425$ ($b_n = 2.524$)} 
\newcommand{\SersicfitMWn}{$n = 1.425$} 

\newcommand{\Sersicfitp}{($R_e = 692.8$ kpc, $\kappa_e = 0.1007$, $n = 2.148$ [$b_n = 3.968$])} 

\newcommand{\cDuffy}{$c_{200} = 3.0^{+1.3}_{-0.9}$ ($c_{vir} = 3.9^{+1.6}_{-1.1}$)}
\newcommand{\crelDuffy}{$c_{200} = 3.4^{+1.4}_{-1.0}$ ($c_{vir} = 4.4^{+1.8}_{-1.3}$)}

\newcommand{\kR}{$\kappa(R)$}
\newcommand{\MR}{$M(<R)$}
\newcommand{\MRE}{$M(<R_E)$}

\newcommand{\kth}{$\kappa(\vec \theta)$}

\slugcomment{Draft}

\shorttitle{LensPerfect Abell 1689 Analysis}
\shortauthors{Coe et al.}

\begin{document}

\title{The Highest Resolution Mass Map of Galaxy Cluster Substructure To Date
Without Assuming Light Traces Mass:
LensPerfect Analysis of Abell 1689}


\author{Dan~Coe\altaffilmark{1}}
\author{Narciso~Ben\'itez\altaffilmark{2}}
\author{Tom~Broadhurst\altaffilmark{3}}
\author{Leonidas~A.~Moustakas\altaffilmark{1}}
\author{Holland~Ford\altaffilmark{4}}
\email{coe@caltech.edu}

\altaffiltext{1}{
  Jet Propulsion Laboratory, California Institute of Technology, 
  4800 Oak Grove Dr, MS 169-327, Pasadena, CA 91109
}
\altaffiltext{2}{
  Instituto de Astrof\'isica de Andaluc\'ia (CSIC), 
  Camino Bajo de Hu\'etor 50, Granada 18008, Spain
}
\altaffiltext{3}{
  School of Physics and Astronomy,
  Tel Aviv University,
  Tel Aviv 69978, Israel
}
\altaffiltext{4}{
  Johns Hopkins University, 
  Dept.~of Physics \& Astronomy, 
  3400 N. Charles St., Baltimore, MD 21218, USA
}

\begin{abstract}
We present a strong lensing mass model of Abell 1689
which resolves substructures $\sim 25$ kpc across
(including about ten individual galaxy subhalos)
within the central $\sim 400$ kpc diameter.
We achieve this resolution by perfectly reproducing 
the observed (strongly lensed) input positions of 
\Nknotimages multiple images of \Nknotss knots residing within
\Nimages images of \Ngals galaxies.
Our model makes no assumptions about light tracing mass,
yet we reproduce the brightest visible structures
with some slight deviations.
A1689 remains one of the strongest known lenses on the sky,
with an Einstein radius of \NFWMRE\ for a lensed source at $z_s = 2$.
We find a single NFW or S\'ersic profile 
yields a good fit simultaneously (with only slight tension)
to both our strong lensing (SL) mass model
and published weak lensing (WL) measurements at larger radius (out to the virial radius).
According to this NFW fit, 
A1689 has a mass of \NFWfitMWMvir\ (\NFWfitMWM)
within the virial radius \NFWfitMWrvir\ (\NFWfitMWr),
and a central concentration \NFWfitMWcvirt\ (\NFWfitMWct).
Our SL model prefers slightly higher concentrations than previous SL models,
bringing our SL+WL constraints in line with other recent derivations.
Our results support those of previous studies
which find A1689 has either an anomalously large concentration
or significant extra mass along the line of sight (perhaps in part due to triaxiality).
If clusters are generally found to have higher concentrations 
than realized in simulations,
this could indicate they formed earlier,
perhaps as a result of early dark energy.



\end{abstract}


\section{Introduction}
\label{LPA1689:intro}

Galaxy clusters are the largest gravitationally bound objects yet to form in our universe
and prove interesting to study
both intrinsically and toward other ends.
Maps of their dark matter and baryons yield insights into structure formation
\citep{Umetsu09,Kawaharada10}
and can even constrain the nature of dark matter particles
\citep[specifically, their self-interacting cross-section;][]{Randall08,Feng09}.
Cosmological constraints can be obtained from 
number counts of clusters with measured masses \citep{Mantz09,Vikhlinin09c,Rozo10}
or from 
the observed gravitational lensing of more distant objects with measured redshifts
\citep{GilmoreNatarajan09}.
As the strongest gravitational lenses on our sky,
clusters can also be used as cosmic telescopes
allowing us to peer further back in time 
to reveal galaxies in the distant ($z \gtrsim 7$) universe
\citep{Bradley08,Richard08}
or study galaxies at ``modest'' redshifts ($z \sim 3$ to 4)
in greater detail \citep{Bunker00,Frye07}.

By mapping the dark matter distributions within galaxy clusters
(as projected on our sky),
gravitational lensing analyses have yielded
many exciting (and at times controversial) findings,
from the Bullet Cluster \citep{Markevitch04,Clowe06,Randall08}
to the ``dark matter ring'' in CL0024 \citep{Jee07}
and the ``cosmic train wreck'' in Abell 520 \citep{Mahdavi07}.
These results are exciting in part because they reveal the distribution of dark matter
{\it without assuming that light traces mass} (hereafter, ``LTM''\footnote{We will
avoid describing models as ``parametric'' or ``non-parametric'',
since strictly speaking all mass models do have parameters.
Here we introduce the more important distinction between
``LTM'' and ``non-LTM'' models.}).

Similarly exciting dark matter maps
can now be obtained in finer detail in galaxy cluster cores
thanks to strong gravitational lensing (hereafter, SL) analysis of high quality data.
The galaxy cluster Abell 1689 
is one of the strongest gravitational lenses on our sky\footnote{See
\cite{Zitrin09b} for the current strongest gravitational lens.}, 
with an Einstein radius of $R_E \sim 47\arcsec$ for a background object at $z_s = 2$.
Deep (20-orbit) multiband ACS GTO observations of this cluster
reveal over 100 strongly-lensed multiple images of 30 background galaxies
\citep{Broadhurst05}, 
a huge leap forward in the number of SL constraints available 
for any single massive body.
Based on these multiple images (and others identified since),
we derive a mass model which resolves substructure 
approximately 25 kpc ($8 \arcsec$) across
within the central $\sim 400$ kpc ($2 \arcmin$) diameter.
This is the highest resolution mass model to date of any galaxy cluster
without assuming LTM.
For comparison, weak lensing (hereafter, WL) mass maps
typically resolve structure on 
$\sim 1 \arcmin$ scales from ground-based imaging
or $\sim 45 \arcsec$ from space \citep{Heymans08}.
Of course WL complements SL nicely,
as WL probes cluster mass distributions to much larger radii 
(the virial radius $r_{vir} \gtrsim 2$ Mpc)
where the lensing strength is weaker.

\begin{deluxetable*}{llccccc}  
\tablecaption{\label{LPA1689:tabprev}Published strong lensing mass reconstructions of A1689 based on the ACS images}
\tablewidth{0pt}
\tablehead{
\colhead{}& 
\colhead{}& 
\colhead{}& 
\colhead{}& 
\colhead{}&
\multicolumn{2}{c}{Average residuals}\\
\colhead{Alias}  & 
\colhead{Paper}  & 
\colhead{Assume LTM}  & 
\colhead{Galaxies}  & 
\colhead{Multiple Images}&
\colhead{source plane}&
\colhead{image plane\tablenotemark{a}}
}
\startdata

B05&
\cite{Broadhurst05}& 
\checkmark& 
30&
106&
\nodata&
$3.2\arcsec$\\

&
\cite{Diego05}& 
$\times$& 
30&
106&
\nodata&
$3\arcsec$\\

Z06&
\cite{Zekser06}& 
\checkmark& 
22--30&
74--106&
$0.7\arcsec$--$1.2\arcsec$&
\nodata\\

&
\cite{Saha06}& 
$\times$& 
7,9&
28,30&
0&
0\\

H06&
\cite{Halkola06}& 
\checkmark& 
31+arc&
107&
\nodata&
$2.7\arcsec$\\

H07&
\cite{Halkola07}& 
\checkmark& 
31+arc&
107&
\nodata&
$2.5\arcsec$\\

L07&
\cite{Limousin07}& 
\checkmark& 
34&
113&
$0.45\arcsec$&
$2.87\arcsec$\\

&
\cite{JulloKneib09}& 
\checkmark$\times$\tablenotemark{b}& 
12&
28&
\nodata&
$0.28\arcsec$\\

LP&
This paper& 
$\times$& 
\Ngals&
\Nimages&
0&
0\\

\vspace{-0.1in}
\enddata
\tablenotetext{a}{Observational uncertainties are $\sim 0.05\arcsec$ in the image plane.}
\tablenotetext{b}{\cite{JulloKneib09} include both LTM and non-LTM substructure components.}
\end{deluxetable*}


Traditional SL analysis methods were not able to fully process
the large numbers of multiple images revealed in the ACS images of A1689.
LTM models produced by 
\citet[hereafter B05]{Broadhurst05},
\citet[hereafter Z06]{Zekser06}, 
\citet[hereafter H06]{Halkola06}, and
\citet[hereafter L07]{Limousin07}
all failed to reproduce the 100+ observed multiple image positions
by $\gtrsim 2.5\arcsec$ (see Table \ref{LPA1689:tabprev}).
This is roughly 50 times the observational uncertainties of one pixel or so
($\sim 0.05\arcsec$).
By failing to match the tight observational constraints,
these models discard a great deal of information available in the quality ACS images.

Despite their greater flexibility, 
non-LTM models were unable to reduce these residuals 
given all 100+ multiple images.
\cite{Diego05}, using SLAP\footnote{Strong Lensing Analysis Package}, 
allowed residuals similar to those in LTM models.
\cite{Saha06}, using PixeLens, 
produced non-LTM models which perfectly reproduce some of the data,
but computational limitations restrict PixeLens 
to fitting only 30 multiple images at a time.
Similarly, \cite{JulloKneib09}, 
using a hybrid scheme combining LTM and non-LTM substructure,
leave relatively small ($0.28\arcsec$) average residuals,
but only for a subset of 28 multiple images.

In this paper, we present non-LTM mass models
which perfectly\footnote{While the observed
image positions do have (small) observational uncertainties 
of a pixel ($0.05\arcsec$) or so,
our solutions do perfectly fit the data as input.}
reproduce the observed (lensed) input positions of 
\Nknotimages multiple images of \Nknotss knots residing within
\Nimages images of \Ngals galaxies.
The resolution of a reconstructed mass model
is given roughly by the spatial density of the multiple image constraints
\citep{Coe08,Coe09decadal}.
Thus our model which incorporates \Nimages multiple images
has $\sim 4\times$ greater spatial resolution
($\sim 2\times$ greater along each axis)
than the aforementioned models which incorporate only $\sim 30$ multiple images.

Our SL analysis method \citep[LensPerfect,][]{Coe08} 
was made possible just recently thanks to the development of mathematical tools
enabling one to obtain
curl-free interpolations of a vector field 
which perfectly reproduce the input data at the input positions
\citep{Fuselier06, Fuselier07}.
We make only minimal assumptions about the physicality of the mass distribution,
including no assumptions about LTM.

As the strongest known gravitational lens on our sky for some time,
A1689 has provided us with an excellent laboratory.
It was the first cluster to have its mass measured via both analyses of 
weak lensing \citep{Tyson90} and
magnification of the background galaxy population \citep{Taylor98}.
Yet before A1689 was observed with ACS,
no multiple images had been identified in the field
except for two fold arcs, our 8ab and 13abc \citep{Miralda-Escude95}.
These arcs were sufficient for a rudimentary strong lensing analysis to be performed
yielding an Einstein radius of $R_E \approx 45\arcsec$,
the largest of any known lens at the time.
With such a large Einstein radius, 
the ACS GTO team devoted 20 orbits to imaging the cluster,
confident that it would reveal many highly magnified background galaxies
and multiple image systems.

The observations (obtained in 2002 June) delivered,
and the 100+ multiple images still far surpass any other gravitational lens.
To date, other clusters observed to the same depth 
have yielded far fewer multiple images:
53, 35, and 33 from A1703, A2218, and CL0024, respectively
\citep{Richard09,Eliasdottir08,Zitrin09a}.
A1689's many SL constraints 
allow for detailed mass modeling,
spawning many publications
and helping to make A1689 one of the best studied galaxy clusters.

\begin{deluxetable*}{llll}  
\tablecaption{\label{LPA1689:tabcprev}Published concentration measurements of A1689}
\tablewidth{0pt}
\tablehead{
\colhead{$c_{200}$\tablenotemark{a}} &
\colhead{Alias}  & 
\colhead{Paper}  & 
\colhead{Constraints\tablenotemark{b}}
}
\startdata
6&
&
\cite{CloweSchneider01}&
WL\\

4.7&
&
\cite{King02}&
WL\\

5.7&
&
\cite{King02b}&
WL (infrared)\\

7.9&
&
\cite{Clowe03}&
WL\\

$3.5^{+0.5}_{-0.3}$ &
&
\cite{Bardeau05}&
WL\\

$5.2\pm0.3$ &
&
\cite{Bardeau07}&
WL\\

$6.5^{+1.9}_{-1.6}$&
B05&
\cite{Broadhurst05}&
SL\\

$10.8^{+1.2}_{-0.8}$ &
&
\cite{Broadhurst05b}&
SL+WL\\

$5.7^{+0.34}_{-0.5}$&
Z06&
\cite{Zekser06}&
SL\\

$6\pm0.5$&
H06&
\cite{Halkola06}&
SL\\

$7.6^{+0.3}_{-0.5}$&
H06&
\cite{Halkola06}&
SL+WL\\

$6.0\pm0.6$\tablenotemark{c}&
L07&
\cite{Limousin07}&
SL\\

$7.6\pm1.6$&
L07&
\cite{Limousin07}&
WL\\

\NFWfitMcval&
LP&
This work&
SL\\

\NFWfitMWcval&
&
This work&
SL+WL\\

$10.7^{+4.5}_{-2.7}$&
&
\cite{UmetsuBroadhurst08}&
WL+counts\\

$10.1^{+0.8}_{-0.7}$&
&
\cite{UmetsuBroadhurst08}&
SL+WL+counts\\

$12.5^{+3.9}_{-2.7}$&
&
\cite{Umetsu09}&
WL\\

$9.8^{+0.7}_{-0.6}$&
&
\cite{Umetsu09}&
SL+WL+counts\\

11.10&
&
\cite{Corless09}&
WL\\

$12.2 \pm 6.7$\tablenotemark{d}&
&
\cite{Corless09}&
WL + RE + priors\\

$7.7^{+1.7}_{-2.6}$&
&
\cite{AnderssonMadejski04}&
X-ray\\

$5.6$&
&
\cite{Riemer-Sorensen09}&
X-ray\\

$5.3^{+1.3}_{-1.2}$&
&
\cite{Peng09}&
X-ray (non-parametric)\\

$6.6\pm0.4$&
&
\cite{Peng09}&
X-ray (parametric)\\

$10.9\pm3.5$&
&
\cite{Fusco-Femiano09}&
X-ray\\

$9.7^{+0.7}_{-0.8}$&
&
\cite{Lemze08}&
SL+WL + X-ray\\

$>10.4$&
&
\cite{Lemze09}&
SL+WL+counts + X-ray + dynamical\\

$4.58\pm0.34$\tablenotemark{e}&
&
\cite{Morandi10}&
SL+WL + X-ray\\


%
\vspace{-0.1in}
\enddata
\tablecomments{For previous compilations of concentrations derived for A1689, see
\citet[Table A1]{ComerfordNatarajan07};
\citet[Table 4]{Limousin07};
\citet[Table 5]{UmetsuBroadhurst08};
\citet[Table 4]{Corless09}.}
\tablenotetext{a}{ Fits to spherical NFW profiles unless indicated otherwise.  Concentrations assuming an overdensity of 200 (converted from quoted values if necessary).}
\tablenotetext{b}{ SL = strong lensing; WL = weak lensing; RE = Einstein radius; counts = magnification depletion; X-ray; dynamical = cluster galaxy velocity measurements}
\tablenotetext{c}{ 3-$\sigma$ uncertainties}
\tablenotetext{d}{ Includes uncertainties due to halo triaxiality}
\tablenotetext{e}{ Fit to ellipsoidal gNFW profile (with variable central slope)}
\end{deluxetable*}


One result that stands out from both lensing and X-ray analyses of A1689
is that its mass appears to be more centrally concentrated 
than predicted by CDM simulations of structure formation.
Simulated dark matter halos have mass profiles
which are generally well described by 
NFW \citep{NFW96}
or Einasto / S\'ersic \citep{Navarro04} 
profiles \citep[e.g.,][]{Navarro10}.
More massive clusters (such as A1689) 
finished forming later when the universe was less dense overall,
and thus are found (at least in simulations) to be less centrally concentrated.
Yet where A1689 is expected \citep{Duffy08} to have a NFW concentration of \cDuffy,
it has been shown observationally (Table \ref{LPA1689:tabcprev})
to have a much higher concentration
$c_{200} \sim 7 - 11$ ($c_{vir} \sim 8 - 12$)\footnote{\cite{Morandi10} 
claim to account for A1689's high concentration
(along with the discrepancy between lensing and X-ray masses)
by fitting an 3-D ellipsoidal gNFW profile (with variable central slope)
simultaneously to the SL, WL, and X-ray data.
While the method is impressive, 
simulated clusters have yet to be analyzed in the same way
(they are normally fit to spherical NFW profiles),
so it is unclear that a direct comparison can be made.}.

As one of the strongest lenses on our sky, 
we might expect A1689's concentration to be on the high side.
However, even accounting for
triaxiality \citep{Oguri05,Corless09}
and selection (lensing) bias \citep{Hennawi07,OguriBlandford09,Meneghetti10},
A1689's high concentration still seems extremely unlikely \citep{BroadhurstBarkana08}.

Lensing analyses of larger samples of clusters 
seem to support the idea that clusters may have higher concentrations 
\citep{Broadhurst08,Oguri09,Sereno10}
and larger Einstein radii 
\citep{BroadhurstBarkana08,Richard10,Zitrin10}
than simulated clusters.
However only a small lensing-biased sample of these 
have been studied in sufficient detail.
We note that joint SL+WL fitting is required 
to constrain mass profiles well 
and concentrations to $\sim 10\%$ according to simulations
\citep{Meneghetti09}.
The addition of X-ray, SZ, and/or velocity dispersion data
can constrain the mass profiles further still \citep{Sand08,Newman09,Lemze09}.

More conclusive results are expected
from the ``CLASH''\footnote{Cluster Lensing And Supernova survey with Hubble,
\href{http://www.stsci.edu/~postman/CLASH/}{http://www.stsci.edu/$\sim$postman/CLASH/}}
Treasury Project,
a large (524 orbit) HST Multi-Cycle Treasury program (P.I. Postman)
to observe 25 X-ray-selected galaxy clusters at $0.18 < z < 0.9$,
each to a depth of 20 orbits, 
or the depth of the ACS GTO images of five clusters including A1689.
Combined with Subaru images and other data,
these observations should yield conclusive results (see \S \ref{LPA1689:future}).

Might baryons, lacking from the simulations discussed above,
be responsible for higher mass concentrations in nature
\citep[e.g.,][]{BarkanaLoeb09}?
Recent hydrodynamical simulations \citep{Duffy10, Mead10} show this is unlikely.
Baryons appear to increase cluster concentrations
only modestly at best ($\sim 10\%$),
though they more likely decrease concentrations slightly
(as found when strong AGN feedback is included in the simulations).

If real clusters in fact have higher concentrations than simulated clusters,
this could imply that clusters formed earlier in nature than in simulations.
One mechanism to explain such early growth
is a small but non-negligible amount of dark energy in the early universe, 
say $\Omega_{DE} \sim 0.10$ at $z = 6$
\citep{FedeliBartelmann07,SadehRephaeli08,Francis09,GrossiSpringel09}.
The additional dark energy actually {\it suppresses} formation of structure,
but this means that structures must have formed earlier 
to reach the abundances observed today.
Perhaps high cluster concentrations 
along with detections of massive halos at $z > 1$ 
\citep{Eisenhardt08,Jee09,Huang09,Papovich10,Schwope10}
are providing observational hints of such early dark energy (EDE).

We note that semi-analytic modeling of cluster formation 
(in a ``standard'' $\Lambda$CDM universe)
suggests that high concentrations ($c \sim 10$) may be fairly common in nature
and that cluster mass profiles may differ slightly 
from the NFW profiles found in simulations \citep{LapiCavaliere09}.

In this paper we concentrate on presenting our revised multiple image identifications,
our mass model, and mass profile fits to SL and WL data,
including measurements of the mass concentration.
In future work we will take greater advantage of our method's main strength:
the ability to map substructure without assuming LTM.

Our outline is as follows.
The observations, object detections, 12-band photometry, and photometric redshifts 
are described in \S \ref{LPA1689:BPZ}. 
In \S \ref{LPA1689:multimages} we discuss the observed multiple image systems,
including our additions and revisions to those previously identified.
In \S \ref{LPA1689:LP} we review our LensPerfect method
and discuss some minor improvements we have made to it.
Our mass maps are presented in \S \ref{LPA1689:results}
and the mass profile is analyzed in \S \ref{LPA1689:profanal}.
We discuss substructure and future work in \S \ref{LPA1689:future}
and summarize in \S\ref{LPA1689:summary}.

We use a concordance cosmology of 
($\Omega_m$, $\Omega_\Lambda$, $h$) = (0.3, 0.7, 0.7).
In this cosmology, $1\arcsec \approx 3.11$ ${\rm kpc} \hunits$ at A1689's redshift 
\citep[measured by][]{Frye07} 
of $z = 0.187$.
Though not always explicitly noted,
distances and masses are given in units of $\hunits$, 
where $H_0 = 70 h_{70}$ km/s/Mpc
and $h = 0.7 h_{70}$.
These values would be 30\% lower if quoted in units of $h^{-1}$ instead.





\section{Photometry and Photometric Redshifts}
\label{LPA1689:BPZ}

Multiband observations provide color information
which is absolutely essential to the identification of multiple images.
They also allow us to obtain photometric redshift estimates
where spectroscopic redshifts are unavailable.
Redshifts are essential to the mass model, 
as lensing deflections scale with redshift.

Abell 1689 is among the best studied galaxy clusters,
having been imaged in 12 broadband filters from the near-ultraviolet to near-infrared.
This allows us to obtain robust photometric redshifts, as we describe below.
By carefully modeling and subtracting the light from most of the cluster galaxies,
we recover many faint objects lost in their glare,
including demagnified central images
which allow us to constrain the mass model at small radius.

Much of the analysis described in this section
was performed for and utilized by B05.

\subsection{Observations}

Abell 1689 
(RA=$13^{h}11^{m}30\fs13$, Dec=$-01\arcdeg20\arcmin16\farcs2$ [J2000])
has been observed in four filters ($g\arcmin r\arcmin i\arcmin z\arcmin$) with ACS
and eight more ($UBVRIJHK_s$) from the ground.
Details are provided in Table \ref{LPA1689:obs}.

The ACS observations are among the deepest to date for any galaxy cluster.
In 2002 June, 20 orbits of HST ACS GTO time were used 
to obtain deep exposures in the
$g_{475}$, $r_{625}$, $i_{775}$, and $z_{850}$ passbands
(4, 4, 5, and 7 orbits, respectively).
More details about the ACS observations can be found in B05.




\begin{deluxetable*}{clccc}  
\tabletypesize{\scriptsize}
\tablecaption{\label{LPA1689:obs}Abell 1689 Observations}
\tablewidth{0pt}
\tablehead{
\colhead{Telescope/Camera/Detector}  & 
\colhead{Filter}  & 
\colhead{FWHM}&
\colhead{Area}&
\colhead{Magnitude Limit\tablenotemark{a}}
}
\startdata
ACS/WFC&
$g\arcmin$&
$0\farcs11$&
$11.8~\sq\arcmin$&
27.24\\
ACS/WFC&
$r\arcmin$&
$0\farcs10$&
$11.8~\sq\arcmin$&
27.00\\
ACS/WFC&
$i\arcmin$&
$0\farcs10$&
$11.8~\sq\arcmin$&
26.92\\
ACS/WFC&
$z\arcmin$&
$0\farcs11$&
$11.8~\sq\arcmin$&
26.50\\
DuPont Telescope @ Las Campanas&
$U$ Johnson&
$0\farcs93$&
$72.3~\sq\arcmin$&
\nodata\\
Nordic Optical Telescope (NOT) @ La Palma&
$B$ NOT&
$1\farcs13$&
$41.0~\sq\arcmin$&
\nodata\\
Keck II / LRIS&
$V$ LRIS&
$0\farcs81$&
$80.1~\sq\arcmin$&
27.17\\
Keck II / LRIS&
$R$ LRIS&
$0\farcs74$&
$93.1~\sq\arcmin$&
26.96\\
Keck II / LRIS&
$I$ LRIS&
$0\farcs68$&
$88.7~\sq\arcmin$&
26.41\\
ESO NTT / SOFI&
$J$ SOFI&
$0\farcs87$&
$50.4~\sq\arcmin$&
25.12\\
ESO NTT / SOFI&
$H$ Johnson&
$0\farcs99$&
$44.9~\sq\arcmin$&
24.22\\
ESO NTT / SOFI&
$K_s$ SOFI&
$0\farcs82$&
$50.4~\sq\arcmin$&
24.06\\
\vspace{-0.1in}
\enddata
\tablenotetext{a}{AB, 5-$\sigma$ within a $0.8\arcsec$ diameter aperture 
(except for ACS, which are 10-$\sigma$, $0.2\arcsec$ diameter aperture)}
\end{deluxetable*}



\subsection{Galaxy Detection}

Objects were detected in an ACS 
$g\arcmin$ + $r\arcmin$ + $i\arcmin$ + $z\arcmin$ detection image 
(with each image normalized to its background RMS)
using SExtractor \citep{SExtractor}.
Our first detection was conservative
({\tt DEBLEND\_MINAREA = 5}, {\tt DETECT\_THRESH = 5})
and specially designed to properly deblend the cluster galaxies
({\tt DEBLEND\_NTHRESH = 32}, {\tt DEBLEND\_MINCONT = 0.005}).

Cluster members were identified by their colors,
and their light was carefully modeled and subtracted
from the images (Zekser et al.,~in prep.).
This aids greatly in the detection of background objects,
especially demagnified multiple images 
located behind the main cluster galaxies.
These central images provide valuable constraints to our lensing mass model.
This galaxy subtraction also improves our photometry measurements
and thus photometric redshifts 
for any background galaxy images 
strongly bathed in the light of a nearby cluster galaxy.

We are now able to re-run SExtractor
and detect many galaxies revealed by the subtraction of the cluster galaxies.
But the detection and object segmentation 
(the art of assigning each pixel to a given object)
are still not perfect.
We inspect all of the object segmentations and edit them ``by hand''
where necessary.
We also add a few multiple images which are predicted by our mass model
and are visible in the images but managed to escape detection,
either due to a bright neighbor or otherwise.
Using the SExSeg package \citep{Coe06}, 
we are able to force our revised object definitions into SExtractor
for photometric analysis.

\subsection{Photometry}

With observations obtained from a wide range of telescopes,
both from space and from the ground,
care must be taken to obtain 
robust PSF-corrected aperture-matched photometry.
This proves especially crucial for the faint lensed background galaxies
we are most interested in.
Thus we use the software package ColorPro,
which we developed and applied previously to obtain robust photometry of galaxies
in the UDF \citep{Coe06}.
As in our UDF analysis, 
we also recalibrate the photometric zeropoints of several of our images.
This procedure is described below.

\subsection{Photometric Redshifts}

Photometric redshifts were obtained for the objects in our Abell 1689 catalog
using an updated version 1.99.2 of the Bayesian Photometric Redshift software BPZ
\citep{Benitez00}.
This version features the recalibrated 
CWW+SB SED (spectral energy distribution) templates
introduced in \citet{Benitez04}
plus the two younger starburst templates added in \citet{Coe06}.

In our UDF analysis \citep{Coe06}, 
we reported a photometric redshift accuracy of $\Delta z = 0.04 (1 + z_{spec})$.
Here we expect to achieve similar, or perhaps slightly worse, precision.
The A1689 exposures are not as deep as the UDF,
although in regions of high magnification ($\mu \gtrsim 6$; $\Delta m \gtrsim 2$)
we can actually detect fainter galaxies than possible in the UDF.
The images are very crowded with cluster galaxies 
whose light may contaminate that of neighbors
(despite our best efforts to model and subtract this light)
affecting the photometry and thus photometric redshifts.

Abell 1689 has been the target of several spectroscopic redshift campaigns,
including
\cite{Teague90, Fort97, Balogh02, Duc02, Frye02, Frye07}.
Of the spectroscopic redshifts published in these works,
113 lie within the ACS FOV.
These are presented in Table \ref{LPA1689:zspecA}.
Of these objects, three are stars, another (\#172) is half outside the ACS FOV,
and for another (\#194) it was unclear which object was being referenced
by the published coordinates.
What remain are 108 galaxies.
In our first attempt to obtain photometric redshifts for these galaxies,
it was clear that our SED templates provided a poor fit
for the observed $U$ and $B$ magnitudes.

To reveal flux miscalibrations in any of the filters,
we apply techniques similar to that used in our study of the UDF \citep{Coe06}.
We select cluster ellipticals based on their spectroscopic redshift,
BPZ spectral type fit, and visual confirmation in the ACS images.
We then use BPZ to fit SEDs to the photometry of these objects,
fixing the redshifts at the spectroscopically-determined values.
We find the following offsets between the observed and predicted magnitudes:
$V$ -0.08,
$R$ -0.10,
$I$ -0.00,
$J$ -0.08,
$H$ -0.06,
$K_s$ +0.01,
$g\arcmin$ -0.15,
$r\arcmin$ +0.03,
$i\arcmin$ -0.04,
$z\arcmin$ +0.02.
We subtract these biases from our measured magnitudes.
Our $U$ and $B$-band images did not arrive well calibrated,
so we used this procedure to calibrate them as well.

\begin{figure}
\plotone{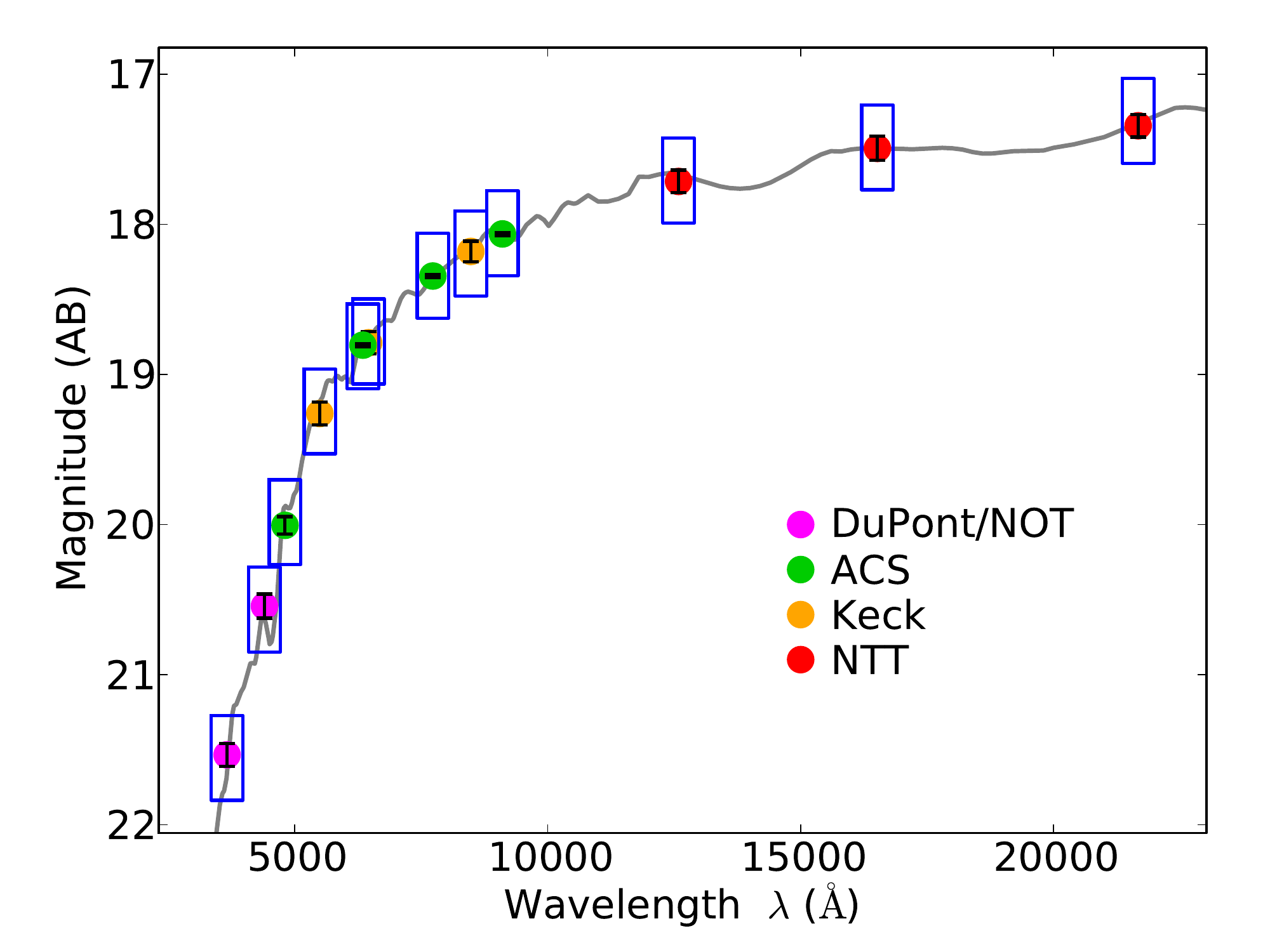}
\caption[Good BPZ SED fit]{
\label{LPA1689:sedfit}
Example of an excellent
BPZ SED fit to the photometry of a cluster elliptical (\#198),
after flux recalibrations of all filters (see text).
The photometric redshift is correctly derived as 
$z = 0.18 \pm 0.12$ (95\% confidence).
The colored circles give the observed AB magnitudes with uncertainties
in 12 filters,
while the blue rectangles give the SED model predicted magnitudes
(with rough uncertainties) in those filters.
The colors serve to distinguish the telescopes / filters:
{\it purple} -- Las Campanas, La Palma ($UB$); 
{\it green}  -- ACS ($g\arcmin r\arcmin i\arcmin z\arcmin$);
{\it orange} -- Keck/LRIS ($VRI$);
{\it red}    -- ESO NTT/SOFI ($JHK_s$).
}
\end{figure}

\begin{figure}
\plotone{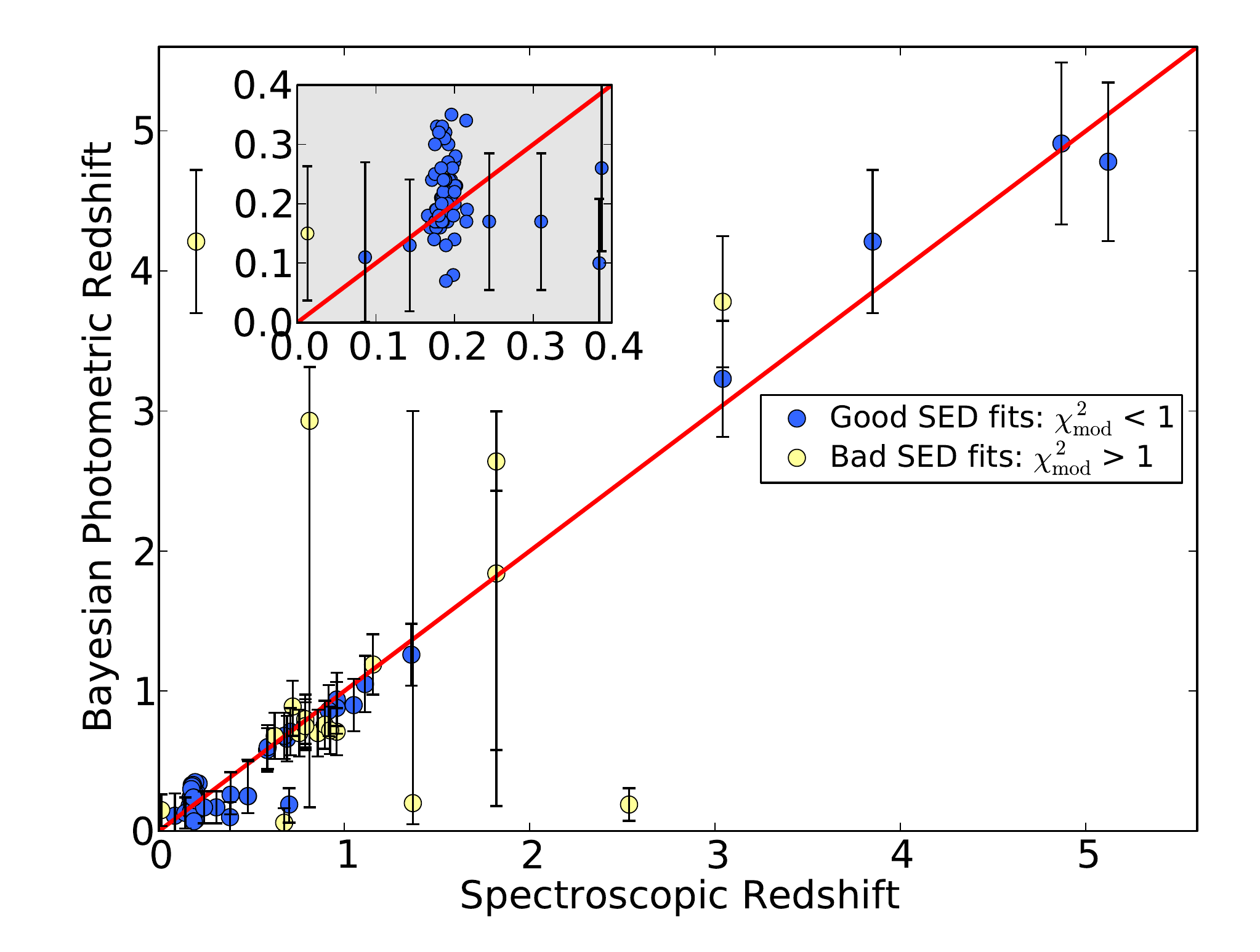}
\caption[Photometric vs.~spectroscopic redshifts]{
\label{LPA1689:bpzvsspecz}
Bayesian photometric redshifts plotted versus
108 spectroscopic redshifts obtained within the ACS FOV 
(Table \ref{LPA1689:zspecA}).
Good BPZ SED fits ($\chi^2_{mod} < 1$) are plotted in blue,
while poorer fits to the photometry are plotted in yellow.
For the 89 galaxies with good SED fits,
the photometric redshifts agree with the spectroscopic values
to within $\Delta z = 0.07 (1 + z_{spec})$.
If we recursively remove outliers with more than three times 
the RMS deviation $\Delta z$ \citep[as in][]{Coe06},
two galaxies are pruned and the rest agree to within
$\Delta z = 0.06 (1 + z_{spec})$.
}
\end{figure}

Given our recalibrated fluxes, we rerun BPZ on all our galaxies
without constraining the redshifts to the correct values as above.
We measure of goodness of fit $\chi^2_{mod}$ between observed and model fluxes
with some uncertainty assigned to the model fluxes \citep[see][]{Coe06}.
An example of an excellent ($\chi^2_{mod} = 0.03$)
and correct ($\Delta z = 0.01$) SED fit to a cluster elliptical
obtained with recalibrated fluxes is shown in Fig.~\ref{LPA1689:sedfit}.
Of the 108 galaxies with spectroscopic redshifts in the ACS FOV,
we obtained good SED fits ($\chi^2_{mod} < 1$) for 89.
For these 89 objects,
our Bayesian photometric redshifts agree with the spectroscopic redshifts
to within $\Delta z = 0.07 (1 + z_{spec})$
(see Fig.~\ref{LPA1689:bpzvsspecz}).
And if we recursively remove outliers 
with more than three times this deviation \citep[as in][]{Coe06},
then two galaxies are pruned
and the remaining 87 agree to within $\Delta z = 0.06 (1 + z_{spec})$.
The two outliers are hardly catastrophic,
having $\Delta z = 0.30 (1 + z_{spec})$ and $\Delta z = 0.21 (1 + z_{spec})$.

Finally, we obtain photometry and photometric redshifts
for our multiple images.
The results are given in Table \ref{LPA1689:tabmultimages}.
For each redshift, 
a 95\% confidence interval is quoted as well as $\chi^2_{mod}$.
Note that in B05 the BPZ prior was modified.
The redshift range $z < 0.7$ was excluded
and the prior allowed for observed fluxes up to $20\times$ magnified.
We have not implemented this modified prior here.

Note that these redshifts were obtained using photometry from all 12 filters.
We have also obtained photometric redshifts
based on photometry obtained in the four ACS images
with the light of cluster galaxies modeled and subtracted.
We are unable to apply this galaxy subtraction in a consistent and robust manner
to the ground-based images due to their significantly worse PSFs.
Yet even with four filters, we obtain improved photometric redshifts for those objects
whose light is significantly contaminated by nearby cluster galaxies.



\section{Multiple Images}
\label{LPA1689:multimages}

Building on previous work,
we present a catalog of \Nimages images of \Ngals background galaxies.
We contribute 20 new candidate multiple images of 8 galaxies,
along with a few tweaks to previous identifications.
We discard three central demagnified images in use since B05,
finding these identifications suspect;
our technique is more sensitive 
to the positions of central images than other techniques 
(see \S\ref{LPA1689:centralimages}).
We add one new central image identification.

\begin{figure*}
\plotone{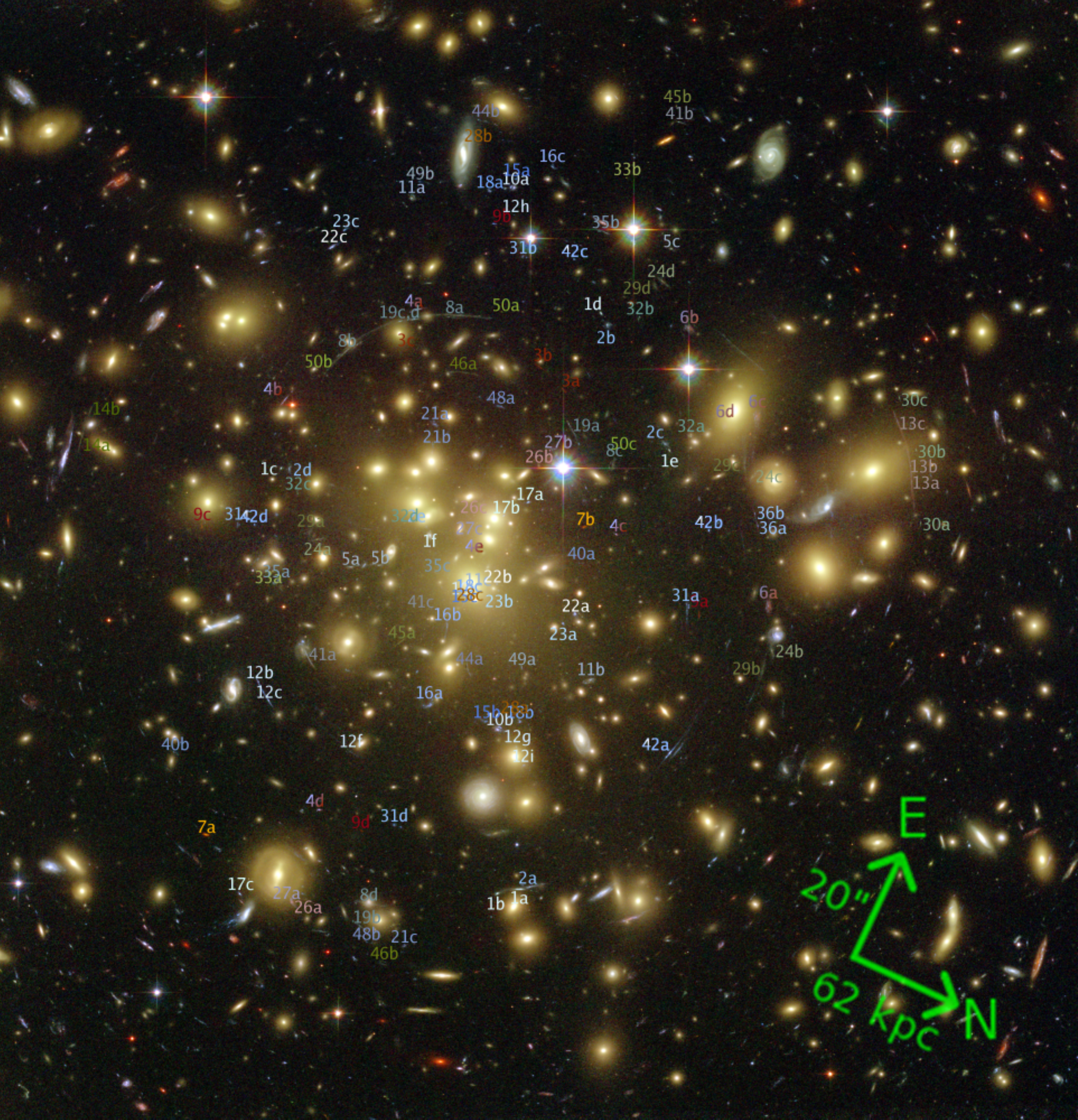}
\caption[Multiple Images in Abell 1689]{\label{LPA1689:labeled}%
Multiple images in Abell 1689 labeled on the 
STScI $3.2\arcmin \times 3.3\arcmin$
ACS $g\arcmin r\arcmin z\arcmin$ color image.
Most of the labels are directly above the images, 
but some have been offset slightly to minimize clutter.
The exact coordinates of each image are given 
in Table \ref{LPA1689:tabmultimages}.
A compass provides the orientation (a $115^\circ$ roll angle),
while the lengths of the arrows provide the scale
($20\arcsec \approx 62$ kpc).
}\end{figure*}

We also identify multiple knots in \Ngals of the galaxy images
which we use as additional constraints.
Constraining the positions of three non-collinear knots in an image
is equivalent to
constraining the precisely-measured shear and magnification of that image.
Our mass map solutions perfectly reproduce the observed positions of 
\Nknotimages multiple images of \Nknotss knots residing within
\Nimages images of \Ngals galaxies.
The final set of multiple images used in this work
is given in Table \ref{LPA1689:tabmultimages} 
and shown in Fig.~\ref{LPA1689:labeled}.
Close-ups of all the images are shown in Fig.~\ref{LPA1689:images}.
Additional knots are visible in these images and listed in Table \ref{LPA1689:tabknots}.
In this section we discuss the multiple image identification
as well as the redshifts for these systems.

\subsection{Multiple Image Identification}

The original analysis of the ACS A1689 images (B05)
yielded 106 multiple images of 30 background galaxies.
This was a truly pioneering effort
as the first multiple image identifications were the most difficult.
The relatively steep mass profile of A1689 near the Einstein radius 
produces relatively thin arcs.
Thus there are no truly obvious systems with thick multiple images as in, say, CL0024
\citep{Tyson98,Zitrin09a}.
Only after careful study of the color image
did B05 discover the image system 1-2,
a pair of pale green and blue specks that repeats five times about the image,
leaving no doubt as to its identification.
Once these first multiple images are identified,
an initial mass model may be obtained,
greatly facilitating the identification of further image systems.

B05 identified many image systems, but they did not attempt to identify all.
Additional systems have since been proposed, and we propose still more in this paper.
Of the 30 image systems proposed by B05,
doubts have been since been raised about three.
We further call into question three of their central image identifications.

Z06 (their \S 8.4) 
experimented with excluding some image systems from their modeling
on the grounds that they yielded larger than average scatter in the delensed positions.
But they stopped short of calling any of B05's systems into question,
as large scatter may also result from shortcomings of the mass model.
Among the problem systems they cited were systems 1-2 and 15
which have ironclad identifications as multiple images from visual inspection.
Images near critical curves such as these 
often prove difficult to fit for conventional modeling methods,
as the predicted image positions 
are unstable with respect to small variations in the mass model.
For this reason, 
L07 also found it necessary to remove the giant arc (system 8)
from their modeling.

Some more systematic changes to the B05 catalog were made by H06 and L07.
They discarded and altered a few systems 
which appear to have been misidentified in the original B05 analysis.
H06 (see discussion in their \S 3.3) split B05's system 12 in two: 
a new streamlined system 12 comprised of just 12b and 12c;
and a new 4-image system 13
comprised of 12a, 12d, and two new counter images.
(Note this offsets their numbering relative to B05 for B05's systems 13--19.)
Splitting 12a and 12b into different systems was a bold proposition,
given that they were known to have the same spectroscopic redshift
of $z = 1.83$.
But L07 concurred with this assessment (see their \S 4.2), 
and so do we. 
We adopt L07's convention of assigning the number 31 to H06's system 13,
thus preserving the B05 numbering system.

B05 may have misidentified the counterimages of the fold arc 12bc,
but this arc almost certainly has other counterimages.
We identified three counter images 12fgh
plus an image 12i which is a counter image of 12g
due to strong galaxy-galaxy lensing.
(Note these designations, 12fghi, were intended to avoid confusion
with B05's original identifications of images 12abe.
We did not intend to suggest this is a 9-image system.
There are but 6 images in our new system 12.)

H06 also discarded B05's 3-image system 20
located in the top-left corner of the image at large radius.
Not only do the images fit poorly with the mass model,
but the images show slightly different morphologies.
L07 concurred, adding that the spectra are also somewhat different.
We did not attempt to add this system to our mass model.


L07 also took issue with B05's system 25,
finding a different counterimage to 25b 
and naming the new system 33.
We concur, finding the new system a much better fit.
As for B05's 25a (a greenish radial arc),
we identify a new counterimage candidate at fairly large radius,
and we rename the system 45.


H06 identified one brand-new system (their 31),
a small pair of blue images along the fold inside the main subclump.
L07 ``rediscovered'' this system, naming it 36.
L07 also propose the following brand-new systems 
with numbers of images given in parentheses:
32 (4), 35 (3), 36 (2), 40 (2).
We include all of these in our mass model.

We have proceeded to identify 8 new candidate systems
containing a total of 20 multiple images.
We have named these (with number of images given):
41 (3), 42 (4), 44 (2), 45 (2), 46 (2), 48 (2), 49 (2), 50 (3).
As mentioned above, we recycled B05's image 25a for use in our system 45,
but the other 19 multiple images are brand new identifications.

We do not have spectroscopic redshifts for our new systems;
thus our two-image systems do not currently provide strong constraints.
In the course of our mass modeling, 
we add our new systems at the end
and find that our two-image systems
have little effect on the mass map.

We also mention in passing H06's ``system'' 32,
a single large but low surface brightness arc
just outside the image pair 6cd in the main subclump.
They were unable to identify a counterimage for the arc and neither were we.
We do not utilize this arc in our modeling.

\subsection{Demagnified Central Images}
\label{LPA1689:centralimages}

The identification of central images is
crucial to constraining the inner mass profile of any lens.
This is generally a difficult task 
both because central images are strongly demagnified
and because this faint light is overwhelmed by the bright
galaxy or galaxies which make up the lens.
We have carefully modeled and subtracted most of the cluster galaxies
from each of the $g\arcmin$, $r\arcmin$, $i\arcmin$, and $z\arcmin$
ACS images (Zekser et al.,~in prep.).
The recombined galaxy-subtracted color image
reveals many colored specks in the central region of the cluster.
A great number of these specks are globular clusters belonging to A1689
\citep{Mieske04,Mieske05}.
The rest may be demagnified multiple images.
Aside from sorting out this confusion,
we also have to contend with residuals which persist
from the galaxy subtraction.
Correct identification of a central multiple image
relies on finding the proper color speck 
(or group of specks for paired image systems)
at approximately the location predicted by the model.
Often we find several specks of approximately the correct color 
in approximately the predicted location.
Thus it should not be surprising 
when we claim that a few mistakes may have been made previously.
Conventional mass modeling has allowed these mistakes to go unnoticed
by B05 and in subsequent studies.

As mentioned above,
conventional modeling methods may be very sensitive to the positions of images
in regions of high magnification.
Conversely, they generally will not be very sensitive 
to the exact positions of demagnified images.
Our model-predicted positions for the central images 7c, 8e, and 19e
are ``only'' offset by $\sim 2\farcs5$, $7\arcsec$, and $2\farcs5$ 
from the B05 positions, respectively.
Thus these offsets may be easily missed
by routines that minimize offsets in the image plane.
(Note that attempts may be made
to normalize the offsets by the local magnification,
but this is often not attempted
as it can lead to instability in the optimization routine.)
If optimization is instead performed in the source plane
(technically inferior but much quicker),
the method might actually be more sensitive to the offsets of central images.
L07 do find higher than average ($0\farcs4$) offsets in the source plane
for systems 7 ($1\farcs09$) and 19 ($0\farcs48$).
And as mentioned above, L07 found such large errors for system 8,
that they excluded it from their analysis.
The incorrect identification of 8e, $7\arcsec$ from the position we predict,
may have contributed to their errors
as much as the unstable model positions of 8ab
(near the critical curve for that redshift).
The high offset for the 3-image system 7
may have raised alarms in L07's analysis.
But they find similarly large offsets for the pair of systems 26-27,
and we find no quarrel with this pair.
The 26-27 pair does, however, 
strongly require asymmetry in the center of the mass map,
which had not been observed by previous authors.
In conventional mass modeling,
it is difficult to say when a larger than average offset is a misidentification
and when it may simply indicate a shortcoming of the mass model.

LensPerfect, on the other hand, 
is extremely sensitive to the relative positions
of all of the central images.
For example, a central image incorrectly identified
to the wrong side of another central image
is generally disastrous for the mass model,
causing the deflection field to get tangled in itself.

We reexamined all of B05's central image identifications.
The pair 1f-2e is confidently identified by
the images' colors and proximity to one another.
As for the rest 
(4e, 7c, 8e, 19e, 10-15-18c, 11c, 22-23b, 26-27c, 32d, 35c)\footnote{Dashes
link objects (e.g., 10, 15, and 18) assumed to be in physical groups
with redshifts constrained to be equal in our model.
We are {\it not} referring to objects 10 through 18, inclusive.},
we purged them all from the image list
and obtained a mass model solution without them.
We then re-added the multiple images one by one to our model.
We found that a few central images (7c, 8e, 19e)
did not fit well with the rest,
producing aphysical models when added
(even after all of the source positions were allowed
to shuffle to new positions in search of a physical model).
We are unable to securely identify 
replacement central images for systems 7, 8, and 19,
as there are too many similarly colored specks in the area 
which confuse the issue.

Red demagnified central images stand out more and are easier to identify.
We identify one new central image candidate 28c
which we incorporate into our mass model.
It fits easily into our model, not significantly affecting it.

\subsection{Additional Knots}
\label{LPA1689:knots}

With LensPerfect we obtain mass map solutions
which delens the input centroids of all images of a given system 
to the exact same position in the source plane.
But this alignment does not guarantee 
that the delensed images will have the same shape or orientation.
Thus we identify additional knots where possible in the multiple images.
These knots are labeled in Fig.~\ref{LPA1689:images}.
If the delensed positions of these knots do not align well naturally
given our mass model,
then we add them as constraints and force them to align.
In Table \ref{LPA1689:tabknots} we give the positions of these additional knots 
which we have constrained in our mass model
(in addition to the centroids / primary knots listed in Table 
\ref{LPA1689:tabmultimages}).
These knots are also labeled in green in Fig.~\ref{LPA1689:delensed}.
In all, our model incorporates 
\Nknotimages observed (lensed) positions of \Nknotss knots.

These additional constraints further improve the accuracy of our mass map
(assuming our multiple image identifications are robust, as we believe they are).
Constraining three non-collinear knots in an image 
effectively uses both the observed shear and relative magnification of that image.
In \cite{Coe08} we compared mass maps of MS1358
given a single multiply-imaged galaxy
both with and without additional knots constrained.
The additional knots add significant detail to the mass map.

\subsection{Redshifts}
\label{LPA1689:redshifts}

In the original B05 analysis,
spectroscopic redshifts were available for 5 systems:
1ad 3.04, 2a 2.54, 7a 4.87, 10a 1.37, and 12ab 1.82.
Since then, system 12 has been split in 2 by H06 (our 12 and 31)
and the redshift of 10a has been called into question by L07,
with the net result being that 
we still had 5 systems with available redshifts.
Our initial analyses made use of these spectroscopic redshifts alone.

L07 have since contributed spectroscopic redshifts for another 19 systems 
(4a, 5ac, 6ab, 10-15-18a, 11a, 14a, 17c, 19d, 22a, 24c, 29d, 30c, 
32b, 33a, 35a, 36a, 40a),
bringing the total to 24 systems with spectroscopic redshifts 
(Table \ref{LPA1689:tabmultimages}).
We have compared our results before and after 
incorporating these redshifts into our mass modeling.
We find the substructure shifts somewhat, 
but overall the mass models appear to be very similar qualitatively.
Our ``optimized'' redshifts (described below)
were generally close to the spectroscopic redshifts,
with a scatter of $\Delta z \approx 0.06 (1 + z_s)$ after pruning one outlier.

For those systems without spectroscopic redshifts, 
we use photometric redshifts (\S \ref{LPA1689:BPZ})
as initial guesses in our optimization routine (\S \ref{LPA1689:optimization}).
We allow these redshifts to wander but they incur a penalty for doing so.
A rough uncertainty ($\Delta z = 0.5, 0.7, 1.0, 2.0$)
is assigned to each redshift according to our relative confidence in it,
and deviations from the input values are divided by these uncertainties.
Given these normalized deviations, we then take the RMS
and add this to our penalty evaluation.
As we are modeling 43 image systems, 
the total RMS will not be very sensitive to single outliers.
This approach will allow individual redshifts to wander unacceptably far.
In order to ``leave no redshift behind,''
we also find the maximum normalized deviation
and add this to our penalty function.
Our input redshifts and uncertainties
along with final optimized redshifts for each system
are given in Table \ref{LPA1689:taborderz}.

We could attempt to penalize redshift deviations more scientifically
by using the redshift probability distributions $P(z)$
returned from BPZ for each object.
However, photo-z uncertainties are often underestimated
by current methods including BPZ \citep{Hildebrandt08}.
Thus we prefer to assign simple and rather generous uncertainties to the redshifts
and allow them to naturally obtain their optimal values based on the mass model.

Some sets of images appear to be physically linked:
10-15-18, 22-23, 24-29, 26-27.
During our optimization procedure, we find that within each of these sets,
all of the redshifts gravitate toward common values
(10-15-18:~2.00-2.14-1.96; 26-27:~1.98; 24-29:~1.91, 22-23:~1.4, 1.46).
Thus we take the liberty of fixing all of the redshifts to be equal 
within each set.
For example, when the redshift of object 10 is optimized, 
the redshifts of objects 15 and 18 are forced to follow.
We believe these systems to be physically linked
but any one of them might instead be a chance alignment.
Object 15, for example, 
did obtain a slightly higher redshift than 10 and 18 when left free
as noted above.

Objects 1-2 also appear to be physically linked,
but our models say otherwise.
The objects do have different spectroscopic redshifts,
but that obtained for object 2 is somewhat uncertain,
as its spectra appears to show absorption from three separate systems
along the line of sight at redshifts 2.53, 2.87, and 3.04 \citep{Frye07}.
We tried fixing the redshift of object 1 to 3.04
and giving object 2 an initial guess of the same redshift ($z = 3.04$)
but allowing it to wander (incurring a small penalty for doing so).
A lower redshift of $\sim 2.5$ was clearly preferred for system 2,
thus supporting the spectroscopic redshift of 2.53 preferred by \cite{Frye07}.

\subsection{Coordinate System}
\label{LPA1689:coordsys}

The final catalog of multiple images used in our modeling
is given in Table \ref{LPA1689:tabmultimages}.
We provide coordinates in both (RA, Dec) and in ($x$, $y$).
Our ($x$, $y$) coordinate system is based on the original APSIS \citep{APSIS}
ACS GTO pipeline reductions.
These images are $4421 \times 4525$ pixels
(1 pixel = $0.05 \arcsec$).
Our bottom-left pixel is centered at (1,1) as in SExtractor and ds9.
North and East are $115^\circ$ clockwise from up and left, respectively.

Based on the central $3853 \times 4000$ area of these images,
STScI released a $g\arcmin r\arcmin z\arcmin$ color image.\footnote{\href{http://hubblesite.org/newscenter/archive/releases/2003/01/}{http://hubblesite.org/newscenter/archive/releases/2003/01/}}
An offset of (350, 232) may be subtracted from our coordinates
to obtain coordinates in the color image.
The ACS images were later reprocessed by APSIS
yielding images $4379 \times 4481$ pixels.
The improvements included 
better correction for the geometrical distortion of the images.
Nevertheless, the offset in coordinates between 
the original and subsequent processed images
is a nearly constant (21, 22) across the entire image,
only deviating by a pixel in $x$
in the top-left and bottom-right corners.



\section{LensPerfect Mass Map Reconstruction}
\label{LPA1689:LP}
\label{LPA1689:optimization}

LensPerfect is a novel approach to gravitational lens mass map reconstruction.
The 100+ strong lensing features produced by Abell 1689
present us with a large puzzle.
We must produce a mass model of A1689 
with the correct amounts of mass in all the right places 
to deflect light from 30+ background galaxies into multiple paths 
such that they arrive at the 100+ positions observed.

Most strong lensing analysis methods construct many possible models
and then iterate to find that which best matches the data.
LensPerfect instead uses direct matrix inversion 
to find perfect solutions to the input data.
Using LensPerfect, we may, for the first time, obtain a mass map solution
which perfectly\footnote{Again, while the image positions
do have (small) observational uncertainties of a pixel ($0.05\arcsec$) or so,
our solutions do perfectly fit the data as input.
We may vary the image positions within this uncertainty 
to produce other valid solutions,
however this is not a significant uncertainty in our models.}
reproduces the input positions of all 100+ multiple images
observed in Abell 1689.

LensPerfect makes no assumptions about light tracing mass.
Non-LTM models are common in analyses of WL or combined SL+WL
\citep[e.g.,][]{Bradac06,Diego07,Deb10}.
Dedicated SL analysis methods are able to process greater numbers of multiple images. 
Non-LTM SL analysis methods include
PixeLens \citep{SahaWilliams04,Coles08},
SLAP \citep{Diego05a,Diego05},
and methods developed by 
\cite{Liesenborgs06,Liesenborgs09} and \cite{JulloKneib09}.
(The latter includes both LTM and non-LTM components.)
We note non-LTM methods are also used in SL modelling 
of extended images lensed by individual galaxies \citep[e.g.,][]{Vegetti09}.

LensPerfect was made possible by a recent advance in the field of mathematics
\citep{Fuselier06, Fuselier07}.
The method was described in detail in \cite{Coe08}.
Here we provide a brief outline of the procedure.

Image deflection by a gravitational lens is governed by 
a few simple equations \citep[e.g.,][]{Wambsganss98}.
Given the bend angle $\vec \alpha$ of light due to mass predicted by \cite{Einstein16},
we can derive the deflection of light due to a mass sheet
with surface density $\kappa$ 
as a function of position $\vec \theta$ in the lens / image plane:
\\
\begin{equation}
  \label{LPA1689:deflkappa}
  \vec \alpha (\vec \theta) = \frac{1}{\pi} \int d^2 \vec \theta \arcmin
  \kappa(\vec \theta \arcmin)
  \frac{\vec \theta - \vec \theta \arcmin}
  {\left \vert \vec \theta - \vec \theta \arcmin \right \vert ^2},
\end{equation}
\\
with the simple corresponding inverse relation:
\\
\begin{equation}
\label{LPA1689:div}
\nabla \cdot \vec \alpha = 2 \kappa.
\end{equation}
The surface density $\kappa = \Sigma / \Sigma_{crit}$ 
is defined in units of the critical density at the epoch of the lens.
The critical density is that generally required 
for multiple images to be produced.
It is a function of source redshift as given by:
\\
\begin{equation}
\label{LPA1689:E_c}
\Sigma_{crit} = \frac{c}{4 \pi G} \frac{D_S}{D_L D_{LS}},
\end{equation}
\\
involving a ratio of the angular-diameter distances from
observer to source $D_S = D_A(0,z_S)$,
observer to lens $D_L = D_A(0,z_L)$,
and lens to source $D_{LS} = D_A(z_L,z_S)$.
For a flat universe ($\Omega = \Omega_m + \Omega_\Lambda = 1$), 
angular-diameter distances are calculated as follows 
\citep[filled beam approximation; see also \citealt{Hoggcosmo}]{Fukugita92}:
\\
\begin{equation}
  D_A(z_1,z_2) = \frac{c}{1 + z_2} \int_{z_1}^{z_2} \frac{dz^\prime}{H(z^\prime)},
\end{equation}
\\
where the Hubble parameter varies with redshift as:
\\
\begin{equation}
  H(z) = H_0 \sqrt{\Omega_m (1 + z)^3 + \Omega_\Lambda}.
\end{equation}
\\
(Formulae for non-flat cosmologies can be found in 
\citealt{CoeMoustakas09}, for example.)

From simple geometry we find that
the deflection angle $\vec \alpha$
(and thus the critical density $\Sigma_{crit}$)
is a function of redshift.
The deflection is greatest ($\vec \alpha_\infty$)
for a source at infinite redshift.
For sources at less than infinite redshift,
this deflection is reduced by the distance ratio:
\\
\begin{equation}
\label{LPA1689:deflscale}
\vec \alpha = \left( \frac{D_{LS}}{D_S}  \right) \vec \alpha_\infty.
\end{equation}
\indent Thus the problem of mass map reconstruction
can be reduced to determining the deflection field
with all deflections scaled to a common redshift (e.g., $\vec\alpha_\infty$),
at which point we simply take the divergence 
and divide by 2 to obtain the mass map (Eq.~\ref{LPA1689:div}).
The deflection field
$\vec \alpha(\vec \theta) = \vec \theta - \vec \beta$
may be measured at the multiple image positions $\vec \theta$
once source positions $\vec \beta$ are determined.
However, in order to take its divergence,
the deflection field must be solved for
as a continuous function of position
(or at least defined on a regular grid).
Our interpolated deflection field must also be curl-free 
\citep[see e.g.,][]{Coe08}.

Only recently were the mathematical tools developed
that enable us to obtain a curl-free interpolation of a vector field
\citep{Fuselier06, Fuselier07}.
The technique uses direct matrix inversion to obtain a solution 
which exactly matches the vectors at the given data points
and interpolates or extrapolates it elsewhere.
The solution is composed of radial basis function (RBFs)
each placed at the position of an observed multiple image.
Each basis function has two free parameters (amplitude and rotation)
{\it equal to the number of constraints} (the $x$ and $y$ coordinates of the image).
While some non-LTM methods have many more free parameters than constraints,
ours does not.

In gravitational lensing, 
our vector (deflection) field is not defined 
until we assume source positions $\vec \beta$ for our lensed galaxies.
As we add each galaxy to our model,
we can obtain a good initial guess for each $\vec \beta$.
We then perturb all of our source positions.
Each arrangement of source positions yields a new mass map.
Iterating over various arrangements of source positions, 
we find a range of mass maps
all of which perfectly reproduce the observed image positions.
Among these, we select the ``most physical'' mass map
using a set of non-restrictive criteria.
Aside from the requirement that the mass map be positive,
these criteria \citep[described in detail in][]{Coe08}
promote mass maps which 
are smooth, 
decrease outward from the center on average,
and are azimuthally symmetric 
(small scatter in radial bins, with extra penalties for ``tunnels'').

Lensing generally constrains the projected mass within the Einstein radius,
or more precisely, the region within the multiple image positions.
This region is known as the ``convex hull''
in the language of our interpolation scheme.
Outside the convex hull, our solutions are ill defined
and in fact drop off to zero (and even negative values) too quickly.
Our mass models should generally be disregarded outside this region.

Since publishing the LensPerfect method paper \citep{Coe08},
we have made small changes in the exact implementation of these penalties
and their relative weights.
And where (for the purposes of calculating penalties only)
we had evaluated the mass map on a $41\times 41$ grid
we now evaluate it on a finer $81\times 81$ grid within the convex hull
and a coarser $21\times 21$ grid outside.
(With the $41\times 41$ grid, 
we found one particular ``tunnel'' was escaping detection.)
Finally, we have added the redshift penalty function
as described already in \S \ref{LPA1689:redshifts}.

We emphasize that our mass models are not ``grid-based''.
The radial basis functions are instead placed at the positions of the multiple images,
as described above.
The resulting mass model has a smooth functional form
and can be calculated at any desired coordinates.
However in order to evaluate and present the results,
we generally calculate the mass model on a regular grid.



\section{Mass Models}
\label{LPA1689:results}

Here we present non-LTM mass models 
which perfectly reproduce the observed positions of
\Nknotimages multiple images of \Nknotss knots within
\Nimages images of \Ngals galaxies.
strongly lensed by A1689.
We stress that there is no unique solution,
and we do obtain a range of solutions
which allow us to estimate our uncertainties (\S\ref{LPA1689:ensemble}).
However first we present the most ``physical'' solution
found by our optimization scheme described above.
This optimization took two weeks to run on a MacBook Pro laptop.
The process runs quickly at first (a few minutes per galaxy added)
but slows as more galaxies are added.
Galaxies were added in the order presented in Table \ref{LPA1689:taborderz}.

\subsection{Most Physical Mass Model}

\begin{figure}
\includegraphics[width=1.07\hsize]{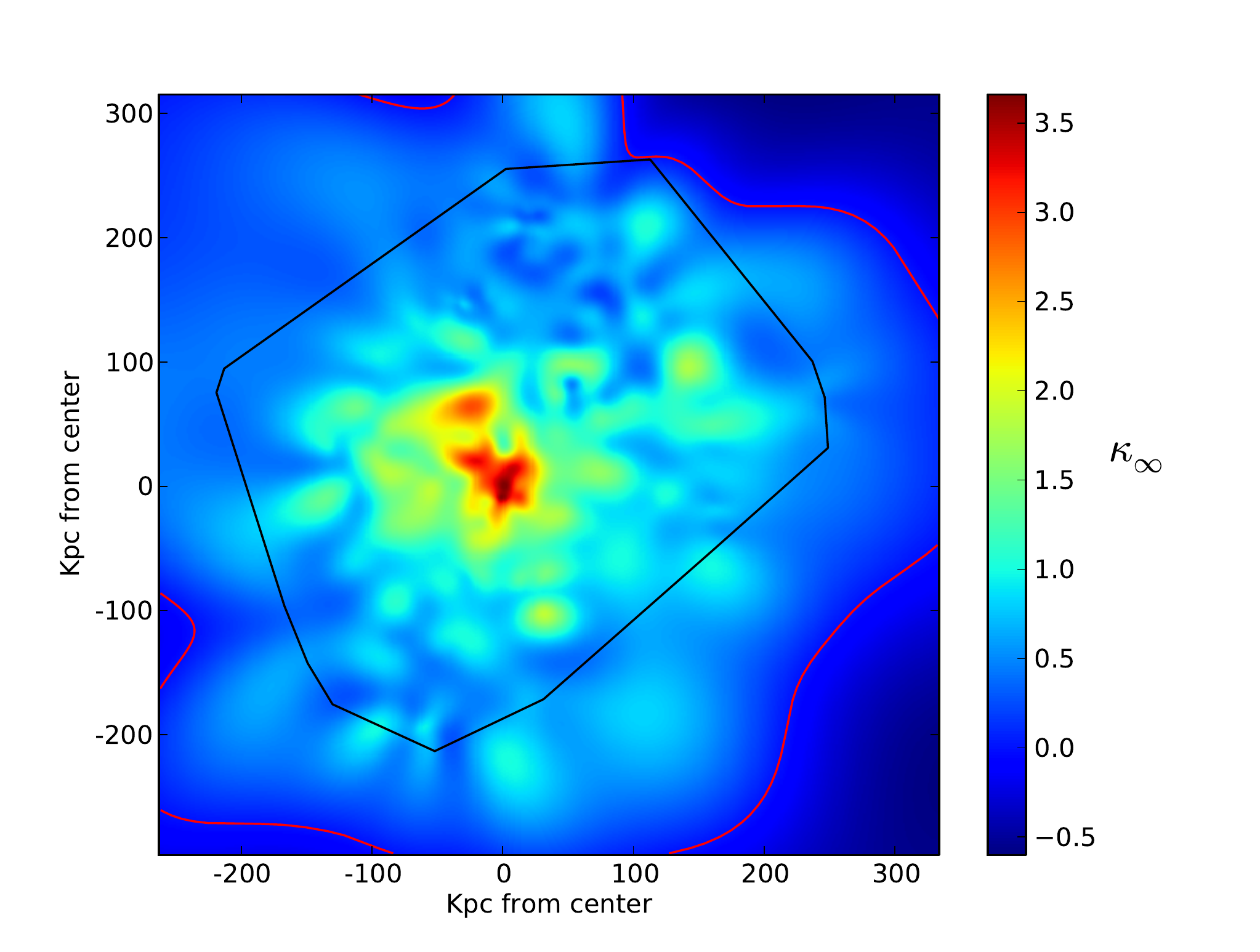}
\caption[Massmap]{\label{LPA1689:massmap}%
Mass map solution for Abell 1689 
which perfectly reproduces the \Nknotimages observed (strongly lensed) positions
of \Nknotss galaxy knots / centroids.
Mass is plotted in units of $\kappa_\infty$
(critical density for a source at $z_s = \infty$)
and cut to the same $3.2\arcmin \times 3.3\arcmin$ field of view 
as the STScI color image
(previous and following figures).
The black line indicates the convex hull.
Multiple images are found within this region.
Outside this region, our solution is highly uncertain
and in fact falls off to zero too quickly.
A red line marks the $\kappa = 0$ contour.
We stress that this solution is not unique,
but had the highest ``physicality'' of all solutions we explored.
Angular diameter distances are given along the axes.
}\end{figure}

\begin{figure*}
\plotone{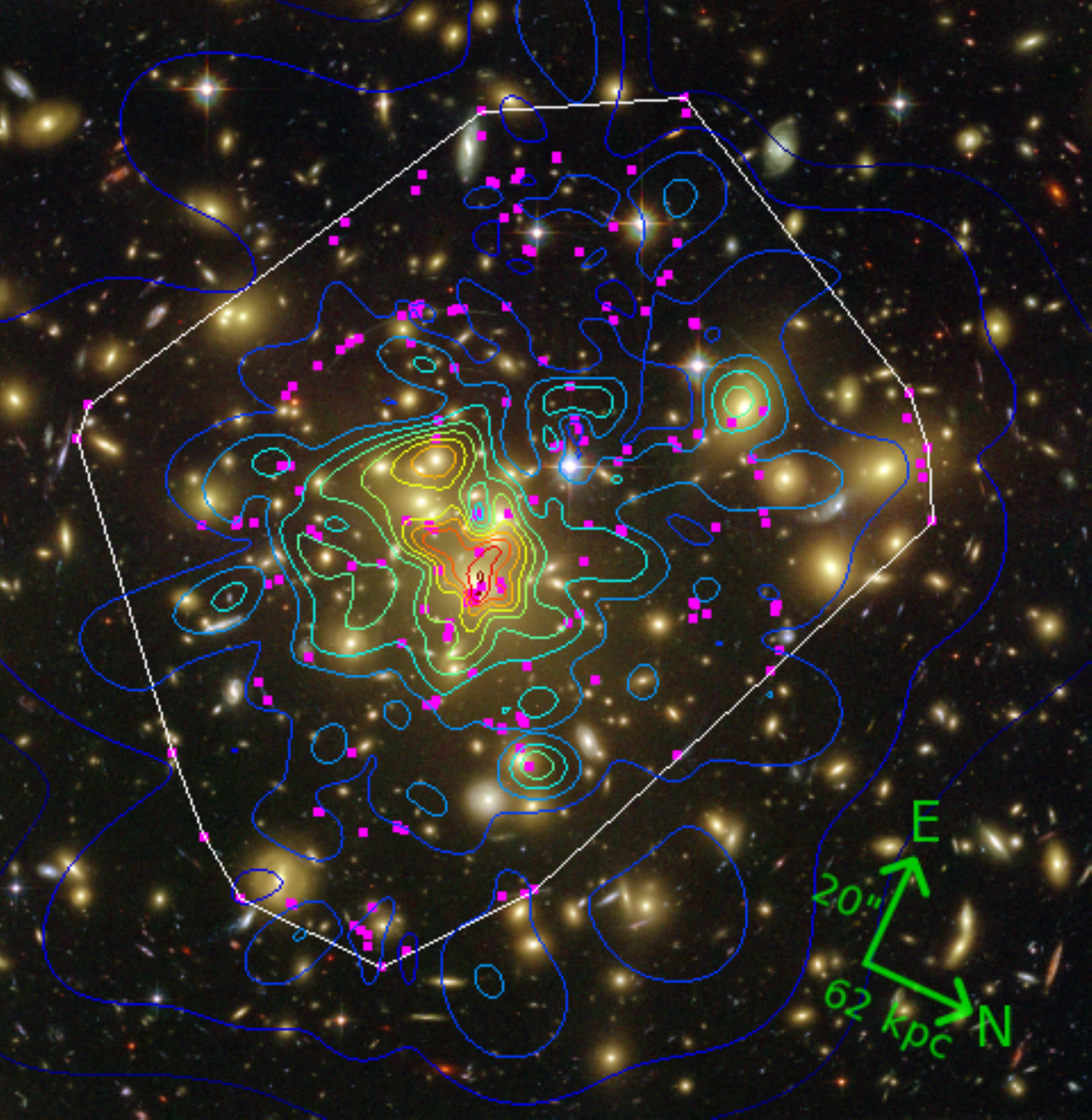}
\caption[Critical Curves]{\label{LPA1689:masscontours}%
Mass map contours in units of $\kappa_\infty = 1/3$ laid over the 
$3.2\arcmin \times 3.3\arcmin$ STScI ACS $g\arcmin r\arcmin z\arcmin$ color image.
The outermost contour, $\kappa_\infty = 0$,
was also plotted in the previous figure.
Pink squares indicate the \Nimages multiple image positions
all perfectly reproduced by our model,
and the white line indicates the convex hull.
Outside this region, our solution should be disregarded.
This solution is not unique but was the ``most physical'' we found.
}\end{figure*}

In Fig.~\ref{LPA1689:massmap} we present
our most physical strong lensing mass model of A1689.
The mass map contours are laid over the 
ACS STScI $g\arcmin r\arcmin z\arcmin$ color image
in Fig.~\ref{LPA1689:masscontours}.
Our model is constrained best near the multiple images (shown in pink),
interpolated between them,
and is highly uncertain where it is extrapolated
outside the outermost multiple images (our ``convex hull''),
traced in
white in Fig.~\ref{LPA1689:massmap} and 
black in Fig.~\ref{LPA1689:masscontours}.
In fact our model falls off to zero too quickly outside the convex hull.

From inspection of Fig.~\ref{LPA1689:masscontours},
we find that our mass model
resolves halos of perhaps ten or so galaxies which are members of A1689.
Although we have made no assumptions about LTM,
much of our mass model's substructure does coincide with luminous galaxies.
Determining masses for these individual galaxy halos 
would be useful though far from trivial,
as we discuss in \S\ref{LPA1689:future}.

There are, however, some potentially interesting offsets
between mass clumps and luminous galaxies.
For example, the mass near the BCG appears more pinched 
than the distribution of galaxies just above and to the left.
Several multiple images (plotted as pink squares) are present in this region
lending confidence to our mass model there.
Furthermore, when we tested our method on simulated lensing 
with a similar mass distribution \citep{Coe08},
our recovered mass maps exhibited no such pinching.

More quantitative and robust conclusions about the alignment of mass and light 
will await future work (\S\ref{LPA1689:future}).
We will perform tests 
to determine how robustly substructure clumps are identified
and their positions determined.

We estimate that our mass map resolves substructures $\sim 23$ kpc across
within $R_E \sim 150$ kpc of the core
(angular diameter distances).
This estimate is based on the density of multiple images,
$N d^2 = \pi R_E^2$,
where $N =$ \Nimages images, 
and we find the average separation among these to be $d \approx 23$ kpc.
Each multiple image provides a constraint on our deflection field
and thus our mass model.
The observed multiple images are more densely packed in some regions;
thus the mass map resolution will be greater there and lesser elsewhere.

This is the highest resolution mass map to date of any galaxy cluster
without assuming LTM.
\cite{Leonard07} present a SL+WL mass map with pixels $\sim 65$ kpc across.
\cite{Saha06} and \cite{JulloKneib09} use smaller pixels but computational issues
limit them to $\sim 30$ multiple images per solution.
We estimate their effective resolution to be $\sim 50$ kpc, 
or about one-fourth our 2-D resolution (half along each axis).
By using four times the number of constraints (multiple images),
we obtain a mass model with four times the resolution.

Finally in Fig~\ref{LPA1689:delensed}
we show our multiple images as delensed to the source plane by our mass model.
Note that each constrained knot in each system
is delensed to the same source position in each multiple image.
To be clear, we do properly model extended images
as the multiple knots in {\it each} image
map back to different locations in the source plane.
For example, in the large arc (8ab), 
three distinct knots are identified and constrained in each image (a and b).
Knots 8a0 and 8b0 map back to one point in the source plane,
8a1 and 8b1 map back to another,
and 8a2 and 8b2 map back to a third point in the source plane.

\subsection{Mass Model Ensemble and Uncertainties}
\label{LPA1689:ensemble}

The mass model solution presented above is not unique.
Using the optimization procedure described in \S \ref{LPA1689:LP},
we explored a wide range of source positions and redshifts,
and found a set which produced this most ``physical'' mass map 
according to our criteria
and without letting the redshifts stray too far from their input values.

We estimate the uncertainties in our mass model
by exploring an ensemble of mass model solutions.
In future work we will develop algorithms to explore this solution space
more thoroughly,
as we describe in \S\ref{LPA1689:future}.

Here we settle for a proxy ensemble of mass models,
based on a broad but non-exhaustive search of our parameter space.
This ensemble consists of \Nensemble solutions which we obtain 
as we add galaxies building up to our final ``best'' solution.
The first model in our ensemble is optimized given systems 1 and 2 only,
and our last includes all \Nknotss knots.

It is unclear whether this technique should be expected to
overestimate or underestimate somewhat our actual uncertainties.
We include too broad a range of solutions 
by including those that only fit some of the data,
yet we may not be thoroughly exploring the solution space.

This technique does capture some of the systematic uncertainties
which would result from adopting various subsets of the multiple image systems.
Previous authors have done just that: used subsets of our multiple images.
In previous work there has also been some variation in these identifications.
One or more of the systems presented in our work may yet prove incorrect,
and we capture some of those uncertainties here.

Our estimated uncertainties on the radial mass profile appear to have the correct form.
Mass enclosed within the Einstein radius $R_E \sim 47\arcsec$
is constrained more tightly than mass within other radii (\S\ref{LPA1689:profilefits}).

The exact structure we resolve in the center is sensitive to
the identifications of demagnified central images,
which can be fairly uncertain.
However we believe our modeling method 
enables robust identification of central images.
As discussed in \S \ref{LPA1689:centralimages}, 
we have reexamined the B05 central image identifications,
purged those which do not fit well with the rest,
and identified a new central image candidate (28c) which we include in our model.

\begin{figure*}
\plottwo{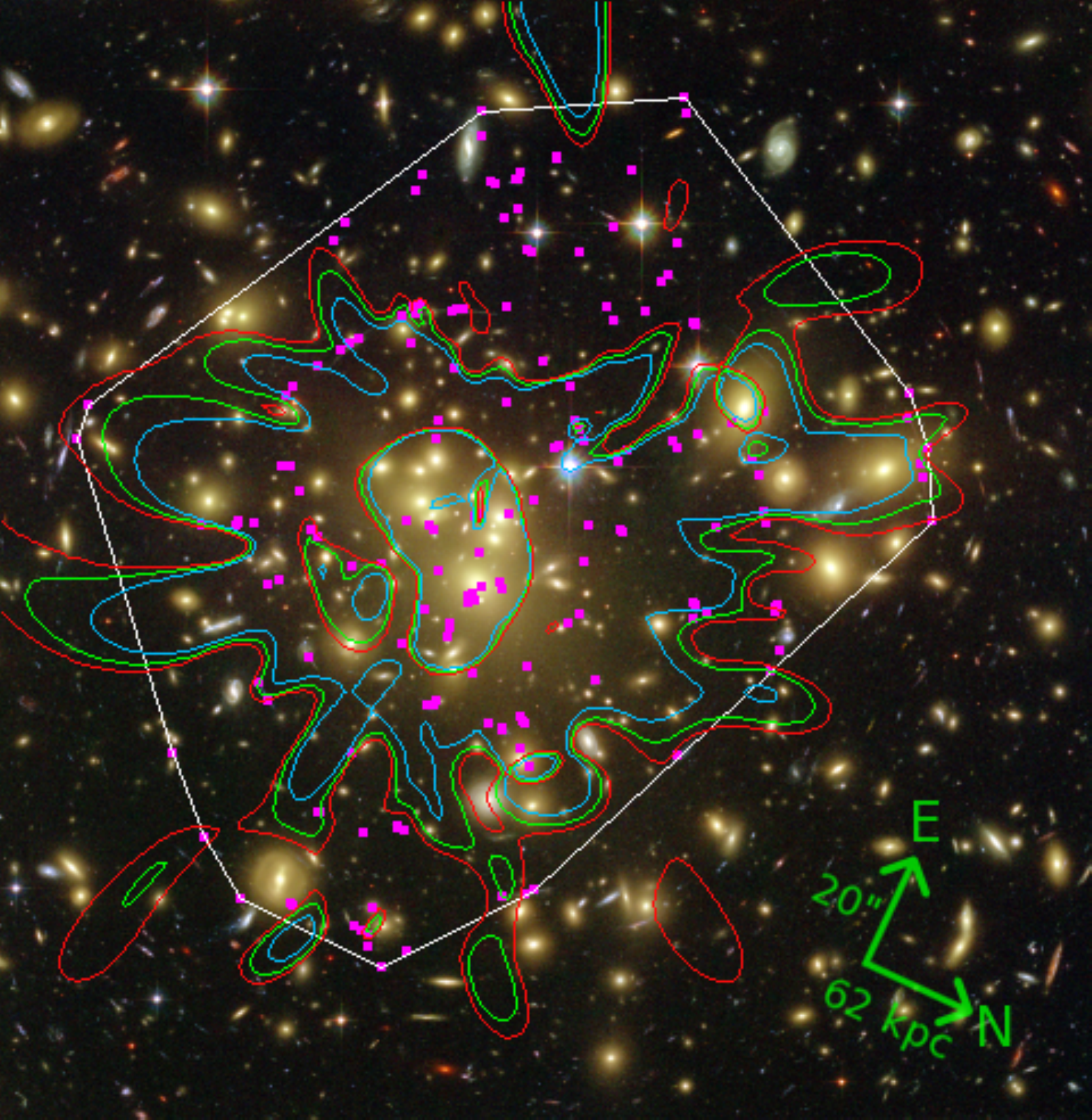}{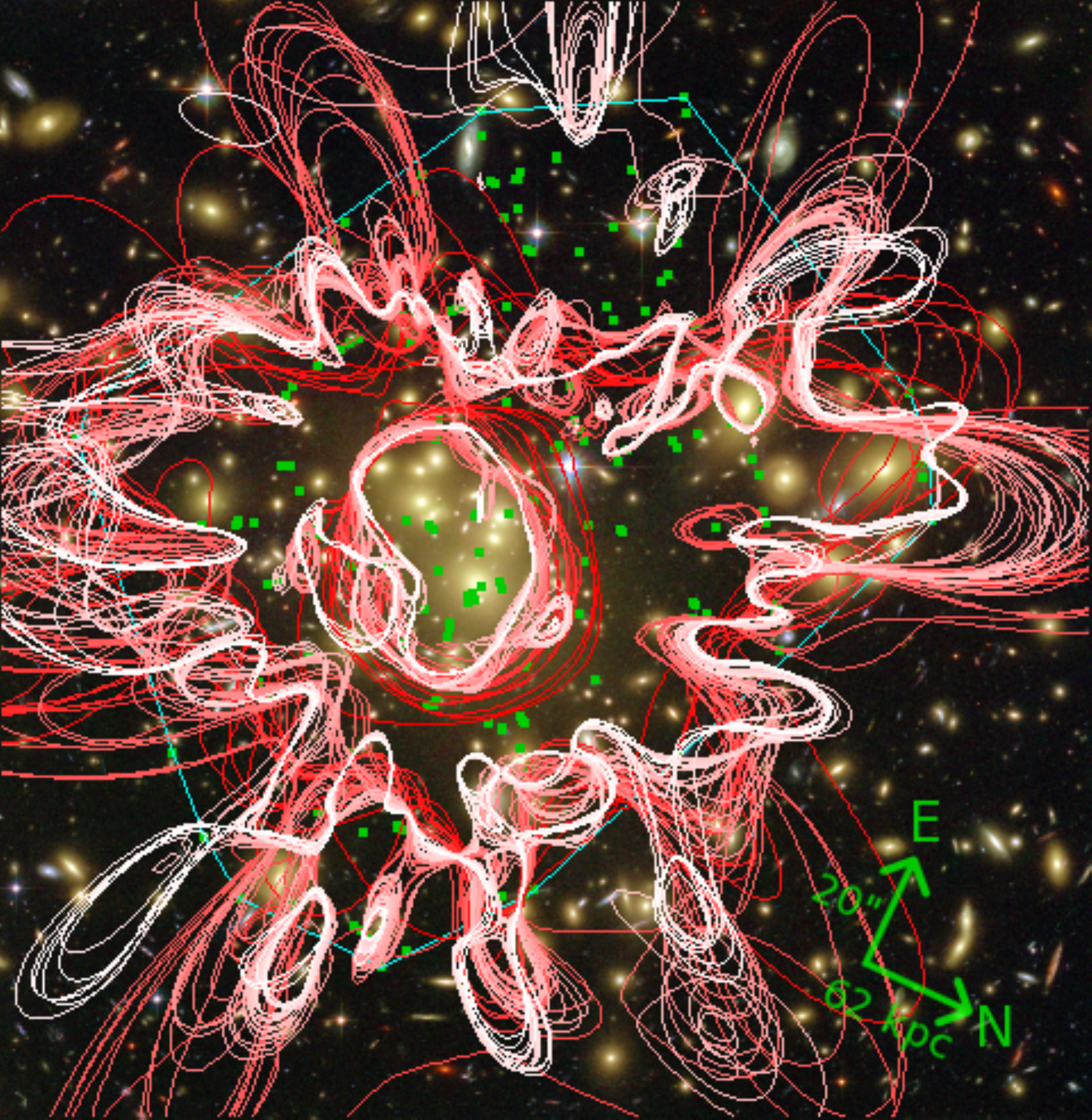}
\caption[Critical Curves]{\label{LPA1689:critcurves}%
{\it Left}: Critical curves for sources at redshifts $z_s = 1$, 2, and 7 laid over the 
$3.2\arcmin \times 3.3\arcmin$ STScI ACS $g\arcmin r\arcmin z\arcmin$ color image.
The loops in the curves are probably too broad, 
an artifact of the insufficiently resolved substructure clumps in our models.
(Compare the curves with the pink squares
indicating the multiple image positions which provide model constraints.)
Given additional multiple images (or a well-informed prior such as LATM),
finer substructure would be resolved,
yielding tighter and more precise critical curves.
{\it Right}: The great freedom in our models 
allows for broad variation in the critical curve shapes.
We plot $z_s = 7$ critical curves for all solutions in our ensemble,
from that which fits two multiple image systems (dark red)
to our final solution (shown in white)
which fits all 55 systems, including multiple knots per galaxy.
The variation in magnification is such that 
points on our final critical curve
should expect to have $\mu > 6$ 68\% of the time.
This would increase for a better constrained model.
}\end{figure*}

In Fig.~\ref{LPA1689:critcurves}, we plot the critical curves
for sources at redshifts $z_s = 1$, 2, and 7.
The exact shapes of the critical curves are not well constrained by the data alone.
They are sensitive to the exact distribution of substructure in the mass map,
which for this purpose is not resolved sufficiently by our models 
given the number of multiple images.
The critical curves would be better constrained by 
additional multiple images or a well-informed prior on the mass distribution.
Even in this limiting case of broad model freedom,
we find that points along our $z_s = 7$ critical curve 
can expect to have $\mu > 6$ 68\% of the time.

We remark that in general,
non-LTM methods probably overestimate uncertainties
while LTM methods probably underestimate them somewhat.
LTM methods do not explore the full range of solutions 
(including asymmetries and other deviations from LTM)
which may reproduce the data.
Furthermore, they are sometimes forced to discriminate between 
a ``best'' solution which reproduces the data at, say, 50-$\sigma$ ($2.5\arcsec$),
and ``deviant'' solutions which reproduce the data at, say, 60-$\sigma$ ($3.0\arcsec$).
Non-LTM methods, on the other hand, may include, 
at worst, a broad range of unphysical solutions
or, at least, solutions 
which do not take advantage of the strong observational priors available,
namely the observed positions of the lensing galaxies.
An ideal method would use LTM as a prior while allowing for deviations
(\S\ref{LPA1689:future}).
This prior might be referred to as ``LATM'', or light {\it approximately} traces mass.





\section{Mass profile and concentration}
\label{LPA1689:profanal}

As discussed in \S\ref{LPA1689:intro},
recent studies show that galaxy clusters formed in nature
may be more centrally concentrated
than their counterparts formed in simulations.
In this section, we obtain new estimates for the mass concentration of A1689.
Our results support previous analyses
which claim A1689's concentration is higher than expected.
We derive these concentration estimates from fits of NFW profiles 
simultaneously to our SL mass model and WL data published elsewhere.
These profiles do simultaneously fit both the SL and WL data well,
however there is a slight tension between the two 
as described in \S\ref{LPA1689:SL+WL}.
First, in \S\ref{LPA1689:profilefits}, we examine our SL mass profile in detail.

\subsection{Profile fits to the strong lensing (SL) data}
\label{LPA1689:profilefits}

\begin{figure}
\plotone{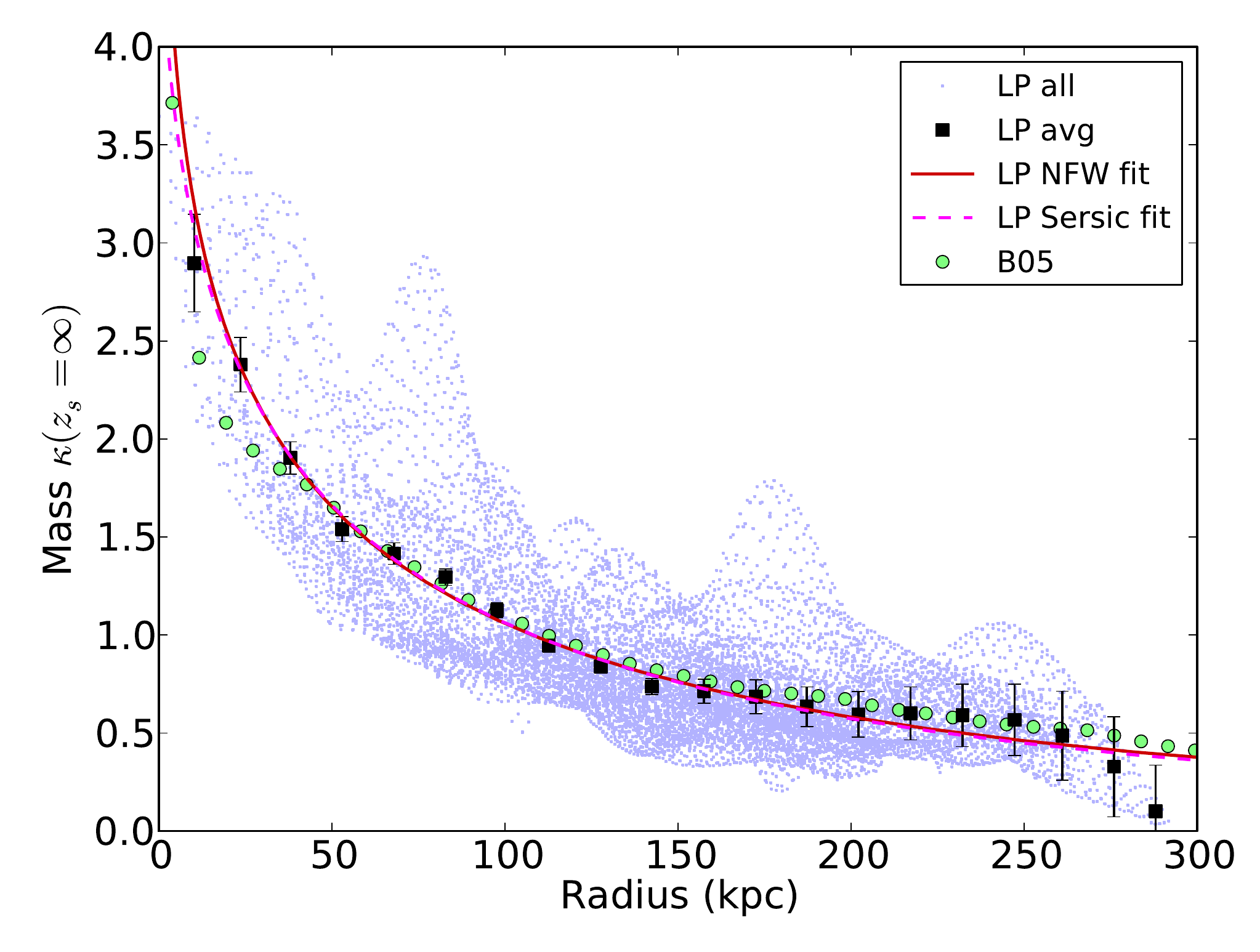}
\caption[Massmap profile fits]{\label{LPA1689:massmapfit}%
Our mass map fit to NFW and S\'ersic profiles.
For all points within our convex hull,
we plot in light blue $\kappa (z_s = \infty)$,
the projected surface mass density in units of critical density
for a source at infinite redshift.
Binned averages are plotted with uncertainties measured as the scatter
among an ensemble of solutions. 
Our best NFW fit is plotted as the red line.
The mass profile obtained by B05 is also plotted for comparison.
}\end{figure}

Clusters formed in simulations
have mass profiles generally well described by 
NFW \citep{NFW96}
or Einasto / S\'ersic \citep{Navarro04} 
profiles \citep[e.g.,][]{Navarro10}.
These profile fitting functions are described in Appendix \ref{LPA1689:profiles}
with further details given in \citep{Coe10DMprofiles}.

\begin{figure}[!ht]
\plotone{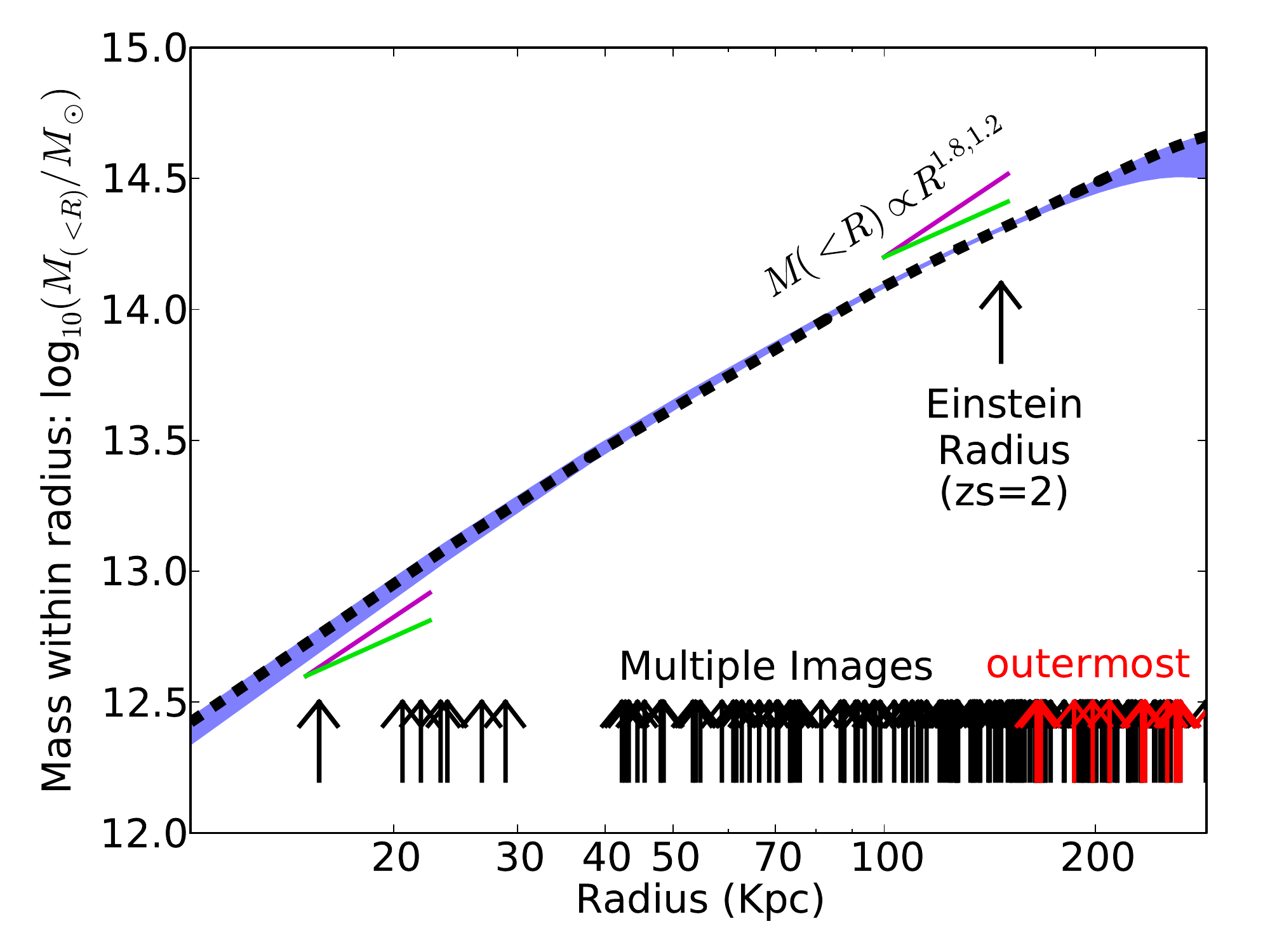}
\caption[Mass within]{\label{LPA1689:Mwithin}%
A1689 projected mass within a cylinder as a function of radius $R$.
The mass profile of our best model is plotted as the dashed line
while the shaded region gives the mean and RMS of the ensemble.
Note the mass is especially well constrained within $R \sim R_E$.
The radii of the multiple images are plotted along the bottom.
Those along the convex hull are plotted in red.
Four central images are outside the range of the plot at small radius.
}\end{figure}

In Fig.~\ref{LPA1689:massmapfit} we plot 
our projected mass density radial profile \kR\ with our origin defined as
the location of our density peak which corresponds to the location of the BCG.
We plot the mass profile from B05 for comparison and find good agreement
even though the analysis methods are very different (including LTM vs.~non-LTM).
For a similar recent comparison, see \citep{Zitrin10A1703}.

Overplotted in Fig.~\ref{LPA1689:massmapfit} are our best NFW
\NFWfitMbestp\ and S\'ersic \Sersicfitp\ fits
to the SL data alone.
The two fits track each other very closely,
although the more flexible S\'ersic profile 
does allow for a slightly shallower central slope.
We could allow the central slope of our NFW profile to vary
with a three-parameter ``generalized NFW profile'' \citep{Zhao96,Wyithe01},
but we do not explore this here.

Rather than fitting to \kR\ as is common,
we can reduce the uncertainties of our NFW fit parameters by fitting to \MR.
The former is projected mass surface density at a given radius $R$,
while the latter is total projected mass {\it enclosed} within a cylinder of radius $R$.

We plot \MR\ from our SL modeling in Fig.~\ref{LPA1689:Mwithin}.
We also plot the uncertainty in \MR\ from our ensemble of models.
This quantity, \MR, is what lensing constrains best,
especially \MRE, the mass within the Einstein radius.
Though our models vary, all give a very consistent amount of mass
within $R_E \sim 47\arcsec$.

We derive an Einstein radius of \NFWMRE\ for a lensed source at $z_s = 2$.
This value increases with redshift
to $R_E \sim 52\arcsec$ for a source at $z_s = 7$.
No cluster is perfectly symmetric,
so the ``correct'' definition of Einstein radius is a bit ambiguous.
As done elsewhere,
we find that radius within which $\langle \kappa \rangle = 1$,
the average mass surface density
is equal to the critical lensing density (Eq.~\ref{LPA1689:E_c}).

In \cite{Coe08}, 
we tested our method's ability to recover a mass distribution similar to A1689
given 93 multiple images of 19 lensed galaxies.
Our recovered mass profile matched the input mass profile extremely well.
In future work (\S\ref{LPA1689:future}),
we will quantify our ability to recover input mass profiles and concentrations.

\begin{figure*}[ht!]
\includegraphics[width=0.49\hsize]{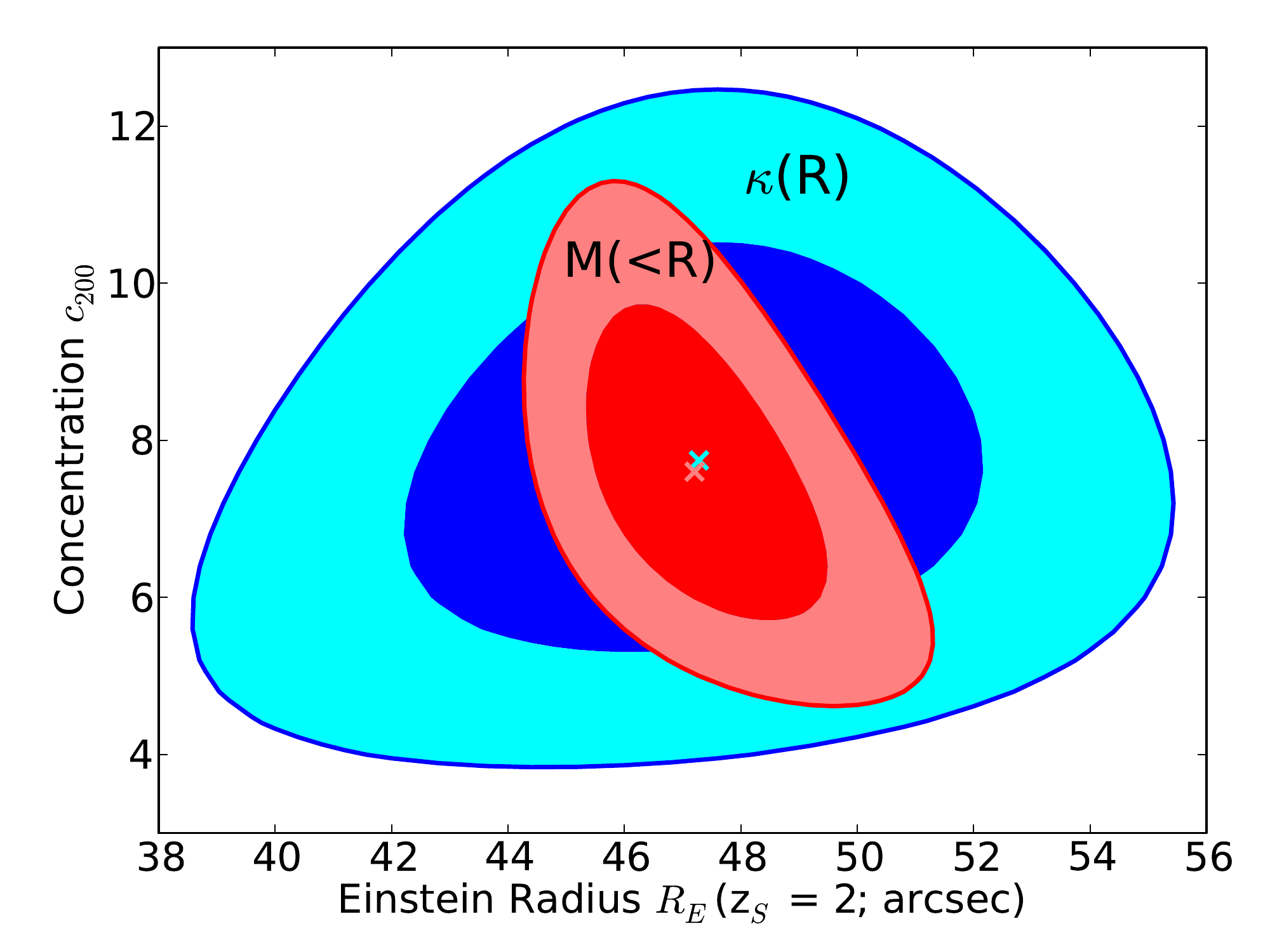}
\includegraphics[width=0.49\hsize]{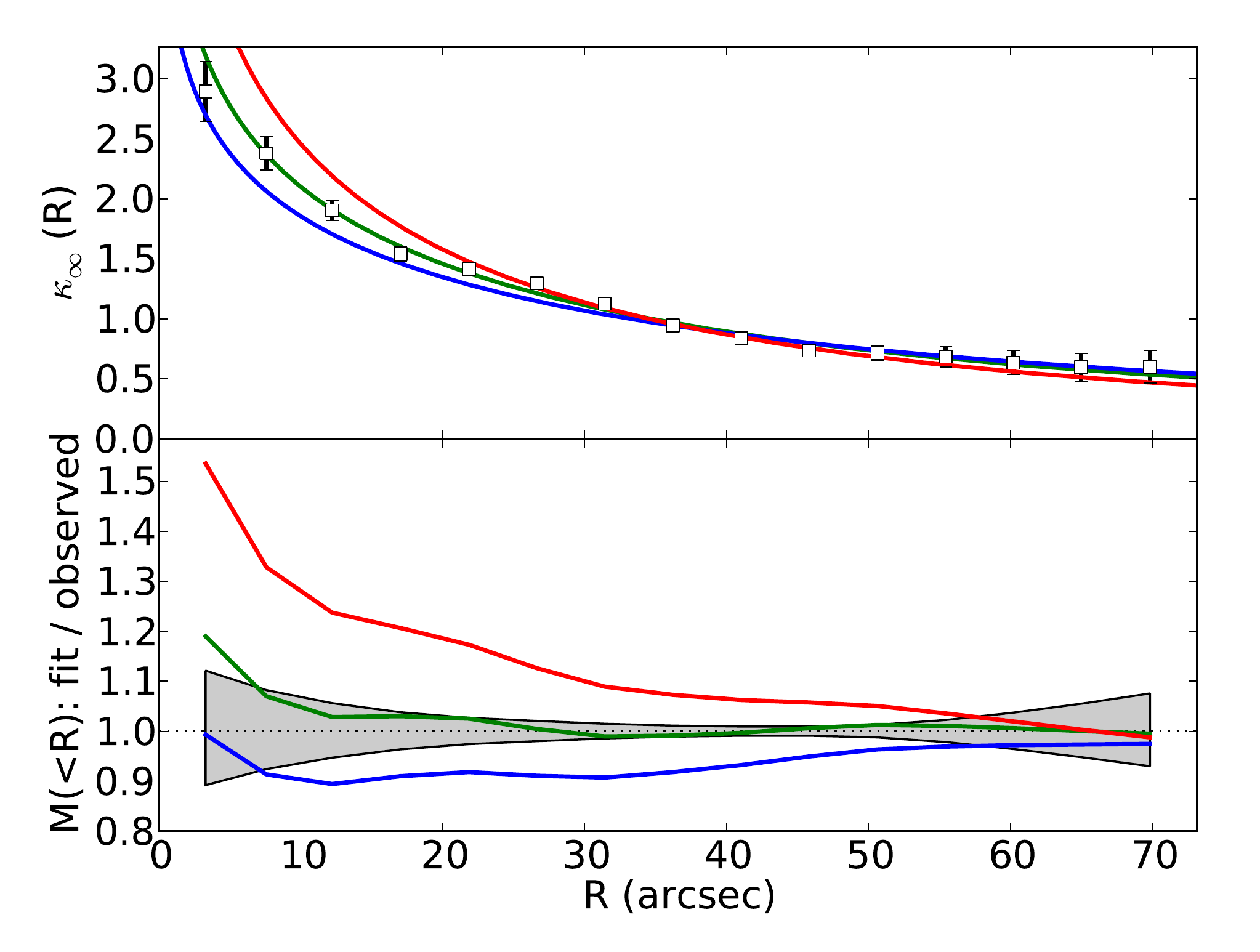}
\caption[NFW fit parameters]{\label{LPA1689:NFWfitcontourscRE}%
Fits to the enclosed mass \MR\ provide tighter constraints than
fits to the mass surface density \kR.
{\it Left}: NFW fit confidence contours (1-$\sigma$ and 2-$\sigma$)
in the plane of Einstein radius $R_E (z_s=2)$ and concentration $c_{200}$.
Best fit values are plotted as X's.
(Note the NFW scale radius $r_s$ is fixed by each ($R_E$, $c$) pair.)
{\it Right}: Illustration of why \MR\ provides tighter constraints than \kR.
{\it Right Top}: Best NFW fit (green) to the observed $\kappa(R)$
and two other fits with reduced $\chi^2 \approx 1$.
While these are all reasonable fits to \kR,
the red (blue) curve significantly
overestimates (underestimates) the observed \MR\ (grey; {\it right bottom}).
}\end{figure*}

In Fig.~\ref{LPA1689:NFWfitcontourscRE} (left), 
we verify that NFW fits to our \MR\ are more tightly constrained
than NFW fits to our \kR.
In the right panel we provide an illustrative explanation.
There is a wider range of NFW profiles which fit \kR\ well ($\chi^2 < 1$)
than fit \MR\ well.\footnote{One might worry 
about correlated uncertainties in our \MR\ bins.
This would be a concern only if \kR\ were a more fundamental observable than \MR.
We believe the opposite to be the case;
lensing most fundamentally constrains \MR.
One should instead worry that derived measurements of \kth\ are correlated;
they must add to produce the correct \MR\ to deflect images to their observed radii.}

Note that mass profiles of simulated galaxy clusters
are often quoted in terms of their 3-D mass density $\rho(r)$.
Lensing alone cannot measure $\rho(r)$
(although the addition of other data can constrain $\rho(r)$, 
as in \citealt{Morandi10}).
Reported measurements of \MR\ in simulations
would enable more direct comparison
between lensing observations and simulations.

Along those lines, \cite{BroadhurstBarkana08} compared observed measurements of
Einstein radii $R_E$ to those measured in simulations.
The Einstein radius is a robust quantity 
which, for an axisymmetric potential,
can be measured roughly by eye from the positions (and redshifts) of large arcs,
then refined by model fitting.
Measurement of $R_E$ yields the equally robust $M(<R_E)$,
as the average surface density within $R_E$ must be equal to the critical lensing density 
$\Sigma_{crit}$ (as a function of redshift).
Our shift from analyzing \kR\ to the more robust \MR\ measurements 
(adopted by other authors as well)
is a natural extension of the shift to robust $R_E$ measurements by
\cite{BroadhurstBarkana08}.

\subsection{Simultaneous fits to the SL and WL data} 
\label{LPA1689:SL+WL}

Lensing-based constraints on mass concentration
are best derived from simultaneous SL + WL analyses.
Together, SL and WL probe a sufficient range of radius from the cluster center,
capturing the profile turnover 
from $\rho \propto r^{-1}$ and $M(<R) \propto R^2$ near the core 
to $\rho \propto r^{-3}$ and $M(<R) \sim$ rising slowly in the outskirts
\citep[see][]{Coe10DMprofiles}.
\cite{Meneghetti09} quantify these statements,
showing that SL+WL analyses of simulated clusters
yield concentration measurements to 
$\sim 3\times$ greater precision than WL-only analyses
(11\% scatter versus 33\% scatter) and
$\sim 5\times$ greater precision than SL-only analyses
(59\% scatter).\footnote{\cite{Meneghetti09} also find that 
concentrations derived from SL-only analyses are biased high by $\sim 60\%$
on average.
They attribute this bias to their multi-component fit.
They include an isothermal BCG component, but the BCG profile is actually steeper,
and the parent halo must compensate with a higher concentration.
Care must be taken in these comparisons however, 
as we and others measure concentration by fitting to the {\it total} mass profile
(parent halo plus galaxy halos),
not the DM halo alone.}

\begin{figure*}
\plottwo{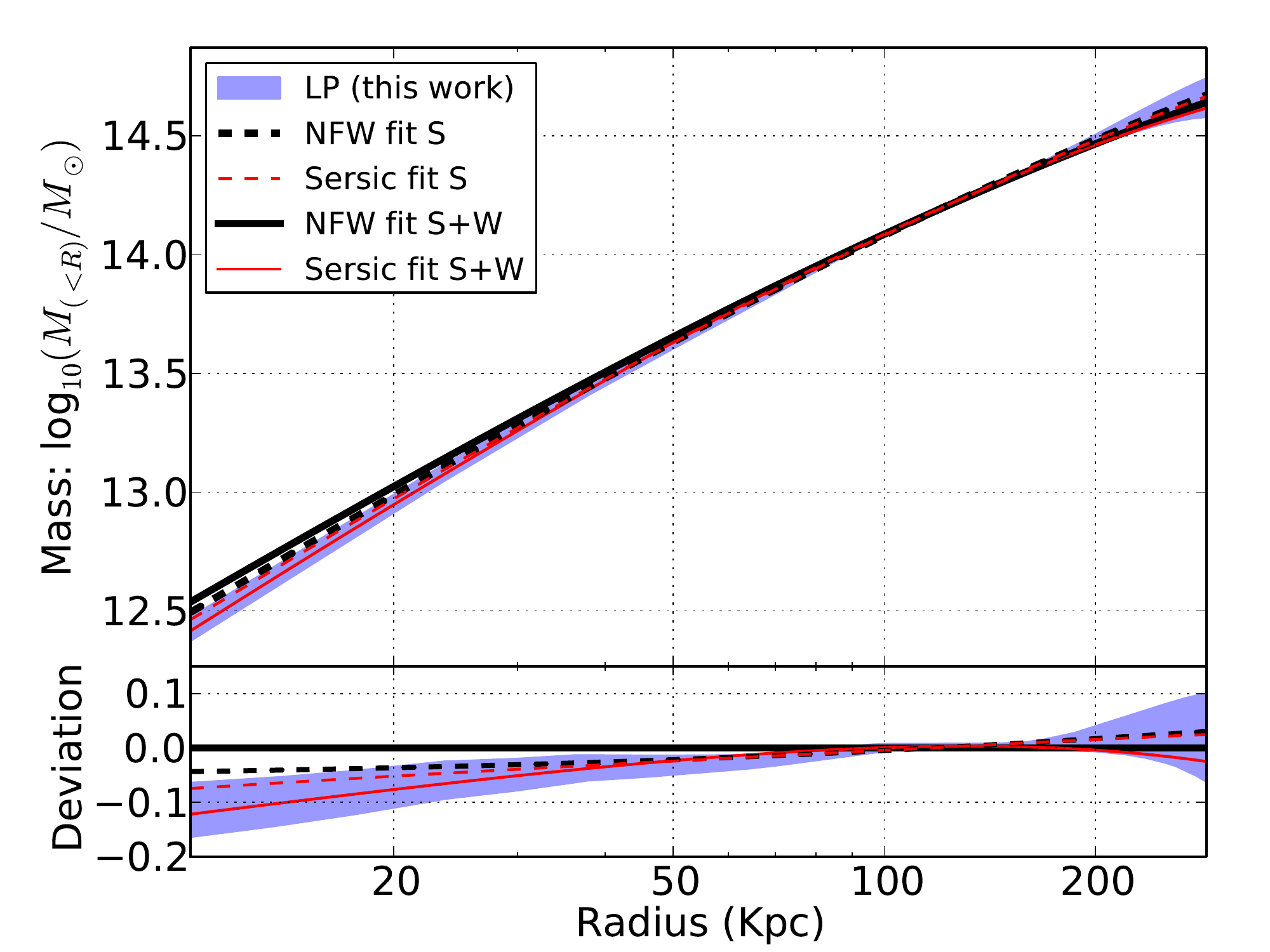}{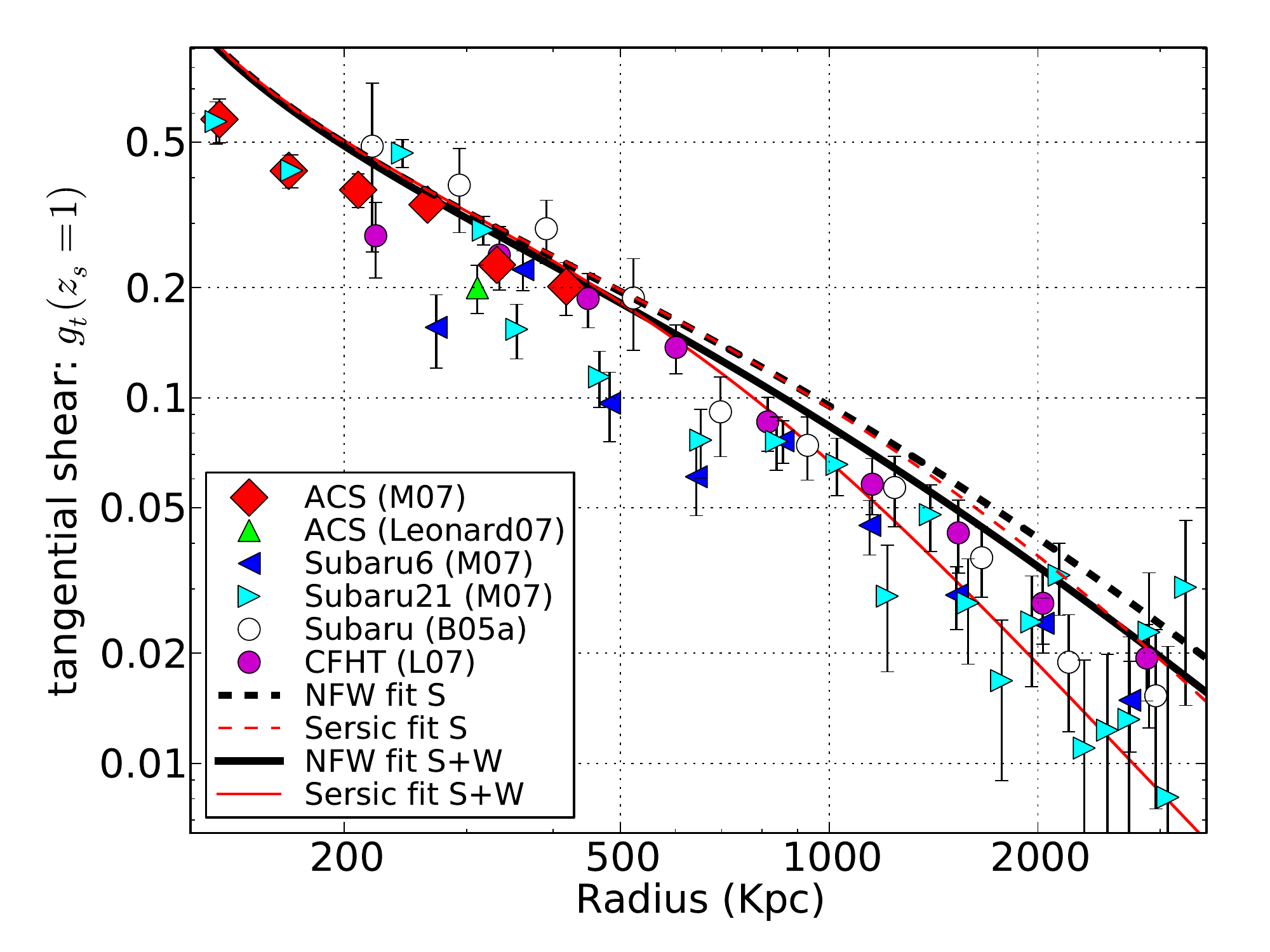}
\caption[NFW fit simultaneously to strong and weak lensing]{
\label{LPA1689:SWfit}
NFW and S\'ersic profiles fit simultaneously to SL and WL data.
Also plotted are fits to the SL data alone.
\textit{Left}: The SL mass profile ($R < 70\arcsec$) 
derived in this work.
\textit{Right}: WL reduced tangential shears ($R > 70\arcsec$) 
measured in ACS (M07, Le07),
Subaru (B05b, M07),
and CFHT images (L07).
}
\end{figure*}

In Fig.~\ref{LPA1689:SWfit} we show NFW and S\'ersic profiles
fit {\it simultaneously} to SL \MR\ from our model
and WL shears measured in previous works (see below).
We obtain an NFW fit of \NFWfitMW\ and a S\'ersic fit of \SersicfitMW.
Also plotted are fits to the SL data only.

Our NFW fits yield 
\NFWfitMWM\ within \NFWfitMWr\ and
\NFWfitMWMvir\ within the virial radius \NFWfitMWrvir.
By definition,
the average overdensities within these regions 
are $\Delta_c = 200$ and 115, respectively,
relative to the critical density to close the universe
(see Appendix \ref{LPA1689:virial}).

We fit to an ensemble of WL data
measured in ACS (\citealt{Medezinski07}, hereafter M07; 
\citealt{Leonard07}, hereafter Le07),  
Subaru (\citealt{Broadhurst05b}, hereafter B05b; M07),
and CFHT images (L07).
We also experiment by fitting to various subsets of this data ensemble.

\begin{figure*}
\includegraphics[width=0.33\hsize]{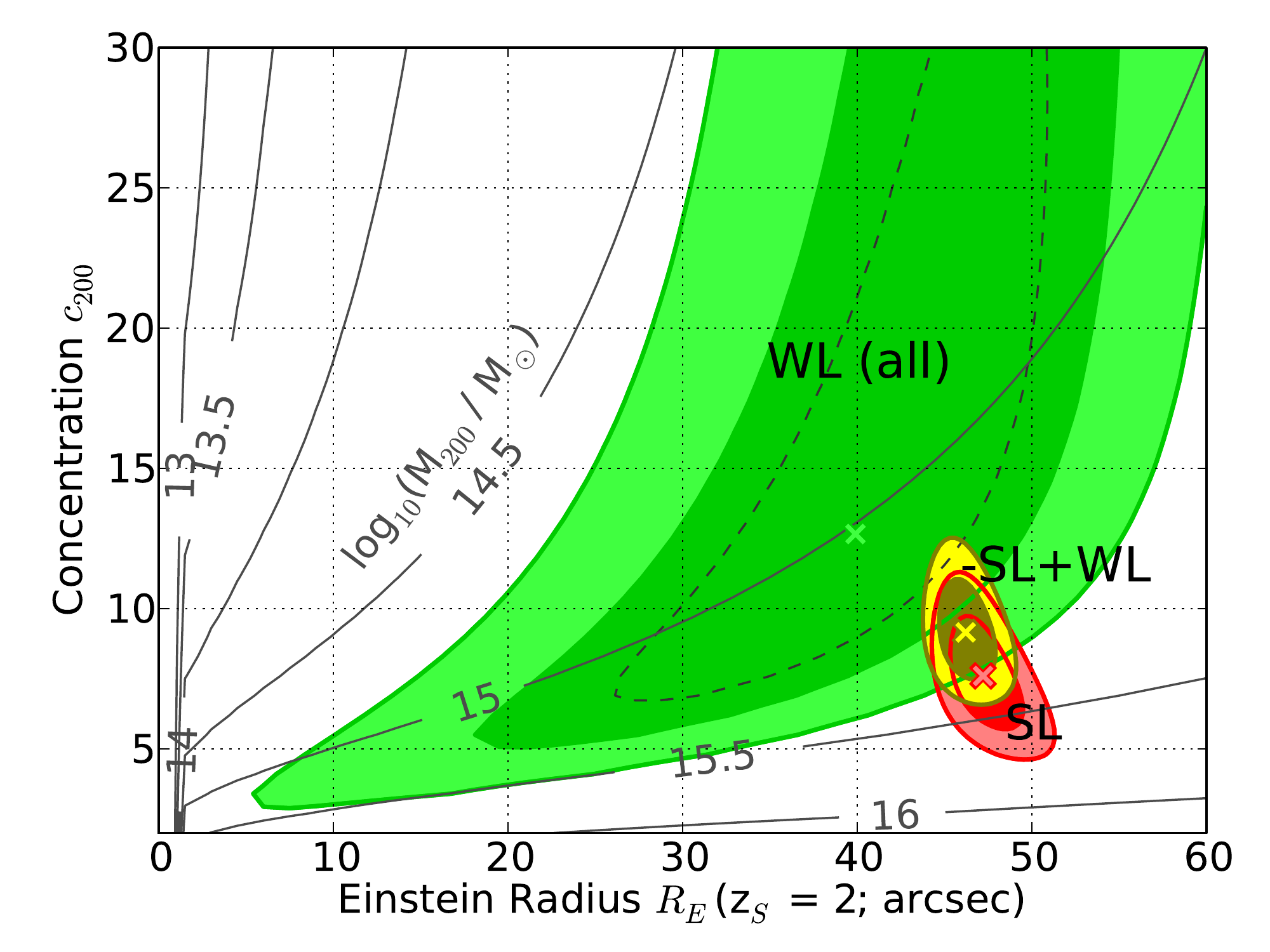}
\includegraphics[width=0.33\hsize]{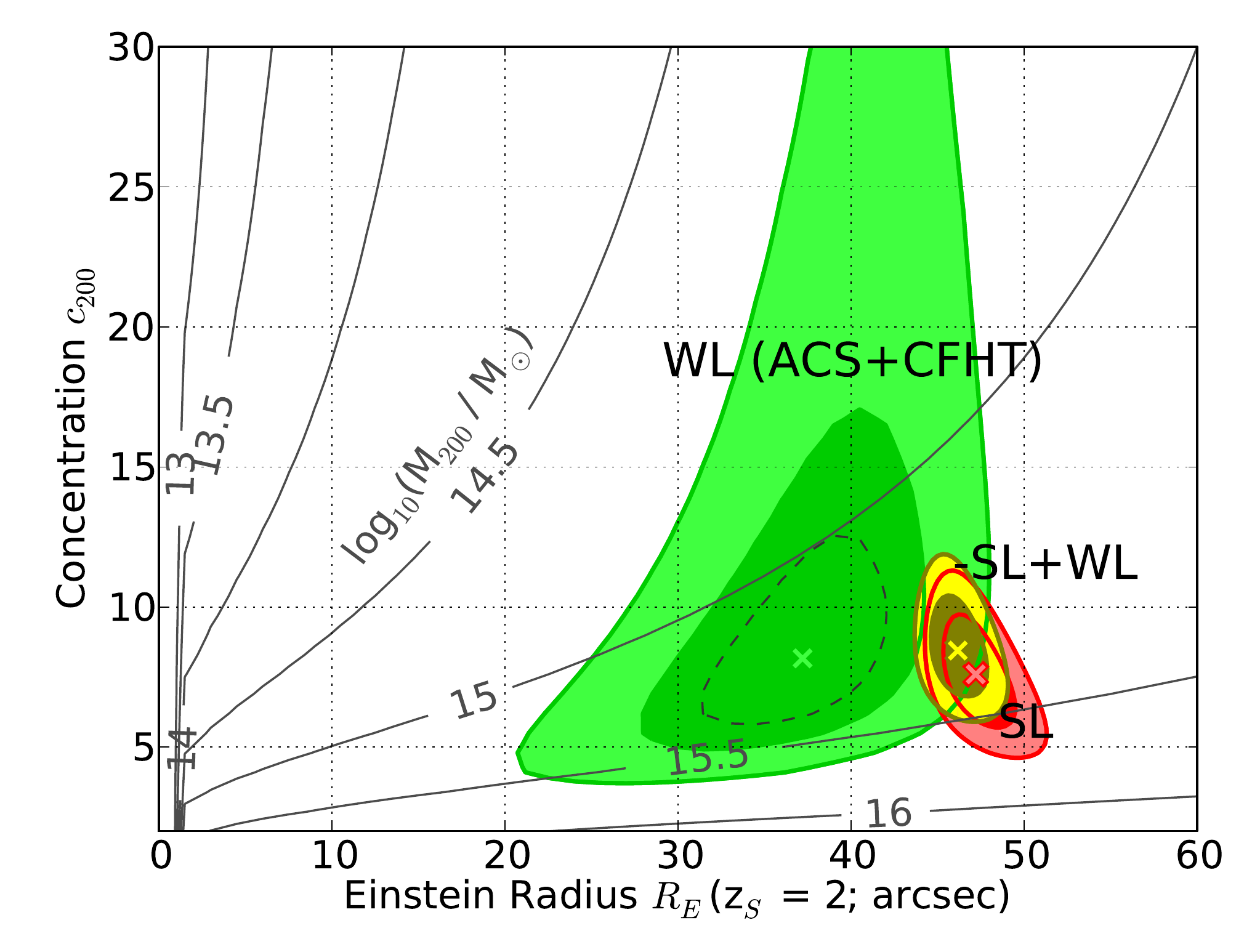}
\includegraphics[width=0.33\hsize]{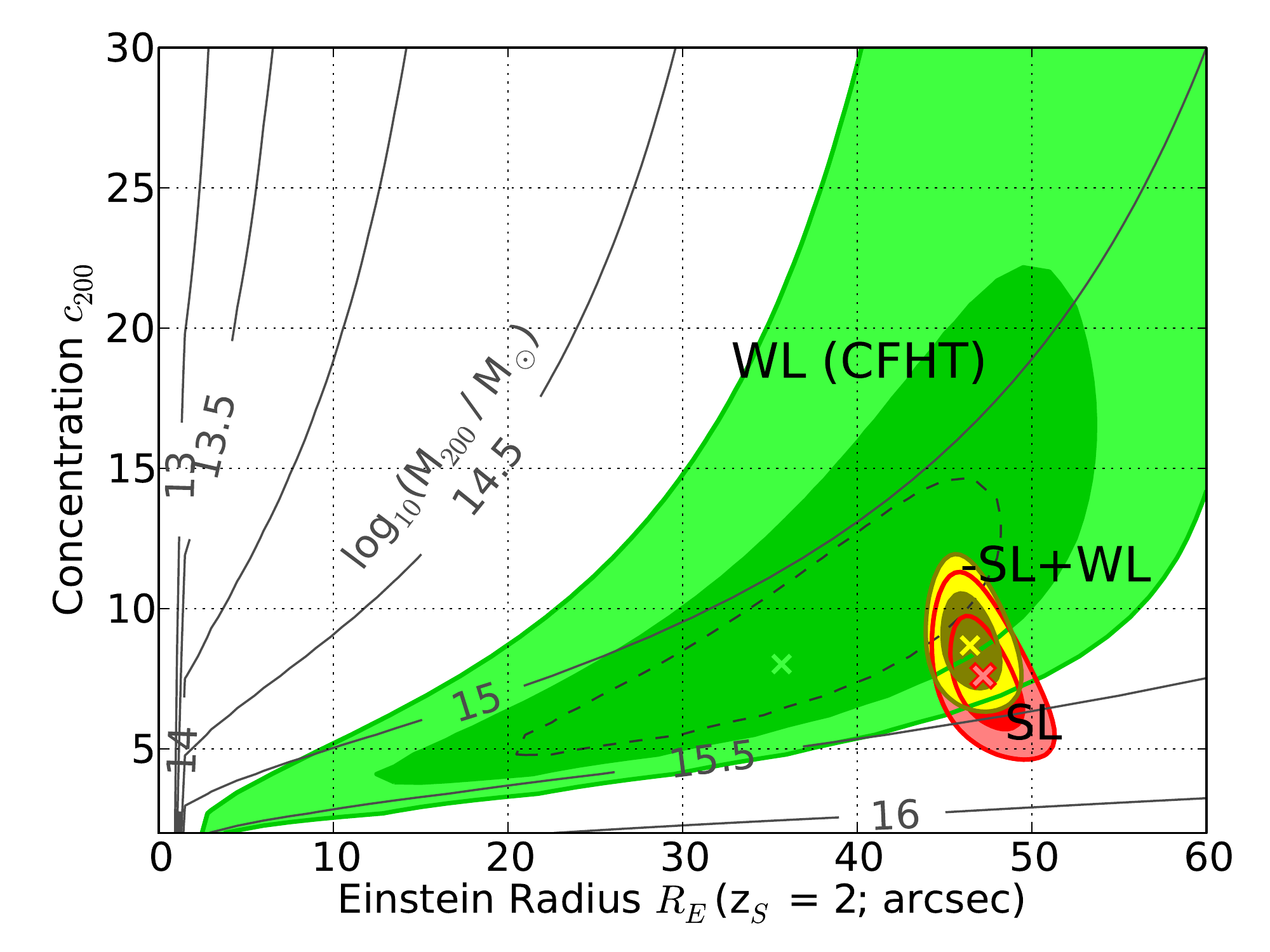}
\caption[NFW fits]{\label{LPA1689:NFWfitcontourscRE_SW}%
Constraints on Einstein radius $R_E$ and concentration $c_{200}$ 
from an NFW profile fit to the observed
SL (red), WL (green), and both combined (yellow).
Three sets of WL data are used -- 
\textit{left}: all;
\textit{middle}: CFHT (L07) \& ACS (M07);
\textit{right}: CFHT (L07).
 Best fits are marked with X's.
Confidence contours are 
1-$\sigma$ (68\%, $\Delta \chi^2 = 2.3$) and 
2-$\sigma$ (95\%, $\Delta \chi^2 = 6.17$).
A black dashed line also indicates the 1-D 1-$\sigma$ WL contours ($\Delta \chi^2 = 1$),
which give the 68\% bounds if one parameter is marginalized over.
Isocontours of $M_{200}$ are overplotted.
}\end{figure*}

\begin{figure*}
\includegraphics[width=0.33\hsize]{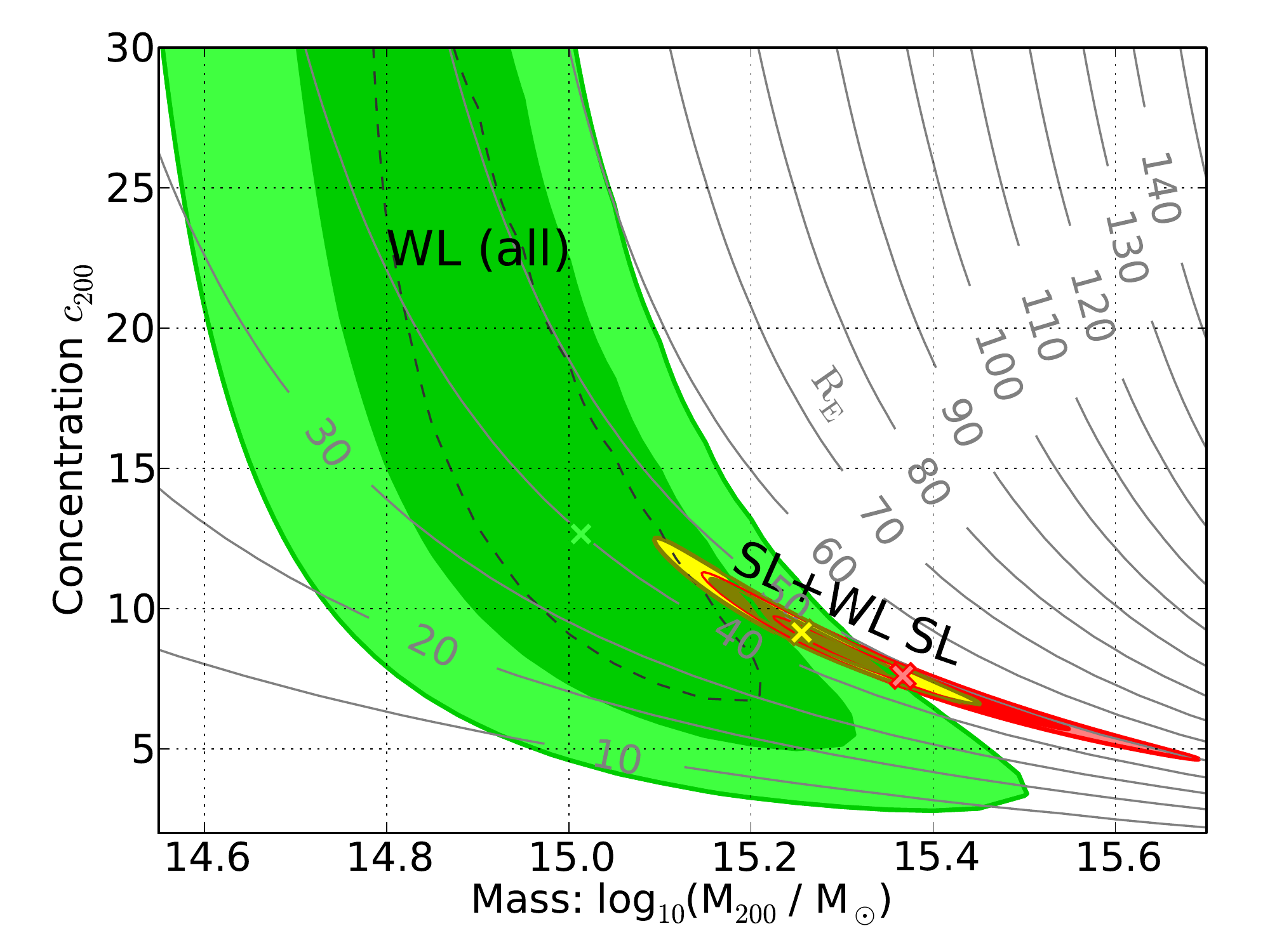}
\includegraphics[width=0.33\hsize]{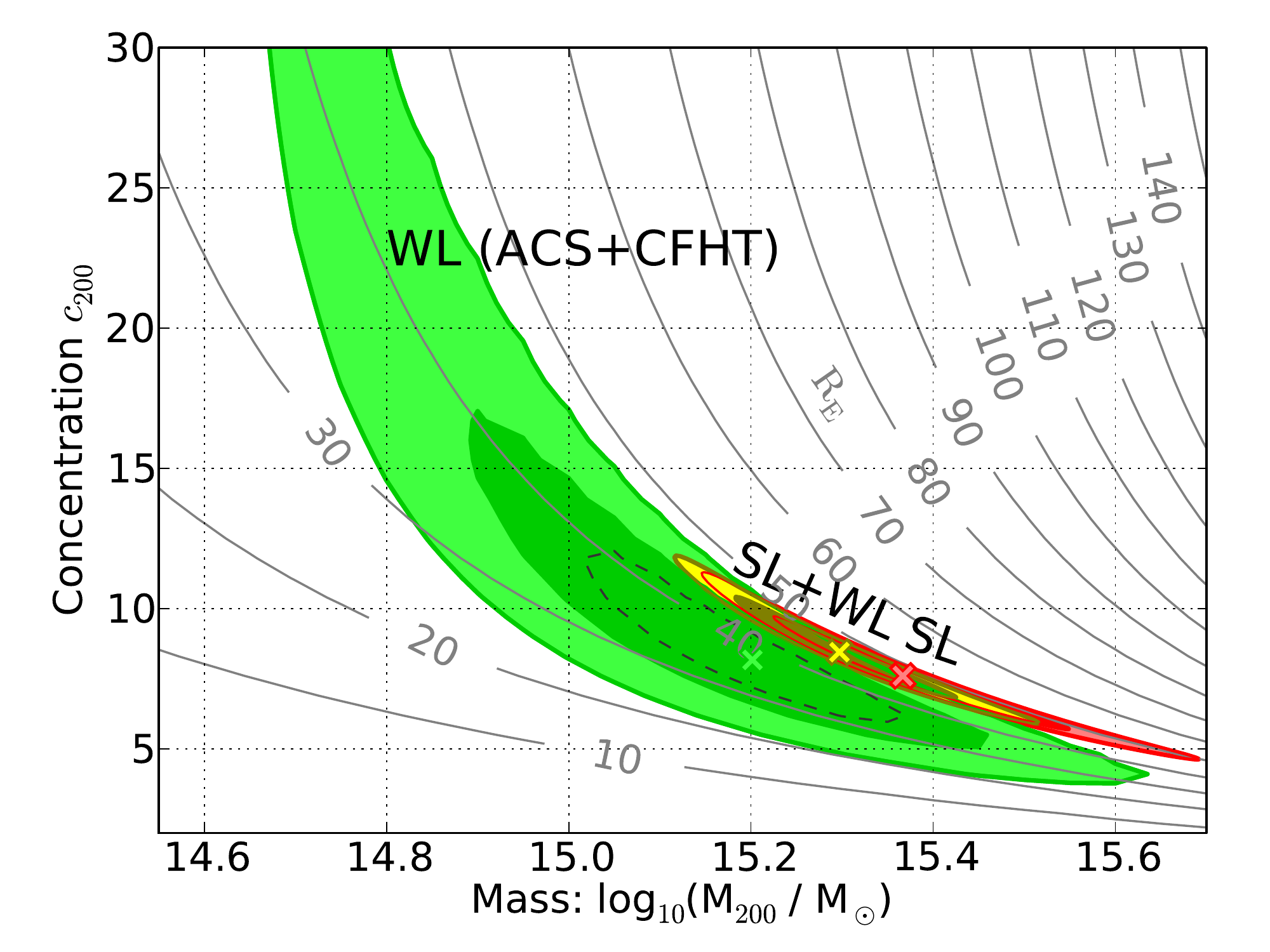}
\includegraphics[width=0.33\hsize]{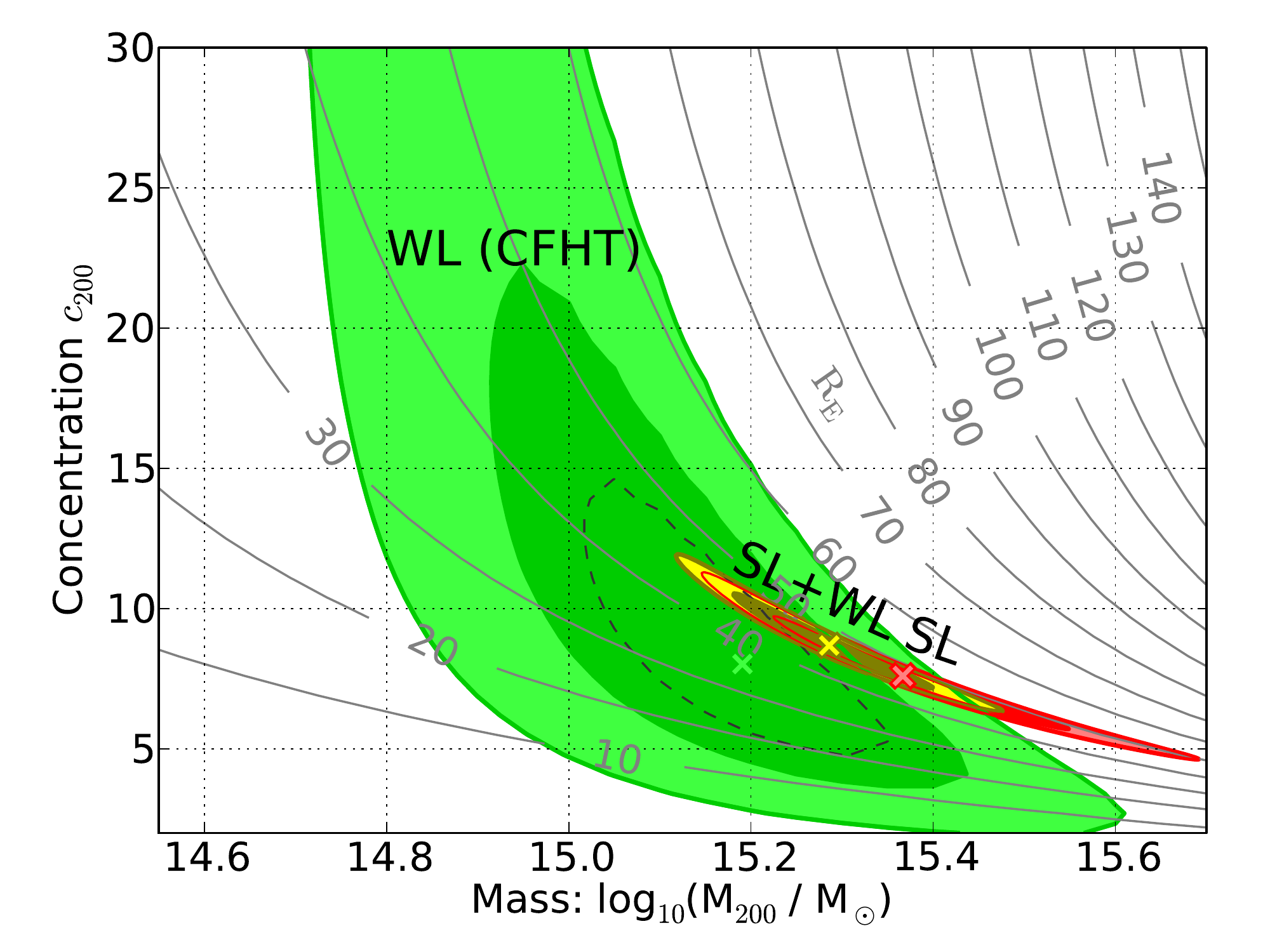}
\caption[NFW fits]{\label{LPA1689:NFWfitcontourscMvir_SW}%
Same as Fig.~\ref{LPA1689:NFWfitcontourscRE_SW} but for mass within the virial radius $M_{200}$
and NFW concentration $c_{200}$.
Isocontours of $R_E$ ($z_s = 2$; arcsec) are overplotted.
}\end{figure*}

Fig.~\ref{LPA1689:NFWfitcontourscRE_SW} compares constraints on ($R_E$, $c_{200}$)
for NFW fits to SL and WL combined
using three different WL data sets.
We consider the L07 CFHT data (right), 
this data combined with the M07 ACS data (middle),
and all data (left).
Fig.~\ref{LPA1689:NFWfitcontourscMvir_SW} is similar 
but shows constraints on ($M_{200}$, $c_{200}$).

Our derived SL+WL parameters are not affected greatly 
by our choice of WL data subset.
However we note that inclusion of the Subaru WL measurements
does increase the concentration slightly.

Note that B05b converted their shear measurements $\gamma(R)$
to mass measurements $\kappa(R)$
using their measurements of magnification $\mu(R)$ 
to break the mass-sheet degeneracy.
For the purposes of fitting an NFW (or S\'ersic) profile to the data,
we prefer to avoid this step which may introduce additional uncertainty.
H06 use the more direct approach which we use here as well.


\subsection{Comparison to Previous Work}

\begin{figure}
\plotone{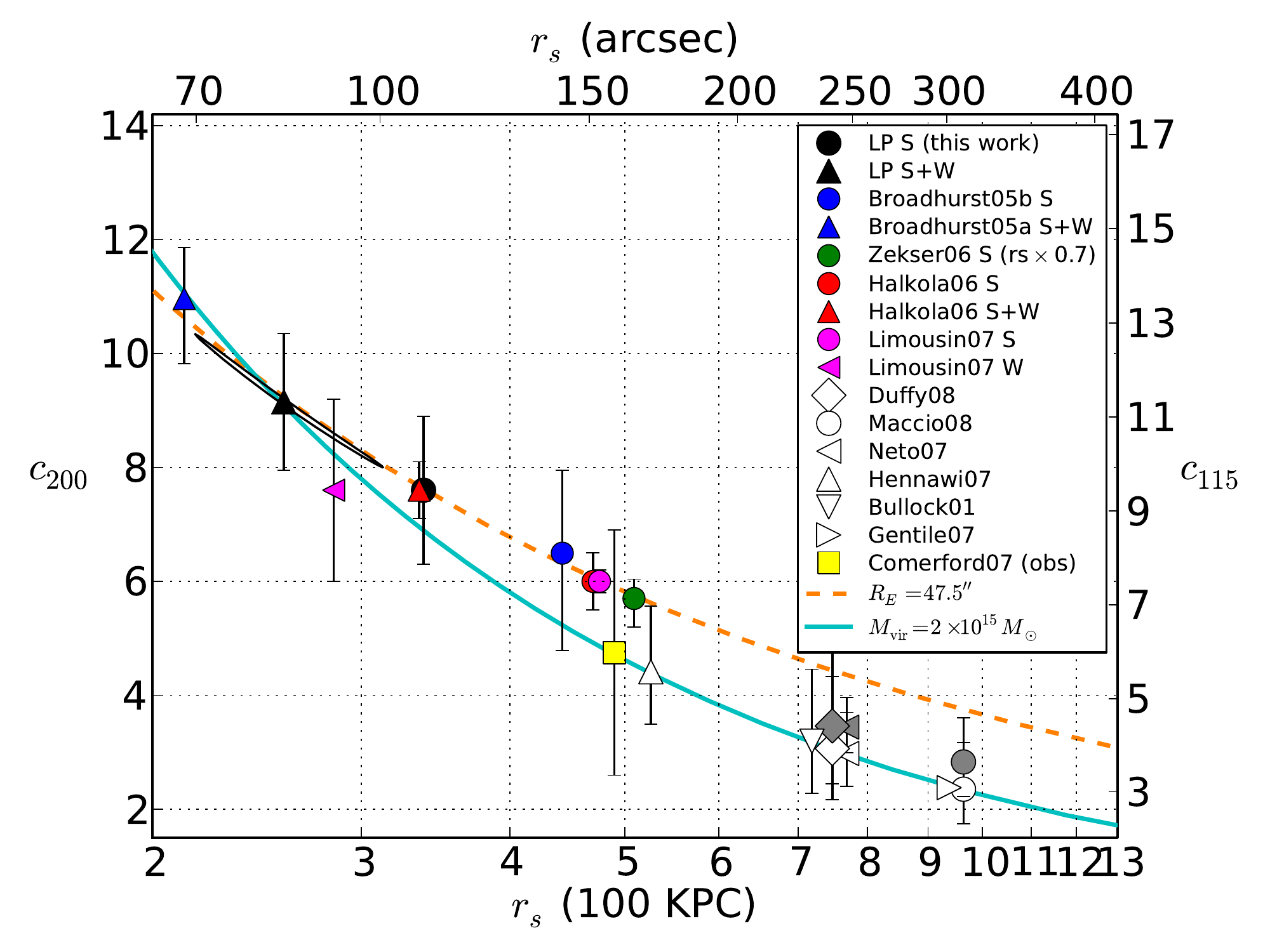}
\caption[NFW fit parameters]{\label{LPA1689:NFWfitcontours}%
NFW fit parameters ($r_s$, $c_{200}$ [$c_{vir}$]) found in this paper and in previous works
compared to those predicted from simulations.
Filled circles give best fit SL parameters from B05, Z06, H06, L07, and ours \NFWfitM.
Filled upright triangles give SL+WL fits from B05b, H06, and ours \NFWfitMW.
For clarity, error bars are plotted vertically, 
though they should actually all tightly follow the 
$R_E = 47\farcs5$ ($z_s = 2$) isocontour,
as shown for our SL+WL fit.
We also plot the WL fit obtained by L07.
This fit, with $R_E = 30\arcsec$ ($z_s = 2$),
is the only fit to fall far from the $R_E = 47\farcs5$ ($z_s = 2$) isocontour.
(Again note the error bar is not properly oriented.)
Along the isocontour $M_{200} = 2 \times 10^{15} M\odot$
(our best fit value),
we have plotted concentrations observed in simulations
\citep{Bullock01, Hennawi07, Neto07, Maccio08, Duffy08},
including the original NFW prescription \citep{NFW96} adapted to the present cosmology
by \cite{Gentile07}.
All assume a scaling of $c \propto (1+z)^{-1}$
except \cite{Duffy08} which find roughly $c \propto (1+z)^{-0.45}$.
Their shallower dependence on $z$ 
happens to bring their lower values back in line with some earlier predictions.
Filled symbols are plotted for relaxed subsets of the full samples.
Again the error bars are plotted vertically, 
but here should follow the $M_{200} = 2 \times 10^{15} M\odot$ isocontour.
Finally, the mass-concentration relation observed in clusters 
\citep{ComerfordNatarajan07} is plotted as the yellow square.
}\end{figure}

In Fig.~\ref{LPA1689:NFWfitcontours} we compare our NFW fit parameters
to published values from other similar studies:
fits to SL, WL, or both simultaneously.
These published values and more can also be found in Table \ref{LPA1689:tabcprev}.

\begin{figure}
\plotone{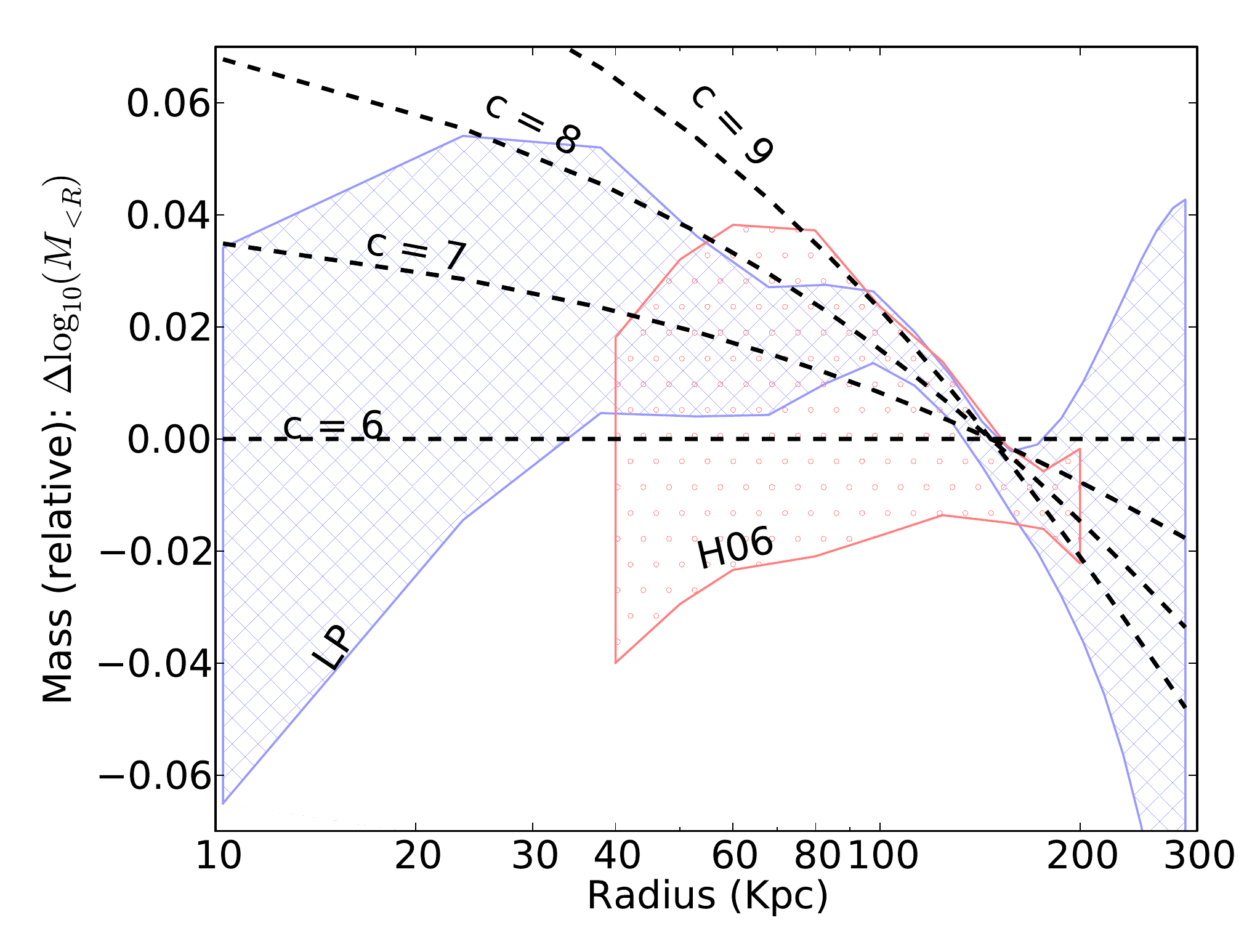}
\caption[Halkola comparison]{\label{LPA1689:Halkola}%
Comparison of our SL model \MR\ to that of H06.
Also plotted are four different NFW profiles with $c_{200} = 6$,7,8,9,
all with the correct Einstein radius.
All of the profiles are normalized to the $c_{200} = 6$ NFW profile.
Our SL model has a higher concentration \NFWfitMc\ than
that of H06 ($c_{200} = 6 \pm 0.5$).
}\end{figure}

Our concentration derived from SL alone 
\NFWfitMc\ is higher than that obtained in previous SL analyses ($c_{200} \sim 6$),
though our values agree roughly within the 1-$\sigma$ uncertainties.
In Fig.~\ref{LPA1689:Halkola}, we compare our SL $M(<R)$ 
to that obtained by \cite{Halkola06}.
Just inside the Einstein radius, our $M(<R)$ is steeper and claims a lower uncertainty.
This budges the concentration value higher.

We have neglected to mask out substructure
which can potentially be problematic for NFW fits \citep[e.g.,][]{Hennawi07}.
However we note the main subclump which we might exclude
is located between $\sim 150$ -- 200 kpc from the cluster center.
Our mass excess relative to H06 is at smaller radius, between $\sim 100$ -- 150 kpc.

From our SL+WL fits, we find \NFWfitMWct.
This lies between the values of
$c_{200} = 7.6^{+0.3}_{-0.5}$ found by H06 and
$c_{200} = 10.8^{+1.2}_{-0.8}$ found by B05b.
We note our results are in good agreement
with more recent analyses that have found values of $c_{200} \approx 10 \pm 0.7$
from SL+WL + number counts \citep{UmetsuBroadhurst08,Umetsu09}
and SL+WL + X-ray data \citep{Lemze08}.

We note our uncertainty contours closely follow
the $R_E = 47\farcs 5$ isocontour of constant Einstein radius.
The best fit NFW parameters obtained in previous SL or SL+WL studies 
all fall along this line,
as the Einstein radius is tightly constrained by SL.

\subsection{Tension between the SL and WL data?}

\begin{figure}
\plotone{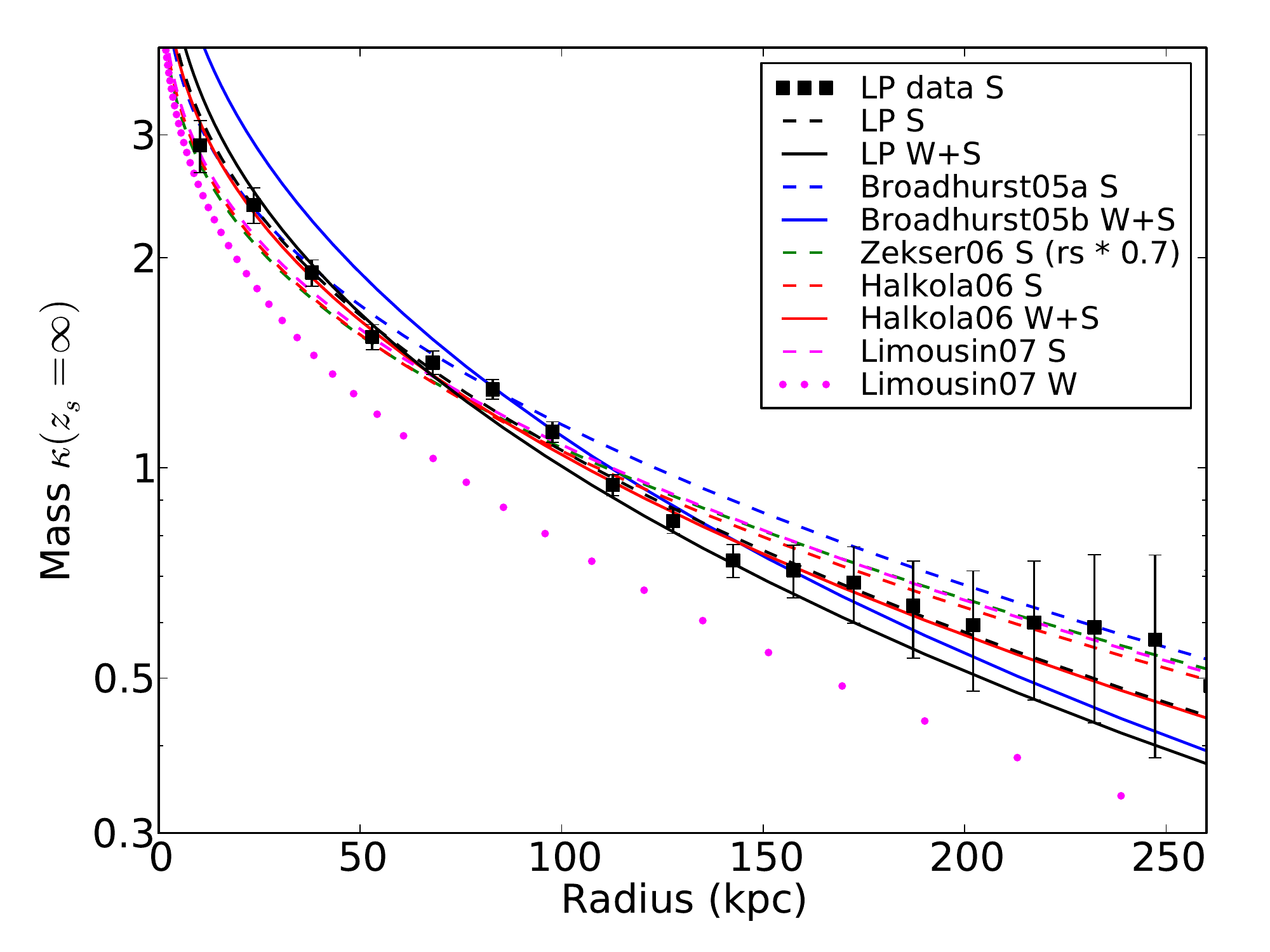}
\caption[NFW fits]{\label{LPA1689:NFWfits}%
Our SL-derived mass density profile \kR\ (black squares),
along with our NFW profile fits and those published elsewhere.
All fits which incorporate SL data
(dashed: SL only; solid: SL+WL)
provide reasonable fits to our SL data at most radii.
The L07 WL-only NFW fit (dotted line) underpredicts the mass surface density 
at all radii plotted here.
}\end{figure}

In every SL+WL study published to date for A1689, including ours,
the WL data prefer a higher concentration and steeper mass slope than the SL data
(Figs.~\ref{LPA1689:NFWfitcontours} and ~\ref{LPA1689:NFWfits}).
While we have found a single NFW profile which provides a decent fit
to both the SL and WL data,
there is a slight tension between the two.
While compatible at 1-$\sigma$ depending on the WL data set chosen
(see Fig.~\ref{LPA1689:NFWfitcontourscRE_SW}),
the WL signal is a bit too low
and/or falls off a bit too quickly relative to the SL data.
Alternatively, the inner SL mass profile is a bit too shallow.
We might suspect the WL signal has been diluted
from contamination of the sample by unlensed (foreground or cluster) objects.
However M07 carefully considered and quantified dilution in their work.

We note that these deviations from NFW
are qualitatively in accord with those proposed by \cite{LapiCavaliere09},
as found in their semi-analytic modeling of cluster formation.
This could prove interesting if similar variations are detected in many clusters.
Slight deviations from NFW are not surprising in a single cluster,
as simulated cluster profiles do exhibit intrinsic variations.

We note that L07 had claimed to resolve the discrepancy 
of SL versus WL concentrations.
They fit NFW profiles separately
to their SL and WL data,
finding concentration parameters of
$c_{200} = 6.0 \pm 0.6$ (3-$\sigma$) and
$c_{200} = 7.6 \pm 1.6$ (1-$\sigma$), respectively.
As the error bars overlap, they claimed agreement.
However as we showed in Fig.~\ref{LPA1689:NFWfitcontourscRE_SW},
SL-only and WL-only NFW fits may yield similar concentration values
but very different Einstein radii.
L07's best fit to the WL data yields $R_E \approx 30\arcsec$ ($z_s=2$),
significantly lower than the value $R_E \approx 47\arcsec$
tightly constrained by SL.
This fit underpredicts mass at all radii within the Einstein radius 
(Fig.~\ref{LPA1689:NFWfitcontours}).
L07 did not attempt to provide a single profile which provides an acceptable fit
to both the SL and WL data simultaneously.

\subsection{Comparison to Predictions}

Our results support findings that A1689 has a higher concentration
than predicted by simulations for a cluster of its mass.
According to \cite{Duffy08}, 
a cluster such as A1689
with $M_{200} \approx 2 \times 10^{15} M_\odot$ at $z = 0.186$
should have \cDuffy.
Relaxed clusters are more symmetric 
yielding better fits to NFW profiles and $\sim 20\%$ higher concentrations
for clusters of this mass: \crelDuffy.
Estimates from \cite{Maccio08} are similar ($\sim 10\%$ lower)
though the normal $c \propto (1+z)^{-1}$ scaling relation \citep{Bullock01}
drags the predictions down another $\sim 10\%$.
\cite{Duffy08} instead find roughly 
$c_{200} \propto (1+z)^{-0.45}$ and
$c_{vir} \propto (1+z)^{-0.70}$.

Predictions from \cite{Bullock01}, \cite{Gentile07}, and
\citet[who analyzed the Millennium simulation]{Neto07}
are also fairly similar to the \cite{Duffy08} predictions
despite concerns about each.
The Millennium simulation used a 
WMAP1 cosmology \citep{Spergel03} including $\sigma_8 = 0.9$.
\cite{Duffy08} and \cite{Maccio08} found this 
results in concentrations $\sim 15\%$ higher than their WMAP5 \citep{WMAP5} input
$\sigma_8 = 0.796$.\footnote{This value is in excellent agreement
with the WMAP 7-year maximum likelihood value $\sigma_8 = 0.803$
\citep{WMAP7}.}
The \cite{Bullock01} simulations used $\sigma_8 = 1.0$ 
and did not produce halos as massive as A1689.
\cite{Zhao03b} cites the dangers of extrapolating these results to higher mass.
\cite{Gentile07} is a reformulation of the original 
\cite{NFW96} prescription to WMAP3 \citep{WMAP3}.
Those early simulations were lower resolution and produced only 19 halos
(compared to $\sim$1,000 and $\sim$10,000
for \citealt{Duffy08} and \citealt{Maccio08}, respectively).

\cite{Hennawi07} measure significantly larger concentrations in their simulations.
For A1689, they predict $c_{200} = 4.4^{+1.2}_{-0.9}$,
$\sim 50\%$ larger than predicted by \cite{Duffy08}.
Their use of $\sigma_8 = 0.95$ probably only results in concentrations
inflated by $\sim 20\%$.
The remaining disagreement may be a result of their halo density fitting procedure
which they claim is better for comparison with lensing measurements.

All of these predictions
\citep[detailed further in][]{Coe10DMprofiles}
are plotted in Fig.~\ref{LPA1689:NFWfitcontours}. 
The predictions are all significantly lower than our SL+WL-derived \NFWfitMWct.

Clusters selected with a lensing bias may have 
measured concentrations biased high by
$\sim 34\%$ \cite{Hennawi07},
$\sim 50\%$ \cite{OguriBlandford09},
or even
$\sim 100\%$ \cite{Meneghetti10}, 
on average.
This results from a combination of 
higher intrinsic (3-D) concentrations
and additional mass along the line of sight
(either due to cluster elongation or otherwise)
resulting in higher projected (2-D) concentrations.
The $c_{200} = 4.4^{+1.2}_{-0.9}$ predicted by \cite{Hennawi07}
plus a $\sim 100\%$ bias might begin to explain measurements of $c_{200} \sim 9$.
However such extremely high biases are expected only for less massive clusters.
Even accounting for such biases, 
the high concentration of A1689 seems unlikely given the results from 
$\Lambda$CDM simulations
\citep{BroadhurstBarkana08,OguriBlandford09}.

There are fewer published fits of S\'ersic profiles to simulated cluster halos.
We do note that \cite{Merritt05} found $n = 2.38 \pm 0.25$ for their cluster sample.\footnote{Alternatively, fitting a deprojected S\'ersic profile 
to the 3D spatial density $\rho(r)$,
\cite{Merritt05} and \cite{Merritt06} found 
$n_d = 2.99 \pm 0.49$ and $n_d = 2.89 \pm 0.49$, respectively.
The deprojection they used \citep{PrugnielSimien97} is approximate
and thus does not yield exactly the same $n$ as fitting directly to the surface density.}
Our best fit \SersicfitMWn\ is much lower 
giving our density profile a more rapidly varying slope.
This allows it to fit both the SL profile and the low WL signal 
measured by some authors at large radius.
We have not experimented with fitting S\'ersic profiles
to various subsets of the WL data.

Recently the Einasto profile has gained popularity over the S\'ersic profile
\citep[e.g.,][]{Navarro10}.
The two have similar forms,
but the former gives density $\rho(r)$ as a function of 3-D radius,
while the latter gives surface density $\Sigma(R)$ as a function of projected 2-D radius.
The former was found \citep{Merritt06} 
to provide better fits to halos of a wide range of masses,
though the latter performed slightly better specifically for cluster halos
(and only slightly worse for galaxy halos).
Mass concentration relations derived from Einasto profile fits 
have been published by \cite{Duffy08,Gao08,HayashiWhite08}.
These fits yield slightly ($< 20\%$) different concentrations than NFW fits 
\citep{Coe10DMprofiles}.
We do not explore Einasto profile fits here.

The S\'ersic profile remains intriguing for the direct comparisons which can be made 
\citep{Merritt05} to many published S\'ersic fits to galaxy luminosity profiles,
\citep[although this may just be coincidental][]{DharWilliams10}.
We also note the possibility of
using well established software such as GALFIT \citep{galfit} 
to derive S\'ersic parameters for surface density mass maps.





\section{Future Work}
\label{LPA1689:future}

In this paper, we have focused on measurement of the radial mass profile of A1689.
However, a key strength of LensPerfect is its ability to map massive substructure
without assuming LTM.
In future work, we will verify in detail our ability to resolve halo subclumps
as well as measure their masses.
The latter proves difficult (both in observed and simulated halos)
as the subhalo masses must be disentangled cleanly from the greater parent halo
\citep[e.g.,][]{Natarajan09,JulloKneib09}.

One novel method developed recently by \cite{Powell09}
demonstrates the ability to detect subclumps in a 2-D mass map
down to $10^{13}$ or even $10^{12}$ $M_\odot$
and measure their masses to within a factor of two.
By weighing subhalos associated with cluster galaxies,
we may provide evidence for galaxy halo stripping in individual galaxies.
This would provide an excellent complement to studies which have measured
stripping ``globally'', averaged over many galaxy halos
\citep{Natarajan98, Natarajan02, Gavazzi04, Limousin07, Halkola07, Natarajan09}.

\cite{WilliamsSaha04} and \cite{Saha07}
have experimented with several methods to identify substructure
in their non-LTM (PixeLens) mass models.
From the 2-D mass density map \kth, 
they have subtracted each of the following:
the average \kR\ in that radial bin;
\kth\ $180^\circ$ across (directly opposite w.r.t.~the center);
or the best fitting NFW profile.
In \cite{Saha07}, they show that the observed substructure
(extended ``meso-structure'')
appears to correlate with the luminous galaxies.

In future work we will develop algorithms to 
thoroughly explore the range of model solutions
which perfectly reproduce all observed multiple image positions.
This method must take care 
not to remain trapped in a local minimum near our ``best'' solution.
Also, we must correctly account for
the larger uncertainties in voids between the multiple images.

In \cite{Coe08}, we discussed our ability to mold the mass map
by adding extra artificial constraints.
These modified mass models would successfully reproduce all of the input data
plus the artificial constraints.
These added constraints squeeze the mass model,
tweaking the positions of subclumps
or increasing the concentration of mass in the desired regions.
We must explore such solutions to accurately account for all uncertainties.
We may also mold our mass models 
in attempts to force mass to follow light more closely.
We might construct that mass model which follows light best, 
as in \cite[for example][]{SahaWilliams97}.

Ultimately a hybrid approach combining non-LTM and flexible-LTM components
may prove ideal.
A prior of LATM (light {\it approximately} traces mass) could be assumed. 
This ideal method would include 
a parent halo, galaxy components, and line of sight structure,
all with {\it sufficient flexibility}.
Each galaxy component might be allowed to vary individually in 
M/L, radial scale, truncation radius, and perhaps position.
Different forms may even be explored:
truncated isothermal ellipsoid versus NFW, for example.
The parent halo should be very flexible (a multi-scale grid perhaps)
to allow for the asymmetries induced
as galaxies infall and their stripped mass is strewn about the cluster.
Additional mass planes behind the lens should also be modeled.

Future deeper observations of galaxy clusters such as A1689 
may reveal hundreds of multiple images.
This wealth of constraints will allow truly high-definition mass models,
which clearly resolve individual galaxy halos and perhaps dark subhalos as well
\citep{Coe09decadal}.
As more multiple images are revealed and greater details are obtained,
line of sight structure will need to be taken into account.
While seemingly a nuisance, this raises the prospect of 
strong lensing tomography -- the mapping of mass in multiple lens planes.

Over the next 2 to 3 years we are looking forward to new HST images
from the CLASH Multi-Cycle Treasury Program.
As mentioned in the introduction, this program will image 25 clusters 
to a depth of 20 orbits each,
equal to the depth of the A1689 images studied here.
Analysis of these ACS and WFC3 images along with supporting data
will yield measurements of mass concentration for a sizable sample of clusters,
selected free of lensing bias.
By comparing these values to the concentrations of simulated halos,
we expect to either show agreement
or detect average deviations as small as 15\% with 99\% confidence.


\section{Summary}
\label{LPA1689:summary}

We have presented a strong lensing mass model of A1689 
which resolves structures down to $\sim 25$ kpc on average
within the central 400 kpc diameter
without assuming light traces mass.
The most luminous galaxies appear to trace the mass distribution fairly well
but with some deviations which may prove interesting, pending verification.
This is the highest resolution mass map of any galaxy cluster to date.
Our mass model perfectly reproduces the observed positions of
\Nknotimages multiple images of \Nknotss knots 
within \Nimages images of \Ngals galaxies.
Included are 20 new candidate multiple images of 8 galaxies 
which we have identified in this work.
We have also tweaked some of the identifications from previous works,
discarding three suspect central images and adding one new one.

Compiling published weak lensing measurements 
from ACS, Subaru, and CFHT images,
we find that a single mass profile, either NFW or S\'ersic,
is able to provide a decent fit simultaneously
to both the observed weak and strong lensing.
However there remains a slight tension 
in that the weak lensing data prefer higher concentrations than the strong lensing data.

Based on simultaneous fitting of the strong and weak lensing data,
we measure a NFW central mass concentration of \NFWfitMWct.
Thus we concur with previous claims that the mass profile of A1689 
appears to be more centrally concentrated
than clusters of similar mass 
($M_{200} \sim 2 \times 10^{15} M_\odot \hunits = 1.4 \times 10^{15} M_\odot h^{-1}$)
formed in CDM simulations ($c_{200} \sim 3$).

In future work, we will perform further simulations
and verify our ability to not only resolve substructure
but measure the masses of these halo subclumps.
Backed by these tests, future analyses of our substructure maps
should yield further evidence for galaxy halo stripping in cluster environments,
test the degree to which light traces mass,
and perhaps even lead to detection of dark subhalos, should they exist.



\acknowledgements{
We would like to thank Marceau Limousin and Elinor Medezinski
for sending us their weak lensing profiles and for useful conversations.
We also thank Angelo Neto for useful conversations 
about the Millennium simulation and their study of halo profiles.
This work was carried out in part at Jet Propulsion Laboratory,
California Institute of Technology, under a contract with NASA.
LAM acknowledges support from the NASA ATFP program.




}

\bibliographystyle{astroads}
\bibliography{paperstrunc}

\LongTables
\begin{deluxetable}{rcccc}
\tablewidth{0pt}
\tablecaption{\label{LPA1689:zspecA}Spectroscopic Redshifts
of galaxies within the ACS field of view of Abell 1689
used for our photometric redshift tests}
\tablehead{
\colhead{ID}&
\multicolumn{2}{c}{RA \& DEC (J2000)\tablenotemark{a}}&
\colhead{Survey\tablenotemark{b}}&
\colhead{$z_{spec}$}
}
\startdata
451&
13:11:32.606&
-1:19:22.920&
Miekse&
$0.0000$\\
366&
13:11:24.802&
-1:20:23.070&
Frye&
$0.0000$\\
374&
13:11:26.919&
-1:20:00.650&
Frye&
$0.0000$\\
216&
13:11:28.344&
-1:18:32.380&
MOS&
$0.0130$\\
213&
13:11:27.224&
-1:20:10.180&
MOS&
$0.0862$\\
220&
13:11:33.034&
-1:21:24.780&
MOS&
$0.1430$\\
74&
13:11:25.382&
-1:20:17.140&
Balogh*&
$0.1660$\\
7&
13:11:32.089&
-1:19:36.600&
Balogh*&
$0.1690$\\
127&
13:11:28.194&
-1:20:43.680&
MOS*&
$0.1712$\\
194&
13:11:30.934&
-1:20:30.680&
Teague*&
$0.1739$\\
387&
13:11:30.508&
-1:19:34.670&
Frye&
$0.1740$\\
378&
13:11:27.851&
-1:20:07.650&
Frye&
$0.1750$\\
192&
13:11:30.234&
-1:20:27.680&
Teague*&
$0.1750$\\
124&
13:11:27.894&
-1:21:36.880&
MOS*&
$0.1752$\\
184&
13:11:26.884&
-1:19:37.180&
Teague*&
$0.1754$\\
43&
13:11:37.981&
-1:20:09.780&
Balogh*&
$0.1760$\\
108&
13:11:24.284&
-1:21:14.480&
MOS*&
$0.1766$\\
198&
13:11:32.094&
-1:21:38.380&
Teague*&
$0.1770$\\
4&
13:11:28.344&
-1:19:58.300&
Balogh*&
$0.1775$\\
154&
13:11:32.464&
-1:22:18.280&
MOS*&
$0.1794$\\
112&
13:11:25.104&
-1:19:31.380&
MOS*&
$0.1797$\\
383&
13:11:28.886&
-1:20:01.940&
Frye&
$0.1800$\\
141&
13:11:29.994&
-1:20:17.580&
MOS*&
$0.1801$\\
152&
13:11:32.104&
-1:19:47.180&
MOS*&
$0.1801$\\
61&
13:11:31.510&
-1:19:24.870&
Balogh*&
$0.1817$\\
9&
13:11:34.058&
-1:21:01.990&
Balogh*&
$0.1826$\\
370&
13:11:26.237&
-1:19:56.450&
Frye&
$0.1830$\\
201&
13:11:35.374&
-1:20:43.080&
Teague*&
$0.1835$\\
64&
13:11:29.386&
-1:18:34.790&
Balogh*&
$0.1836$\\
172&
13:11:37.834&
-1:19:20.880&
MOS*&
$0.1839$\\
375&
13:11:27.148&
-1:18:48.440&
Frye&
$0.1840$\\
135&
13:11:29.474&
-1:20:28.080&
MOS*&
$0.1842$\\
118&
13:11:27.084&
-1:18:48.880&
MOS*&
$0.1852$\\
153&
13:11:32.204&
-1:22:10.980&
MOS*&
$0.1855$\\
188&
13:11:29.074&
-1:21:37.380&
Teague*&
$0.1858$\\
132&
13:11:28.724&
-1:19:02.980&
MOS*&
$0.1859$\\
450&
13:11:32.761&
-1:19:48.930&
Miekse&
$0.1859$\\
63&
13:11:29.869&
-1:20:15.230&
Balogh*&
$0.1868$\\
165&
13:11:35.344&
-1:21:33.480&
MOS*&
$0.1870$\\
69&
13:11:28.023&
-1:18:43.890&
Balogh*&
$0.1870$\\
110&
13:11:24.434&
-1:21:11.180&
MOS*&
$0.1870$\\
147&
13:11:31.064&
-1:21:27.980&
MOS*&
$0.1872$\\
384&
13:11:29.100&
-1:19:46.920&
Frye&
$0.1880$\\
6&
13:11:31.400&
-1:19:32.840&
Balogh*&
$0.1884$\\
196&
13:11:31.074&
-1:20:52.780&
Teague*&
$0.1885$\\
419&
13:11:38.051&
-1:19:58.250&
Frye&
$0.1890$\\
390&
13:11:31.472&
-1:21:05.940&
Frye&
$0.1890$\\
163&
13:11:34.754&
-1:20:59.480&
MOS*&
$0.1895$\\
189&
13:11:29.114&
-1:21:55.480&
Teague*&
$0.1908$\\
131&
13:11:28.594&
-1:20:26.780&
MOS*&
$0.1909$\\
70&
13:11:26.812&
-1:19:43.090&
Balogh*&
$0.1910$\\
146&
13:11:30.564&
-1:20:43.980&
MOS*&
$0.1918$\\
1&
13:11:25.354&
-1:20:37.060&
Balogh*&
$0.1922$\\
133&
13:11:29.294&
-1:19:16.980&
MOS*&
$0.1932$\\
187&
13:11:28.994&
-1:21:16.980&
Teague*&
$0.1947$\\
71&
13:11:25.925&
-1:19:51.950&
Balogh*&
$0.1950$\\
119&
13:11:27.094&
-1:21:43.080&
MOS*&
$0.1955$\\
12&
13:11:36.616&
-1:19:42.800&
Balogh*&
$0.1960$\\
148&
13:11:31.124&
-1:21:25.380&
MOS*&
$0.1972$\\
123&
13:11:27.834&
-1:21:13.080&
MOS*&
$0.1977$\\
115&
13:11:25.974&
-1:19:35.680&
MOS*&
$0.1983$\\
140&
13:11:29.984&
-1:22:07.580&
MOS*&
$0.1985$\\
11&
13:11:35.595&
-1:20:12.400&
Balogh*&
$0.1995$\\
128&
13:11:28.324&
-1:18:45.180&
MOS*&
$0.1999$\\
395&
13:11:33.234&
-1:19:17.000&
Frye&
$0.2000$\\
139&
13:11:29.974&
-1:20:40.480&
MOS*&
$0.2002$\\
158&
13:11:32.704&
-1:19:32.080&
MOS*&
$0.2009$\\
143&
13:11:30.044&
-1:20:42.980&
MOS*&
$0.2012$\\
452&
13:11:29.905&
-1:20:05.430&
Miekse&
$0.2014$\\
157&
13:11:32.654&
-1:19:58.980&
MOS*&
$0.2022$\\
2&
13:11:27.109&
-1:20:58.420&
Balogh*&
$0.2147$\\
82&
13:11:23.065&
-1:21:17.280&
Balogh*&
$0.2150$\\
3&
13:11:27.681&
-1:21:07.190&
Balogh*&
$0.2158$\\
398&
13:11:33.555&
-1:19:01.500&
Frye&
$0.2440$\\
221&
13:11:33.724&
-1:19:39.180&
MOS&
$0.3100$\\
214&
13:11:27.844&
-1:18:52.980&
MOS&
$0.3840$\\
399&
13:11:33.621&
-1:22:02.190&
Frye&
$0.3870$\\
376&
13:11:27.204&
-1:18:49.920&
Frye&
$0.4800$\\
365&
13:11:24.652&
-1:20:03.380&
Frye&
$0.4810$\\
405&
13:11:35.088&
-1:21:26.010&
Frye&
$0.5840$\\
406&
13:11:35.258&
-1:20:30.170&
Frye&
$0.5870$\\
418&
13:11:37.699&
-1:19:50.030&
Frye&
$0.6250$\\
388&
13:11:30.677&
-1:18:55.500&
Frye&
$0.6760$\\
402&
13:11:34.328&
-1:19:05.100&
Frye&
$0.6760$\\
389&
13:11:30.751&
-1:21:38.780&
Frye&
$0.6910$\\
379&
13:11:28.221&
-1:20:50.930&
Frye&
$0.7030$\\
380&
13:11:28.325&
-1:18:27.500&
Frye&
$0.7100$\\
367&
13:11:24.960&
-1:19:36.610&
Frye&
$0.7220$\\
364&
13:11:24.609&
-1:19:20.830&
Frye&
$0.7570$\\
411&
13:11:36.536&
-1:19:25.030&
Frye&
$0.7900$\\
396&
13:11:33.028&
-1:19:14.650&
Frye&
$0.7900$\\
217&
13:11:28.764&
-1:21:43.580&
MOS&
$0.7900$\\
414&
13:11:37.183&
-1:20:16.890&
Frye&
$0.8130$\\
362&
13:11:24.240&
-1:19:52.680&
Frye&
$0.8570$\\
363&
13:11:24.398&
-1:19:36.880&
Frye&
$0.8950$\\
409&
13:11:36.052&
-1:19:24.710&
Frye&
$0.9160$\\
413&
13:11:37.089&
-1:19:26.100&
Frye&
$0.9240$\\
373&
13:11:26.683&
-1:19:36.770&
Frye&
$0.9590$\\
393&
13:11:32.056&
-1:21:55.410&
Frye&
$0.9590$\\
356&
13:11:22.472&
-1:20:39.220&
Frye&
$0.9600$\\
401&
13:11:34.242&
-1:19:23.970&
Frye&
$1.0510$\\
377&
13:11:27.191&
-1:18:26.550&
Frye&
$1.1120$\\
361&
13:11:24.186&
-1:19:56.550&
Frye&
$1.1550$\\
400&
13:11:33.950&
-1:19:15.750&
Frye&
$1.3620$\\
306&
13:11:33.981&
-1:20:50.950&
Broadhurst&
$1.3700$\\
302&
13:11:27.359&
-1:20:54.900&
Broadhurst&
$1.8200$\\
301&
13:11:30.294&
-1:19:50.920&
Broadhurst&
$1.8200$\\
371&
13:11:26.541&
-1:19:55.700&
Frye&
$2.5360$\\
307&
13:11:33.065&
-1:20:27.450&
Broadhurst&
$3.0410$\\
308&
13:11:26.441&
-1:19:56.940&
Broadhurst&
$3.0410$\\
385&
13:11:29.969&
-1:19:14.880&
Frye&
$3.8500$\\
368&
13:11:25.447&
-1:20:51.740&
Frye&
$4.8680$\\
404&
13:11:34.994&
-1:19:51.170&
Frye&
$5.1200$\\
\vspace{-0.1in}
\enddata
\tablenotetext{a}{Slight modifications have been made
to the previously published coordinates where necessary.}
\tablenotetext{b}{Redshift surveys are 
\cite{Balogh02}, MOS \citep{Duc02}, \cite{Teague90}, 
\cite{Broadhurst05}, and \cite{Frye02, Frye07}.
Asterisks (*) indicate galaxies designated as 
``spectroscopically-confirmed cluster members'' in each survey.}
\end{deluxetable}



\clearpage
\begin{deluxetable}{rlcrrccccclr}
\tablecaption{\label{LPA1689:tabmultimages}Multiple Images Produced by Abell 1689}
\tablehead{
\multicolumn{2}{c}{}&
\colhead{ID\tablenotemark{a}}&
\colhead{$x$\tablenotemark{b}}&
\colhead{$y$\tablenotemark{b}}&
\colhead{R.A.}&
\colhead{Decl.}&
\colhead{$i\arcmin_{775}$}&
\colhead{}&
\colhead{}&
\colhead{}&
\colhead{}\\
\multicolumn{2}{c}{ID}&
\colhead{B05}&
\colhead{(pixel)}&
\colhead{(pixel)}&
\colhead{(J2000.0)}&
\colhead{(J2000.0)}&
\colhead{(AB mag)}&
\colhead{$z_{spec}$\tablenotemark{c}}&
\colhead{Ref.\tablenotemark{d}}&
\colhead{$z_b$\tablenotemark{e}}&
\colhead{$\chi^2_{mod}$\tablenotemark{f}}
}
\startdata
1&
a&
''&
2213&
1055&
13 11 26.452&
-1 19 56.75&
$23.309 \pm 0.008$&
3.04&
2,3&
$3.21^{+0.65}_{-0.41}$&
  1.06\\
 &
b&
''&
2130&
1040&
13 11 26.289&
-1 20 00.19&
$23.595 \pm 0.009$&
(3.04)&
\nodata&
$3.88 \pm 0.48$&
  1.08\\
 &
c&
''&
1356&
2557&
13 11 29.773&
-1 21 07.43&
$24.509 \pm 0.017$&
3.04&
2,3&
$3.79 \pm 0.47$&
  3.46\\
 &
d&
''&
2499&
3112&
13 11 33.066&
-1 20 27.47&
$24.033 \pm 0.011$&
(3.04)&
\nodata&
$3.76 \pm 0.47$&
  2.37\\
 &
e&
''&
2745&
2621&
13 11 31.932&
-1 20 05.91&
$24.638 \pm 0.018$&
(3.04)&
\nodata&
$3.17^{+0.72}_{-0.41}$&
  0.57\\
 &
f&
''&
1890&
2333&
13 11 29.852&
-1 20 38.50&
$24.942 \pm 0.050$&
(3.04)&
\nodata&
$0.51^{+2.66}_{-0.41}$&
  2.22\\
2&
a&
''&
2245&
1064&
13 11 26.524&
-1 19 55.49&
$23.238 \pm 0.013$&
2.53&
3&
$0.16^{+0.30}_{-0.11}$&
  0.27\\
 &
b&
''&
2522&
3069&
13 11 32.969&
-1 20 25.51&
$23.912 \pm 0.015$&
(2.53)&
\nodata&
$0.20^{+0.25}_{-0.12}$&
  0.47\\
 &
c&
''&
2728&
2644&
13 11 31.978&
-1 20 07.17&
$24.285 \pm 0.024$&
(2.53)&
\nodata&
$2.69^{+0.36}_{-2.48}$&
  9.83\\
 &
d&
''&
1386&
2556&
13 11 29.812&
-1 21 06.05&
$24.351 \pm 0.025$&
2.5&
4&
$2.66^{+0.41}_{-2.46}$&
  0.90\\
 &
e&
''&
1876&
2349&
13 11 29.881&
-1 20 39.48&
$24.806 \pm 0.041$&
(2.53)&
\nodata&
$0.31^{+2.90}_{-0.21}$&
  2.02\\
3&
a&
''&
2372&
2832&
13 11 32.041&
-1 20 27.27&
$26.099 \pm 0.056$&
\nodata&
\nodata&
$0.96 \pm 0.19$&
  4.30\\
 &
b&
''&
2279&
2921&
13 11 32.178&
-1 20 33.37&
$26.287 \pm 0.054$&
\nodata&
\nodata&
$5.22 \pm 0.61$&
  1.03\\
 &
c&
''&
1809&
2984&
13 11 31.703&
-1 20 55.99&
$27.243 \pm 0.129$&
\nodata&
\nodata&
$0.81^{+4.32}_{-0.35}$&
  3.72\\
4&
a&
''&
1844&
3124&
13 11 32.175&
-1 20 57.37&
$24.172 \pm 0.017$&
1.1&
4&
$0.79^{+0.47}_{-0.18}$&
  0.09\\
 &
b&
''&
1369&
2801&
13 11 30.528&
-1 21 12.02&
$23.534 \pm 0.015$&
(1.1)&
\nodata&
$1.02^{+0.53}_{-0.20}$&
  1.91\\
 &
c&
''&
2553&
2322&
13 11 30.758&
-1 20 08.25&
$25.021 \pm 0.027$&
(1.1)&
\nodata&
$1.06^{+0.20}_{-0.34}$&
  0.77\\
 &
d&
''&
1492&
1338&
13 11 26.285&
-1 20 35.40&
$24.610 \pm 0.020$&
(1.1)&
\nodata&
$0.78^{+0.51}_{-0.17}$&
  0.80\\
 &
e&
''&
2052&
2252&
13 11 29.837&
-1 20 29.45&
$27.171 \pm 0.325$&
(1.1)&
\nodata&
$0.68^{+2.84}_{-0.49}$&
  0.89\\
5&
a&
''&
1606&
2205&
13 11 29.064&
-1 20 48.64&
$24.290 \pm 0.030$&
2.6&
4&
$2.91^{+0.38}_{-0.48}$&
  0.03\\
 &
b&
''&
1706&
2211&
13 11 29.224&
-1 20 44.24&
$24.676 \pm 0.041$&
(2.6)&
\nodata&
$0.25^{+2.91}_{-0.14}$&
  1.68\\
 &
c&
''&
2751&
3343&
13 11 34.120&
-1 20 20.96&
$25.103 \pm 0.027$&
2.6&
4&
$3.00 \pm 0.39$&
  1.18\\
6&
a&
''&
3097&
2066&
13 11 30.755&
-1 19 38.19&
$23.521 \pm 0.015$&
1.1&
4&
$1.17 \pm 0.21$&
  0.03\\
 &
b&
''&
2811&
3058&
13 11 33.345&
-1 20 12.20&
$23.624 \pm 0.016$&
1.1&
4&
$1.36 \pm 0.23$&
  0.73\\
 &
c&
''&
3055&
2744&
13 11 32.742&
-1 19 54.49&
$22.131 \pm 0.006$&
(1.1)&
\nodata&
$0.98 \pm 0.19$&
  1.48\\
 &
d&
''&
2943&
2709&
13 11 32.478&
-1 19 58.81&
$21.827 \pm 0.003$&
(1.1)&
\nodata&
$0.53^{+0.25}_{-0.15}$&
  4.20\\
7&
a&
''&
1087&
1250&
13 11 25.446&
-1 20 51.87&
$23.292 \pm 0.007$&
4.87&
1&
$4.93 \pm 0.58$&
  0.05\\
 &
b&
''&
2439&
2349&
13 11 30.678&
-1 20 13.99&
$24.128 \pm 0.014$&
4.8&
4&
$4.86 \pm 0.57$&
  1.75\\
8&
a&
''&
1974&
3105&
13 11 32.302&
-1 20 51.09&
$22.011 \pm 0.009$&
\nodata&
\nodata&
$0.15^{+0.31}_{-0.11}$&
  0.30\\
 &
b&
''&
1596&
2984&
13 11 31.402&
-1 21 05.63&
$21.587 \pm 0.007$&
\nodata&
\nodata&
$0.48^{+0.15}_{-0.39}$&
  1.36\\
 &
c&
''&
2541&
2572&
13 11 31.495&
-1 20 14.10&
$25.550 \pm 0.050$&
\nodata&
\nodata&
$0.22^{+2.86}_{-0.12}$&
  4.42\\
 &
d&
''&
1674&
1001&
13 11 25.526&
-1 20 20.01&
$22.278 \pm 0.005$&
\nodata&
\nodata&
$0.50 \pm 0.15$&
  1.22\\
9&
a&
''&
2850&
2032&
13 11 30.303&
-1 19 48.65&
$25.661 \pm 0.041$&
\nodata&
\nodata&
$4.80 \pm 0.57$&
  4.18\\
 &
b&
''&
2141&
3430&
13 11 33.519&
-1 20 50.42&
$27.269 \pm 0.087$&
\nodata&
\nodata&
$0.79^{+4.33}_{-0.18}$&
  1.01\\
 &
c&
''&
1072&
2347&
13 11 28.737&
-1 21 15.83&
$25.560 \pm 0.036$&
\nodata&
\nodata&
$0.76^{+3.94}_{-0.17}$&
  3.99\\
 &
d&
''&
1645&
1264&
13 11 26.279&
-1 20 26.90&
$26.941 \pm 0.072$&
\nodata&
\nodata&
$4.95^{+0.58}_{-4.23}$&
  1.40\\
10&
a&
''&
2189&
3560&
13 11 33.980&
-1 20 51.01&
$23.011 \pm 0.009$&
1.83&
4&
$0.26^{+2.32}_{-0.21}$&
  0.49\\
 &
b&
''&
2130&
1625&
13 11 28.055&
-1 20 12.61&
$22.655 \pm 0.009$&
(1.83)&
\nodata&
$0.12^{+0.60}_{-0.11}$&
  0.61\\
 &
c&
''&
2012&
2098&
13 11 29.316&
-1 20 27.99&
$23.605 \pm 0.018$&
(1.83)&
\nodata&
$0.56^{+0.15}_{-0.37}$&
  6.72\\
11&
a&
''&
1824&
3522&
13 11 33.349&
-1 21 06.73&
$23.794 \pm 0.015$&
2.5&
4&
$2.89 \pm 0.38$&
  0.15\\
 &
b&
''&
2462&
1801&
13 11 29.056&
-1 20 01.31&
$23.402 \pm 0.015$&
(2.5)&
\nodata&
$2.82^{+0.37}_{-2.52}$&
  0.07\\
 &
c&
''&
2062&
2135&
13 11 29.498&
-1 20 26.51&
$23.729 \pm 0.015$&
(2.5)&
\nodata&
$0.53^{+0.20}_{-0.22}$&
  2.30\\
12&
b&
''&
1275&
1796&
13 11 27.361&
-1 20 54.94&
$24.178 \pm 0.019$&
1.83&
2,4&
$0.60^{+2.01}_{-0.45}$&
  0.64\\
 &
c&
''&
1311&
1730&
13 11 27.213&
-1 20 51.91&
$23.845 \pm 0.015$&
(1.83)&
\nodata&
$0.60 \pm 0.16$&
  0.42\\
 &
f&
\nodata&
1602&
1547&
13 11 27.072&
-1 20 34.86&
$25.403 \pm 0.075$&
(1.83)&
\nodata&
$2.10^{+0.48}_{-0.60}$&
  9.22\\
 &
g&
\nodata&
2197&
1563&
13 11 27.963&
-1 20 08.26&
$22.833 \pm 0.005$&
(1.83)&
\nodata&
$1.64^{+0.26}_{-1.23}$&
  3.48\\
 &
h&
\nodata&
2190&
3463&
13 11 33.688&
-1 20 48.91&
$28.078 \pm 0.150$&
(1.83)&
\nodata&
$2.00^{+1.26}_{-1.78}$&
  7.13\\
 &
i&
\nodata&
2224&
1497&
13 11 27.802&
-1 20 05.64&
$23.806 \pm 0.010$&
(1.83)&
\nodata&
$0.19^{+0.37}_{-0.12}$&
  6.36\\
13&
a&
''&
3610&
2512&
13 11 32.828&
-1 19 24.44&
$23.632 \pm 0.020$&
\nodata&
\nodata&
$1.03 \pm 0.20$&
  1.50\\
 &
b&
''&
3605&
2567&
13 11 32.986&
-1 19 25.83&
$23.764 \pm 0.018$&
\nodata&
\nodata&
$0.37 \pm 0.13$&
  0.88\\
 &
c&
''&
3559&
2725&
13 11 33.398&
-1 19 31.27&
$23.602 \pm 0.018$&
\nodata&
\nodata&
$1.16^{+0.33}_{-0.21}$&
  1.60\\
14&
a&
''&
 639&
2648&
13 11 29.033&
-1 21 41.82&
$25.141 \pm 0.040$&
3.4&
4&
$0.55^{+2.66}_{-0.40}$&
115.58\\
 &
b&
''&
 677&
2772&
13 11 29.461&
-1 21 42.73&
$25.700 \pm 0.052$&
(3.4)&
\nodata&
$3.37^{+0.44}_{-0.43}$&
  3.74\\
15&
a&
15.2&
1356&
2557&
13 11 29.773&
-1 21 07.43&
$25.508 \pm 0.044$&
1.8&
4&
$1.91^{+0.45}_{-1.63}$&
  2.50\\
 &
b&
15.1&
2213&
1055&
13 11 26.452&
-1 19 56.75&
$24.722 \pm 0.040$&
(1.8)&
\nodata&
$1.73^{+0.84}_{-1.60}$&
  2.87\\
 &
c&
''&
2009&
2074&
13 11 29.239&
-1 20 27.62&
$25.622 \pm 0.060$&
(1.8)&
\nodata&
$0.43^{+1.96}_{-0.24}$&
  1.15\\
16&
a&
''&
1892&
1715&
13 11 27.990&
-1 20 25.29&
$23.307 \pm 0.015$&
\nodata&
\nodata&
$1.89^{+0.44}_{-0.28}$&
  0.71\\
 &
b&
''&
1950&
1991&
13 11 28.905&
-1 20 28.53&
$24.541 \pm 0.049$&
\nodata&
\nodata&
$0.87^{+1.52}_{-0.71}$&
  4.84\\
 &
c&
''&
2326&
3635&
13 11 34.400&
-1 20 46.40&
$24.992 \pm 0.031$&
\nodata&
\nodata&
$0.25^{+2.72}_{-0.12}$&
  0.96\\
17&
a&
''&
2240&
2437&
13 11 30.662&
-1 20 24.87&
$24.133 \pm 0.021$&
(2.6)&
\nodata&
$2.72^{+0.36}_{-0.37}$&
  0.91\\
 &
b&
''&
2152&
2389&
13 11 30.392&
-1 20 27.83&
$23.077 \pm 0.010$&
(2.6)&
\nodata&
$0.40^{+0.14}_{-0.25}$&
  2.93\\
 &
c&
''&
1209&
1038&
13 11 24.979&
-1 20 41.84&
$24.451 \pm 0.017$&
2.6&
4&
$2.69 \pm 0.36$&
  7.18\\
18&
a&
18.2&
2499&
3112&
13 11 33.066&
-1 20 27.47&
$24.774 \pm 0.024$&
1.8&
4&
$0.16^{+0.34}_{-0.11}$&
  1.86\\
 &
b&
18.1&
2130&
1040&
13 11 26.289&
-1 20 00.19&
$24.711 \pm 0.030$&
(1.8)&
\nodata&
$0.60^{+2.30}_{-0.46}$&
  4.06\\
 &
c&
''&
2029&
2106&
13 11 29.364&
-1 20 27.39&
$25.437 \pm 0.050$&
(1.8)&
\nodata&
$0.69^{+3.52}_{-0.21}$&
 12.28\\
19&
a&
''&
2404&
2682&
13 11 31.634&
-1 20 22.64&
$24.423 \pm 0.026$&
(2.6)&
\nodata&
$0.20^{+2.43}_{-0.12}$&
  4.43\\
 &
b&
''&
1637&
 924&
13 11 25.241&
-1 20 20.05&
$24.666 \pm 0.026$&
(2.6)&
\nodata&
$0.20^{+3.17}_{-0.12}$&
  0.75\\
 &
c&
''&
1780&
3082&
13 11 31.958&
-1 20 59.38&
$23.472 \pm 0.011$&
(2.6)&
\nodata&
$0.21^{+0.12}_{-0.14}$&
  1.39\\
 &
d&
''&
1825&
3088&
13 11 32.040&
-1 20 57.47&
$23.842 \pm 0.017$&
2.6&
4&
$0.15 \pm 0.11$&
  4.28\\
21&
a&
''&
1907&
2714&
13 11 31.027&
-1 20 45.82&
$24.470 \pm 0.034$&
\nodata&
\nodata&
$1.76^{+0.52}_{-1.52}$&
  0.50\\
 &
b&
''&
1897&
2650&
13 11 30.819&
-1 20 44.91&
$23.667 \pm 0.015$&
\nodata&
\nodata&
$0.58^{+0.16}_{-0.29}$&
  2.06\\
 &
c&
''&
1797&
 852&
13 11 25.250&
-1 20 11.28&
$25.340 \pm 0.036$&
\nodata&
\nodata&
$1.78 \pm 0.27$&
  7.52\\
22&
a&
''&
2407&
2038&
13 11 29.694&
-1 20 08.84&
$23.676 \pm 0.010$&
1.7&
4&
$0.07^{+0.38}_{-0.07}$&
  0.63\\
 &
b&
''&
2127&
2144&
13 11 29.617&
-1 20 23.76&
$24.180 \pm 0.040$&
(1.7)&
\nodata&
$1.73^{+0.27}_{-0.30}$&
  4.48\\
 &
c&
''&
1539&
3348&
13 11 32.420&
-1 21 15.94&
$23.190 \pm 0.009$&
(1.7)&
\nodata&
$0.07^{+0.36}_{-0.07}$&
  0.64\\
23&
a&
''&
2364&
2005&
13 11 29.533&
-1 20 10.08&
$24.640 \pm 0.023$&
\nodata&
\nodata&
$0.09^{+0.51}_{-0.09}$&
  1.34\\
 &
b&
''&
2134&
2121&
13 11 29.558&
-1 20 22.96&
$24.888 \pm 0.048$&
\nodata&
\nodata&
$0.44^{+1.78}_{-0.32}$&
  1.52\\
 &
c&
''&
1582&
3408&
13 11 32.662&
-1 21 15.26&
$24.537 \pm 0.018$&
\nodata&
\nodata&
$0.58^{+1.99}_{-0.15}$&
  1.31\\
24&
a&
''&
1485&
2304&
13 11 29.192&
-1 20 56.22&
$25.065 \pm 0.048$&
(2.6)&
\nodata&
$0.34^{+2.68}_{-0.14}$&
  0.91\\
 &
b&
''&
3039&
2525&
13 11 32.059&
-1 19 50.56&
$23.810 \pm 0.020$&
(2.6)&
\nodata&
$3.63 \pm 0.45$&
  2.53\\
 &
c&
''&
3110&
1906&
13 11 30.290&
-1 19 34.21&
$24.083 \pm 0.019$&
2.6&
4&
$0.22^{+2.86}_{-0.16}$&
  1.22\\
 &
d&
''&
2719&
3225&
13 11 33.719&
-1 20 19.91&
$24.417 \pm 0.024$&
(2.6)&
\nodata&
$2.94 \pm 0.39$&
  0.07\\
26&
a&
''&
1396&
1008&
13 11 25.153&
-1 20 32.74&
$24.880 \pm 0.019$&
\nodata&
\nodata&
$1.31 \pm 0.23$&
  0.09\\
 &
b&
''&
2317&
2621&
13 11 31.326&
-1 20 25.28&
$25.334 \pm 0.032$&
\nodata&
\nodata&
$0.92 \pm 0.19$&
  3.45\\
 &
c&
''&
2046&
2389&
13 11 30.242&
-1 20 32.63&
$26.525 \pm 0.200$&
\nodata&
\nodata&
$0.93^{+2.70}_{-0.70}$&
  8.06\\
27&
a&
''&
1391&
1017&
13 11 25.173&
-1 20 33.16&
$24.976 \pm 0.019$&
\nodata&
\nodata&
$1.08^{+0.20}_{-0.90}$&
  1.17\\
 &
b&
''&
2334&
2627&
13 11 31.369&
-1 20 24.64&
$24.998 \pm 0.034$&
\nodata&
\nodata&
$1.11^{+0.21}_{-0.35}$&
  2.58\\
 &
c&
''&
2034&
2378&
13 11 30.192&
-1 20 32.94&
$27.038 \pm 0.225$&
\nodata&
\nodata&
$1.12^{+2.07}_{-0.85}$&
  1.00\\
28&
a&
''&
2193&
1677&
13 11 28.301&
-1 20 10.86&
$26.717 \pm 0.090$&
\nodata&
\nodata&
$5.13^{+0.60}_{-4.36}$&
  1.22\\
 &
b&
''&
2062&
3713&
13 11 34.262&
-1 21 00.01&
$25.424 \pm 0.036$&
\nodata&
\nodata&
$0.72 \pm 0.17$&
  6.83\\
 &
c&
\nodata&
2034&
2080&
13 11 29.293&
-1 20 26.61&
$25.610 \pm 0.069$&
\nodata&
\nodata&
$0.80^{+0.40}_{-0.28}$&
  0.69\\
29&
a&
''&
1456&
2329&
13 11 29.226&
-1 20 58.06&
$25.805 \pm 0.062$&
(2.5)&
\nodata&
$2.61^{+0.38}_{-2.47}$&
  1.75\\
 &
b&
''&
3076&
1833&
13 11 30.022&
-1 19 34.20&
$24.874 \pm 0.035$&
(2.5)&
\nodata&
$3.58 \pm 0.45$&
  0.67\\
 &
c&
''&
3012&
2577&
13 11 32.177&
-1 19 52.89&
$25.169 \pm 0.073$&
(2.5)&
\nodata&
$3.61 \pm 0.45$&
  1.56\\
 &
d&
''&
2694&
3206&
13 11 33.626&
-1 20 20.64&
$25.393 \pm 0.041$&
2.5&
4&
$3.21 \pm 0.41$&
  0.85\\
30&
a&
''&
3642&
2362&
13 11 32.420&
-1 19 19.80&
$25.938 \pm 0.055$&
(3.0)&
\nodata&
$3.36^{+0.46}_{-3.00}$&
  1.12\\
 &
b&
''&
3626&
2623&
13 11 33.185&
-1 19 26.07&
$26.032 \pm 0.050$&
(3.0)&
\nodata&
$3.28^{+0.42}_{-3.09}$&
  1.10\\
 &
c&
''&
3566&
2809&
13 11 33.662&
-1 19 32.73&
$25.814 \pm 0.045$&
3.0&
4&
$3.05^{+0.70}_{-0.40}$&
  2.37\\
31&
a&
12.1&
1890&
2333&
13 11 29.852&
-1 20 38.50&
$23.985 \pm 0.024$&
1.83&
2,4&
$0.70^{+0.21}_{-0.17}$&
  0.71\\
 &
b&
HL&
2219&
3312&
13 11 33.274&
-1 20 44.39&
$25.373 \pm 0.037$&
(1.83)&
\nodata&
$2.00^{+0.35}_{-1.82}$&
  1.94\\
 &
c&
12.4&
2245&
1064&
13 11 26.524&
-1 19 55.49&
$25.580 \pm 0.053$&
(1.83)&
\nodata&
$0.23^{+2.43}_{-0.14}$&
  1.16\\
 &
d&
HL&
1760&
1282&
13 11 26.496&
-1 20 22.08&
$25.063 \pm 0.030$&
(1.83)&
\nodata&
$2.35^{+0.33}_{-2.08}$&
  4.51\\
32&
a&
L&
2821&
2671&
13 11 32.191&
-1 20 03.53&
$24.083 \pm 0.013$&
(3.0)&
\nodata&
$0.15 \pm 0.11$&
  1.97\\
 &
b&
L&
2637&
3097&
13 11 33.216&
-1 20 20.90&
$25.562 \pm 0.036$&
3.0&
4&
$3.71^{+0.46}_{-0.47}$&
  1.92\\
 &
c&
L&
1416&
2468&
13 11 29.589&
-1 21 02.83&
$26.145 \pm 0.066$&
(3.0)&
\nodata&
$3.37^{+0.43}_{-3.15}$&
  0.47\\
 &
d&
L&
1796&
2361&
13 11 29.804&
-1 20 43.35&
$25.126 \pm 0.054$&
(3.0)&
\nodata&
$0.48^{+3.61}_{-0.24}$&
  5.90\\
33&
a&
L&
1310&
2140&
13 11 28.449&
-1 21 00.66&
$27.552 \pm 0.106$&
4.58&
4&
$0.61^{+3.65}_{-0.45}$&
  2.62\\
 &
b&
25.2&
2745&
2621&
13 11 31.932&
-1 20 05.91&
$26.397 \pm 0.046$&
(4.58)&
\nodata&
$3.79 \pm 0.47$&
  0.38\\
35&
a&
L&
1348&
2159&
13 11 28.560&
-1 20 59.35&
$24.632 \pm 0.029$&
1.9&
4&
$2.95 \pm 0.39$&
  1.91\\
 &
b&
L&
2521&
3397&
13 11 33.958&
-1 20 32.52&
$24.841 \pm 0.031$&
(1.9)&
\nodata&
$0.03^{+3.01}_{-0.03}$&
  9.35\\
 &
c&
L&
1906&
2186&
13 11 29.431&
-1 20 34.66&
$25.685 \pm 0.105$&
(1.9)&
\nodata&
$1.52^{+0.98}_{-1.04}$&
 20.65\\
36&
a&
L&
3060&
2352&
13 11 31.566&
-1 19 45.94&
$25.639 \pm 0.045$&
3.0&
4&
$1.04^{+0.98}_{-0.96}$&
  0.75\\
 &
b&
L&
3049&
2397&
13 11 31.686&
-1 19 47.39&
$25.997 \pm 0.057$&
(3.0)&
\nodata&
$1.84^{+0.80}_{-0.74}$&
  0.46\\
40&
a&
L&
2421&
2219&
13 11 30.260&
-1 20 12.04&
$27.034 \pm 0.118$&
2.52&
4&
$1.78^{+1.18}_{-1.58}$&
  0.68\\
 &
b&
L&
 973&
1545&
13 11 26.176&
-1 21 03.29&
$25.876 \pm 0.050$&
(2.52)&
\nodata&
$2.88 \pm 0.38$&
  1.26\\
41&
a&
\nodata&
1453&
1882&
13 11 27.872&
-1 20 48.71&
$23.062 \pm 0.010$&
\nodata&
\nodata&
$0.55^{+0.15}_{-0.47}$&
  1.57\\
 &
b&
\nodata&
2777&
3795&
13 11 35.522&
-1 20 29.38&
$25.143 \pm 0.033$&
\nodata&
\nodata&
$3.12 \pm 0.40$&
  0.66\\
 &
c&
\nodata&
1856&
2054&
13 11 28.962&
-1 20 34.12&
$99.000 \pm 28.718$&
\nodata&
\nodata&
$1.33^{+3.83}_{-0.95}$&
  1.48\\
42&
a&
\nodata&
2745&
1541&
13 11 28.672&
-1 19 42.98&
$22.969 \pm 0.015$&
\nodata&
\nodata&
$0.53^{+2.19}_{-0.23}$&
  0.43\\
 &
b&
\nodata&
2885&
2336&
13 11 31.270&
-1 19 53.52&
$24.769 \pm 0.021$&
\nodata&
\nodata&
$2.68 \pm 0.36$&
  1.28\\
 &
c&
\nodata&
2403&
3304&
13 11 33.510&
-1 20 35.89&
$25.657 \pm 0.029$&
\nodata&
\nodata&
$2.81 \pm 0.39$&
  2.80\\
 &
d&
\nodata&
1261&
2357&
13 11 29.035&
-1 21 07.49&
$25.556 \pm 0.034$&
\nodata&
\nodata&
$2.52^{+0.35}_{-1.86}$&
  0.56\\
44&
a&
\nodata&
2030&
1825&
13 11 28.517&
-1 20 21.38&
$24.487 \pm 0.034$&
\nodata&
\nodata&
$2.58^{+0.36}_{-0.55}$&
  1.40\\
 &
b&
\nodata&
2060&
3803&
13 11 34.531&
-1 21 02.01&
$25.308 \pm 0.042$&
\nodata&
\nodata&
$2.27^{+0.68}_{-2.11}$&
  0.51\\
45&
a&
25.1&
2522&
3069&
13 11 32.969&
-1 20 25.51&
$25.138 \pm 0.058$&
\nodata&
\nodata&
$3.70 \pm 0.46$&
  1.23\\
 &
b&
\nodata&
2769&
3852&
13 11 35.682&
-1 20 30.95&
$26.066 \pm 0.041$&
\nodata&
\nodata&
$0.62 \pm 0.16$&
  5.33\\
46&
a&
\nodata&
1964&
2900&
13 11 31.669&
-1 20 47.19&
$25.121 \pm 0.040$&
\nodata&
\nodata&
$0.76^{+2.37}_{-0.59}$&
135.43\\
 &
b&
\nodata&
1711&
 796&
13 11 24.959&
-1 20 13.98&
$25.674 \pm 0.053$&
\nodata&
\nodata&
$0.20^{+2.50}_{-0.12}$&
  1.50\\
48&
a&
\nodata&
2144&
2779&
13 11 31.558&
-1 20 36.47&
$25.215 \pm 0.047$&
\nodata&
\nodata&
$1.41^{+1.19}_{-1.27}$&
  0.23\\
 &
b&
\nodata&
1660&
 868&
13 11 25.104&
-1 20 17.82&
$27.085 \pm 0.111$&
\nodata&
\nodata&
$2.47^{+0.66}_{-2.10}$&
  1.25\\
49&
a&
\nodata&
2216&
1849&
13 11 28.853&
-1 20 13.47&
$25.602 \pm 0.049$&
\nodata&
\nodata&
$0.20^{+2.73}_{-0.12}$&
  0.65\\
 &
b&
\nodata&
1852&
3582&
13 11 33.569&
-1 21 06.73&
$26.285 \pm 0.060$&
\nodata&
\nodata&
$2.60^{+0.40}_{-2.32}$&
  0.90\\
50&
a&
\nodata&
2145&
3117&
13 11 32.580&
-1 20 43.60&
$26.691 \pm 0.061$&
\nodata&
\nodata&
$4.08^{+0.50}_{-0.52}$&
  0.42\\
 &
b&
\nodata&
1485&
2910&
13 11 31.021&
-1 21 09.08&
$26.595 \pm 0.047$&
\nodata&
\nodata&
$3.25^{+0.99}_{-0.42}$&
  0.29\\
 &
c&
\nodata&
2570&
2613&
13 11 31.660&
-1 20 13.66&
$28.007 \pm 0.182$&
\nodata&
\nodata&
$2.23^{+1.74}_{-2.02}$&
  1.71
\\
\vspace{-0.1in}
\enddata
\tablenotetext{a}{Quotation marks indicate agreement with the B05 numbering system (where our 1a is their 1.1, etc.). ``L'' indicates a system added by L06, and ``$\cdots$'' a system added in this paper. ``HL'' indicates systems that were added by H06, but for which we use L06's numbering.}
\tablenotetext{b}{Pixel coordinates ($0\farcs 05 / {\rm pix}$) in the original APSIS-reduced ACS images (see \S\ref{LPA1689:coordsys}).}
\tablenotetext{c}{Values in parentheses are assumed from other multiple image(s).}
\tablenotetext{d}{Reference for $z_{spec}$: 1) \cite{Frye02}; 2) B05; 3) \cite{Frye07}; 4) L07}
\tablenotetext{e}{Bayesian photometric redshift and 95\% confidence interval. Based on fits to photometry across 12 filters without subtraction of cluster galaxies.}
\tablenotetext{f}{Poorness of BPZ fit: observed vs.~model fluxes with uncertainties}
\end{deluxetable}


\begin{deluxetable}{rlcrrcc}
\tablewidth{0pt}  
\tablecaption{\label{LPA1689:tabknots}Extra Knots Identified in the Multiple Images}
\tablehead{
\multicolumn{3}{c}{}&
\colhead{$x$\tablenotemark{b}}&
\colhead{$y$\tablenotemark{b}}&
\colhead{R.A.}&
\colhead{Decl.}\\
\multicolumn{3}{c}{ID\tablenotemark{a}}&
\colhead{(pixel)}&
\colhead{(pixel)}&
\colhead{(J2000.0)}&
\colhead{(J2000.0)}
}
\startdata
4&
a&
1&
1818&
3116&
13 11 32.114&
-1 20 58.38\\
 &
b&
1&
1395&
2832&
13 11 30.658&
-1 21 11.50\\
 &
c&
1&
2545&
2328&
13 11 30.765&
-1 20 08.74\\
 &
d&
1&
1485&
1336&
13 11 26.269&
-1 20 35.67\\
6&
a&
1&
3093&
2082&
13 11 30.798&
-1 19 38.71\\
 &
a&
2&
3092&
2075&
13 11 30.775&
-1 19 38.61\\
 &
b&
1&
2803&
3067&
13 11 33.361&
-1 20 12.75\\
 &
b&
2&
2801&
3058&
13 11 33.331&
-1 20 12.65\\
8&
a&
1&
1956&
3098&
13 11 32.255&
-1 20 51.75\\
 &
a&
2&
1996&
3108&
13 11 32.342&
-1 20 50.15\\
 &
b&
1&
1628&
3000&
13 11 31.495&
-1 21 04.52\\
 &
b&
2&
1560&
2961&
13 11 31.281&
-1 21 06.77\\
10&
a&
1&
2179&
3564&
13 11 33.978&
-1 20 51.55\\
 &
b&
1&
2134&
1634&
13 11 28.088&
-1 20 12.62\\
16&
a&
1&
1897&
1732&
13 11 28.048&
-1 20 25.43\\
 &
a&
2&
1864&
1712&
13 11 27.941&
-1 20 26.50\\
 &
b&
1&
1943&
1958&
13 11 28.795&
-1 20 28.14\\
 &
b&
2&
1946&
2005&
13 11 28.942&
-1 20 29.00\\
 &
c&
1&
2327&
3643&
13 11 34.426&
-1 20 46.53\\
18&
a&
1&
2095&
3555&
13 11 33.832&
-1 20 55.16\\
 &
b&
1&
2206&
1663&
13 11 28.277&
-1 20 09.97\\
19&
a&
1&
2384&
2716&
13 11 31.708&
-1 20 24.27\\
 &
a&
2&
2421&
2644&
13 11 31.543&
-1 20 21.06\\
 &
b&
1&
1657&
 910&
13 11 25.227&
-1 20 18.84\\
 &
b&
2&
1613&
 941&
13 11 25.258&
-1 20 21.49\\
 &
d&
1&
1834&
3090&
13 11 32.059&
-1 20 57.11\\
31&
a&
2&
2807&
2022&
13 11 30.212&
-1 19 50.39\\
 &
a&
3&
2814&
2070&
13 11 30.366&
-1 19 51.09\\
 &
b&
3&
2233&
3311&
13 11 33.290&
-1 20 43.73\\
 &
c&
2&
1199&
2349&
13 11 28.923&
-1 21 10.12\\
 &
c&
3&
1207&
2358&
13 11 28.961&
-1 21 09.95\\
 &
d&
2&
1762&
1288&
13 11 26.517&
-1 20 22.11\\
 &
d&
3&
1785&
1276&
13 11 26.513&
-1 20 20.82\\
\vspace{-0.1in}
\enddata
\tablenotetext{a}{ID numbers as in Table \ref{LPA1689:tabmultimages}, with the third column giving the ``knot number''.}
\tablenotetext{b}{Pixel coordinates ($0\farcs 05 / {\rm pix}$) in the original APSIS-reduced ACS images (see \ref{LPA1689:coordsys}).}
\end{deluxetable}


\clearpage
\begin{deluxetable}{rccrccll}
\tablecaption{\label{LPA1689:taborderz}Multiple Image Systems}
\tablehead{
\colhead{}&
\colhead{Number}&
\colhead{Knots}&
\colhead{Input}&
\colhead{Spectroscopic}&
\colhead{}&
\colhead{Input}&
\colhead{Output}\\
\colhead{ID}&
\colhead{of images}&
\colhead{per image}&
\colhead{order}&
\colhead{redshift}&
\colhead{Ref.\tablenotemark{a}}&
\colhead{redshift\tablenotemark{b}}&
\colhead{redshift\tablenotemark{c}}
}
\startdata
1&
6&
1&
1&
3.04&
2,3&
$3.04$&
3.04\\
2&
5&
1&
2&
2.53&
3&
$2.54$&
2.54\\
3&
3&
1&
23&
\nodata&
\nodata&
$5.47 \pm 0.7$&
5.490\\
4&
5&
2&
7&
1.1&
4&
$1.1 $&
1.1\\
5&
3&
1&
13&
2.6&
4&
$2.6 $&
2.6\\
6&
4&
3&
12&
1.1&
4&
$1.1 $&
1.1\\
7&
2&
1&
5&
4.87&
1&
$4.87$&
4.87\\
8&
4&
3&
10&
\nodata&
\nodata&
$1.8  \pm 0.5$&
2.224\\
9&
4&
1&
4&
\nodata&
\nodata&
$5.16 \pm 0.5$&
4.707\\
10&
3&
2&
18&
1.83&
4&
$1.8 $&
1.8\\
11&
3&
1&
25&
2.5&
4&
$2.5 $&
2.5\\
12&
6&
1&
30&
1.83&
2,4&
$1.83$&
1.83\\
13&
3&
1&
31&
\nodata&
\nodata&
$1.02 \pm 0.5$&
1.532\\
14&
2&
1&
32&
3.4&
4&
$3.4 $&
3.4\\
15&
3&
1&
19&
1.8&
4&
(10)&
1.8\\
16&
3&
3&
14&
\nodata&
\nodata&
$2.01 \pm 0.5$&
1.635\\
17&
3&
1&
26&
2.6&
4&
$2.6 $&
2.6\\
18&
3&
2&
20&
1.8&
4&
(10)&
1.8\\
19&
4&
3&
11&
2.6&
4&
$2.6 $&
2.6\\
21&
3&
1&
15&
\nodata&
\nodata&
$1.78 \pm 0.5$&
1.065\\
22&
3&
1&
16&
1.7&
4&
$1.7 $&
1.7\\
23&
3&
1&
17&
\nodata&
\nodata&
(22)&
1.7\\
24&
4&
1&
8&
2.6&
4&
$2.6 $&
2.6\\
26&
3&
1&
21&
\nodata&
\nodata&
(27)&
2.388\\
27&
3&
1&
22&
\nodata&
\nodata&
$1.74 \pm 0.5$&
2.388\\
28&
3&
1&
24&
\nodata&
\nodata&
$5.45 \pm 0.7$&
4.451\\
29&
4&
1&
9&
2.5&
4&
$2.5 $&
2.5\\
30&
3&
1&
33&
3.0&
4&
$3.0 $&
3.0\\
31&
4&
3&
3&
1.83&
2,4&
$1.8 $&
1.8\\
32&
4&
1&
6&
3.0&
4&
$3.0 $&
3.0\\
33&
2&
1&
27&
4.58&
4&
$4.5 $&
4.5\\
35&
3&
1&
28&
1.9&
4&
$1.9 $&
1.9\\
36&
2&
1&
34&
3.0&
4&
$3.0 $&
3.0\\
40&
2&
1&
29&
2.52&
4&
$2.5 $&
2.5\\
41&
3&
1&
37&
\nodata&
\nodata&
$2.5  \pm 2.0$&
5.357\\
42&
4&
1&
35&
\nodata&
\nodata&
$2.0  \pm 2.0$&
2.222\\
44&
2&
1&
38&
\nodata&
\nodata&
$2.0  \pm 2.0$&
3.210\\
45&
2&
1&
39&
\nodata&
\nodata&
$2.5  \pm 2.0$&
5.359\\
46&
2&
1&
40&
\nodata&
\nodata&
$2.5  \pm 2.0$&
1.714\\
48&
2&
1&
41&
\nodata&
\nodata&
$2.0  \pm 2.0$&
1.456\\
49&
2&
1&
42&
\nodata&
\nodata&
$2.0  \pm 2.0$&
1.645\\
50&
3&
1&
36&
\nodata&
\nodata&
$2.5  \pm 2.0$&
3.029
\\
\vspace{-0.1in}
\enddata
\tablecomments{Our mass model was built by iteratively adding our multiple image systems in the order given here. Any additional knots were added immediately after inclusion of the ``main knot'' for that system.}
\tablenotetext{a}{Reference for $z_{spec}$: 1) \cite{Frye02}; 2) B05; 3) \cite{Frye07}; 4) L07}
\tablenotetext{b}{Input redshifts with rough uncertainties where allowed. (Spectroscopic redshifts are given zero uncertainty.) Redshifts of some systems are tied to other systems with numbers given in parentheses. (For example, the redshift of system 26 is tied to that of system 27 in the optimization.)}
\tablenotetext{c}{Final model redshift, optimized if input value included uncertainty.}
\end{deluxetable}



\appendix
\chapter{Dark Matter Halo Mass Profiles}
\label{LPA1689:virial}
\label{LPA1689:profiles}

Clusters formed in simulations
have mass profiles generally well described by 
NFW \citep{NFW96}
or Einasto / S\'ersic \citep{Navarro04} 
profiles \citep[e.g.,][]{Navarro10}.
The latter were found to yield superior fits to simulated halos
\citep{Navarro04,Merritt05,Merritt06,Navarro10}.
Details on these fitting forms can be found in \cite{Coe10DMprofiles}.

The NFW profile gives mass density as a function of radius in three dimensions:
\\
\begin{equation}
\label{LPA1689:NFWEQ}
  \rho(r) = \frac{\rho_s}{(r/r_s)(1+r/r_s)^2}.
\end{equation}
\\
The logarithmic slope increases from -1 at the core to -3 in the outskirts.
The NFW concentration parameter is given as $c = r_{vir} / r_s$.
By definition, the average density within the virial radius $r_{vir}$ is
$\Delta_c \approx 115$ times $\rho_{crit} = 3 H^2(z) / (8\pi G)$
for a collapsed virialized sphere at $z = 0.186$,
as evaluated according to formulae given by \cite{NakamuraSuto97}.
For historical reasons, $\Delta_c \approx 200$ is often used,
so we report concentration values for both overdensities.
Conversion between the two is given by $c_{115} \approx 1.234 c_{200} + 0.172$.
Lensing properties of the NFW profile can be found in \cite{GolseKneib02}.

The \cite{Sersic68} profile has a continuously varying slope that asymptotes 
toward zero at very small radius.
Its use as a mass profile is described by
\cite{GrahamDriver05} and \cite{TerzicGraham05}
with lensing properties calculated by \cite{Cardone04} and \cite{Eliasdottir07}.
These details are summarized in \cite{Coe10DMprofiles}.
Our S\'ersic profile of {\it projected} mass is given as:
\\
\begin{equation}
\label{LPA1689:SersicEq}
  \Sigma(R) = \Sigma_e \exp\left\{
    -b_n \left[ \left( \frac{R}{R_e} \right)^{1/n} - 1 \right]
  \right\}.
\end{equation}
\\
There are three free parameters: $\Sigma_e$, $R_e$, and $n$,
with $b_n$ being a function of $n$.
This profile was found \citep{Navarro04,Merritt05,Merritt06}
to yield better fits to a wide range of simulated dark matter halos
than did the generalized NFW profile \citep{Zhao96,Wyithe01},
which has an equal number (3) of free parameters, including the central slope.

A distinction has been made between the S\'ersic profile described above
and the Einasto profile.
The latter has a similar form but is defined as a function of 3-D density $\rho(r)$
rather than projected surface density $\Sigma(R) = \kappa(R) \Sigma_{crit}$.
Performance of the two forms is similar
with the S\'ersic profile providing a slightly better fit to cluster halos \citep{Merritt06}.
These and other mass profiles are described in more detail in \cite{Coe10DMprofiles}.




\chapter{Multiple Images}
\label{LPA1689:stamps}

\begin{figure*}
\subfigure{
\includegraphics[width=0.155\hsize]{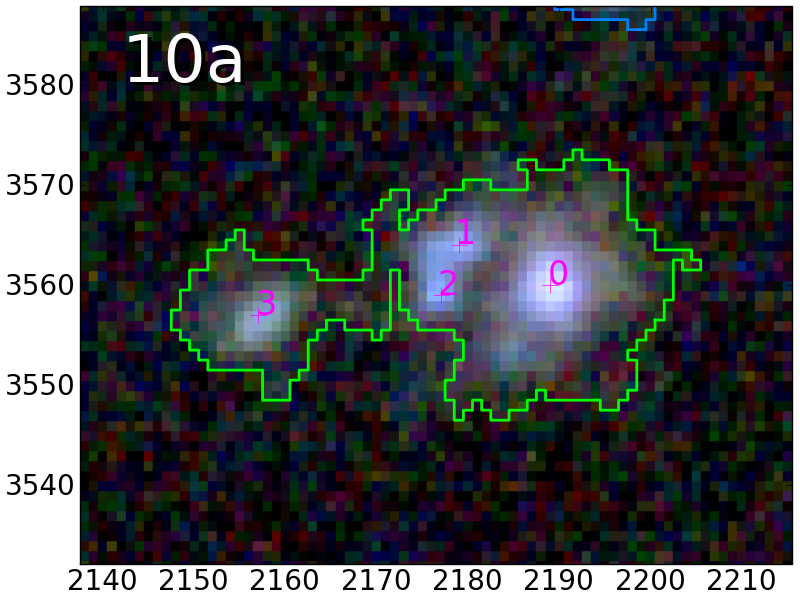}
\includegraphics[width=0.155\hsize]{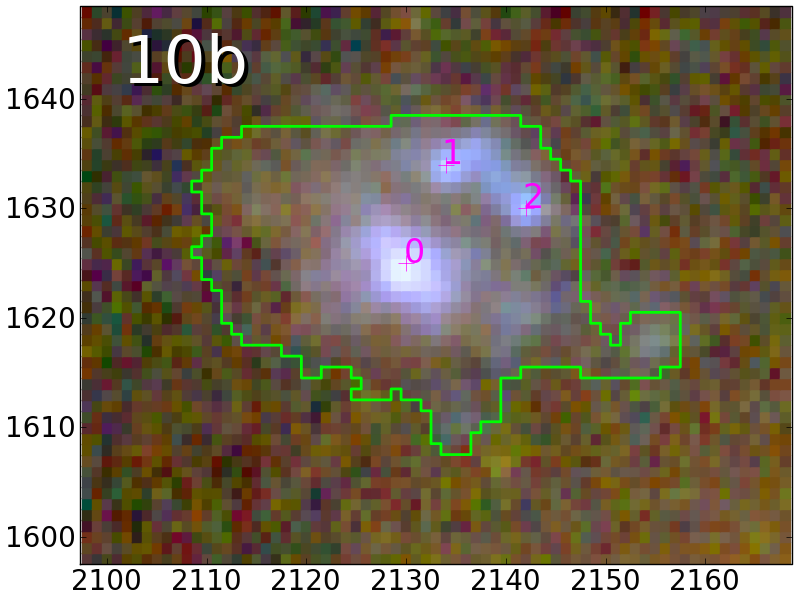}
\includegraphics[width=0.155\hsize]{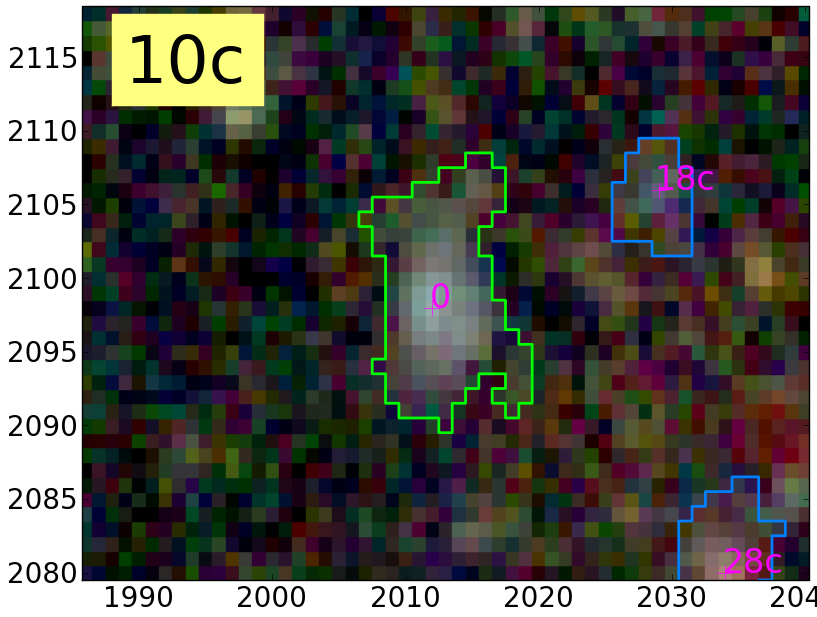}
\hspace{0.03\hsize}
\includegraphics[width=0.155\hsize]{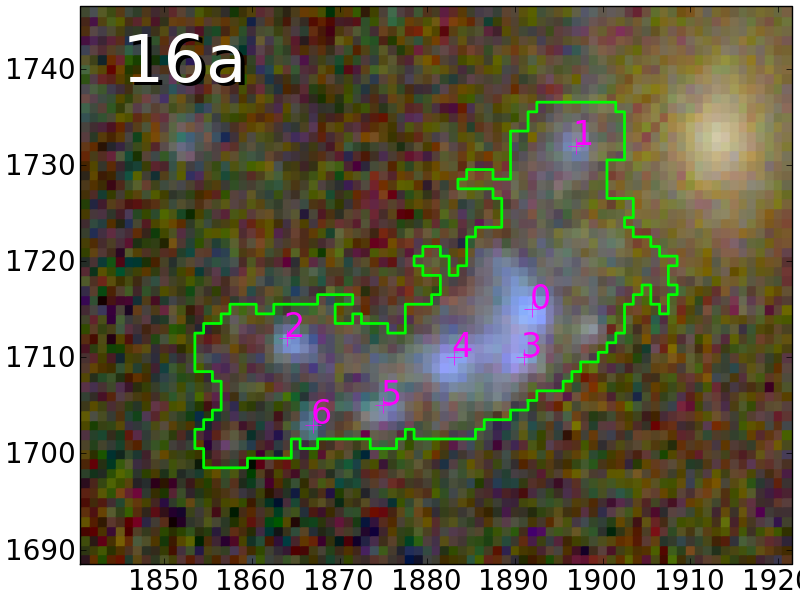}
\includegraphics[width=0.155\hsize]{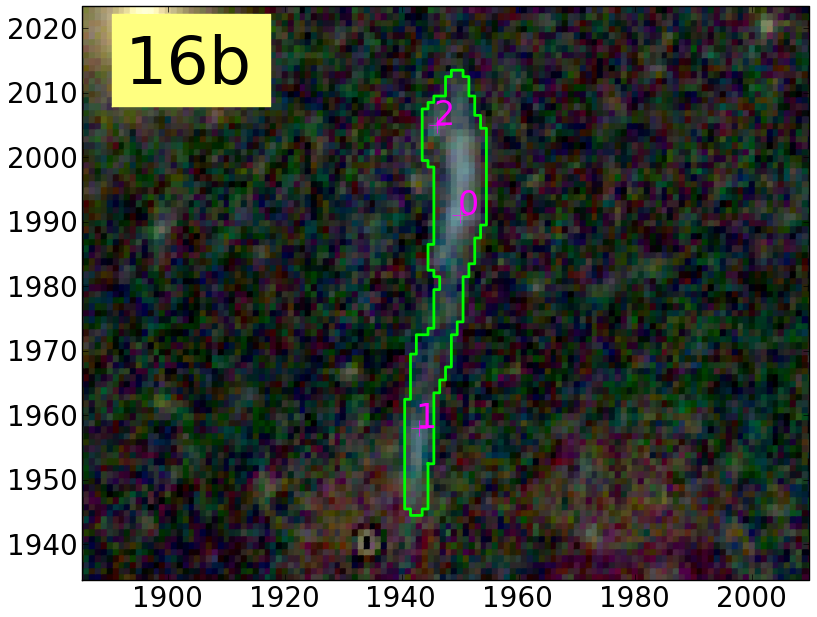}
\includegraphics[width=0.155\hsize]{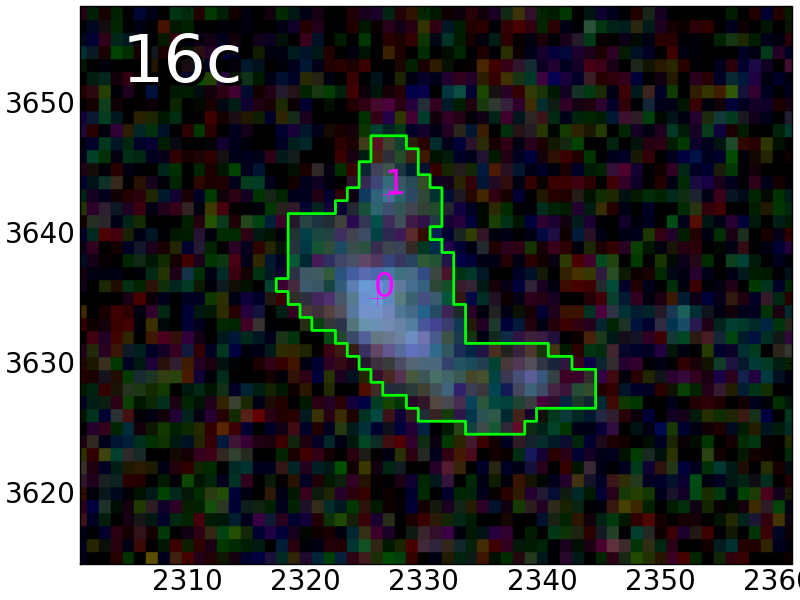}
}
\vspace{0.1in}
\caption[Multiple images in Abell 1689
]{\label{LPA1689:images}Multiple images in Abell 1689.
For each image, identified knots are labeled with numbers and crosses.
Green lines are isophotal apertures, altered where necessary.
Yellow boxes outlining the IDs
indicate color image stamps that were produced
from our galaxy-subtracted images.
All other color stamps are taken from the STScI $g\arcmin r\arcmin z\arcmin$ color image.
Coordinates in pixels are given along the axes.
Each pixel measures $0\farcs 05$ across.
}
\end{figure*}

\begin{figure*}
\subfigure{
\includegraphics[width=0.155\hsize]{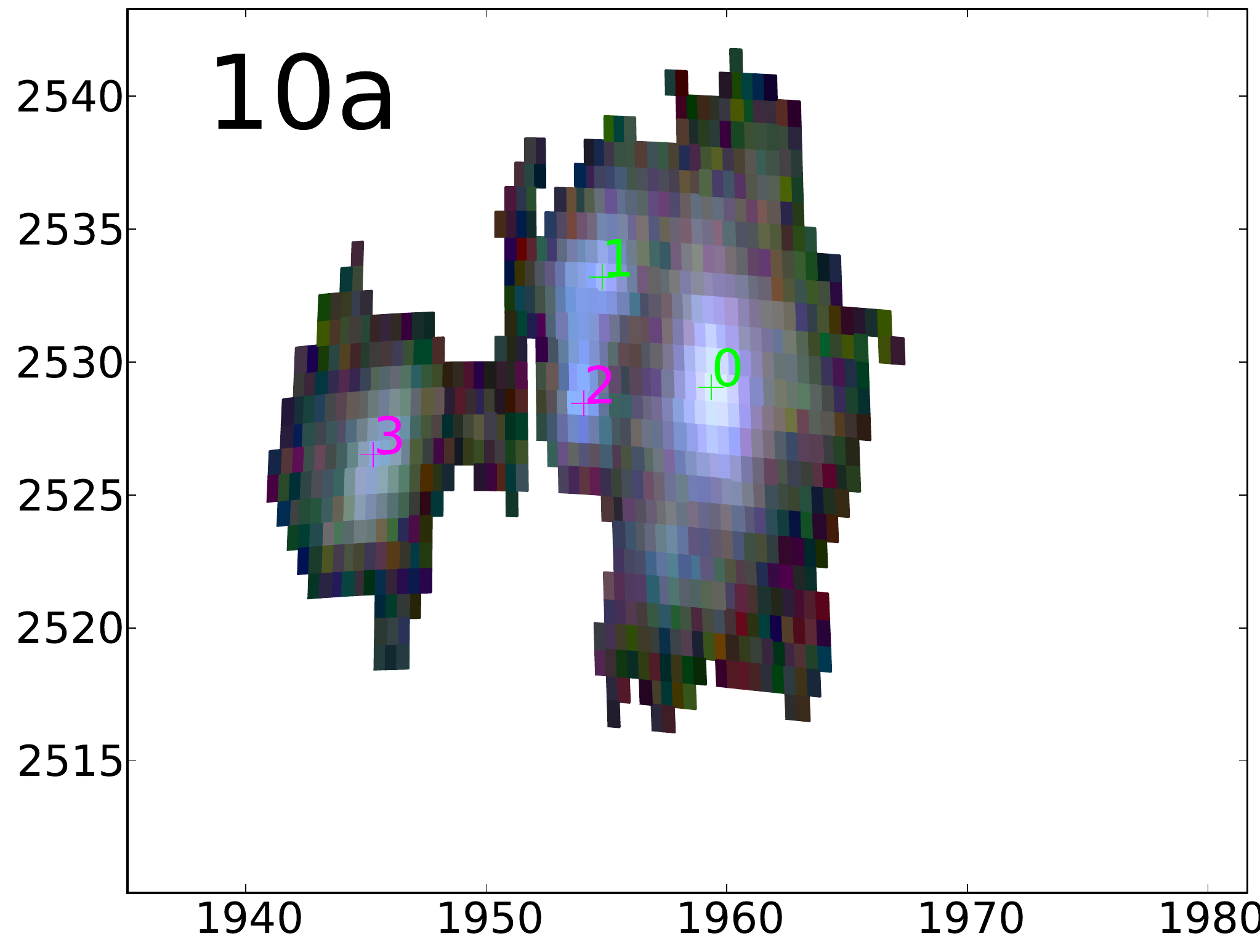}
\includegraphics[width=0.155\hsize]{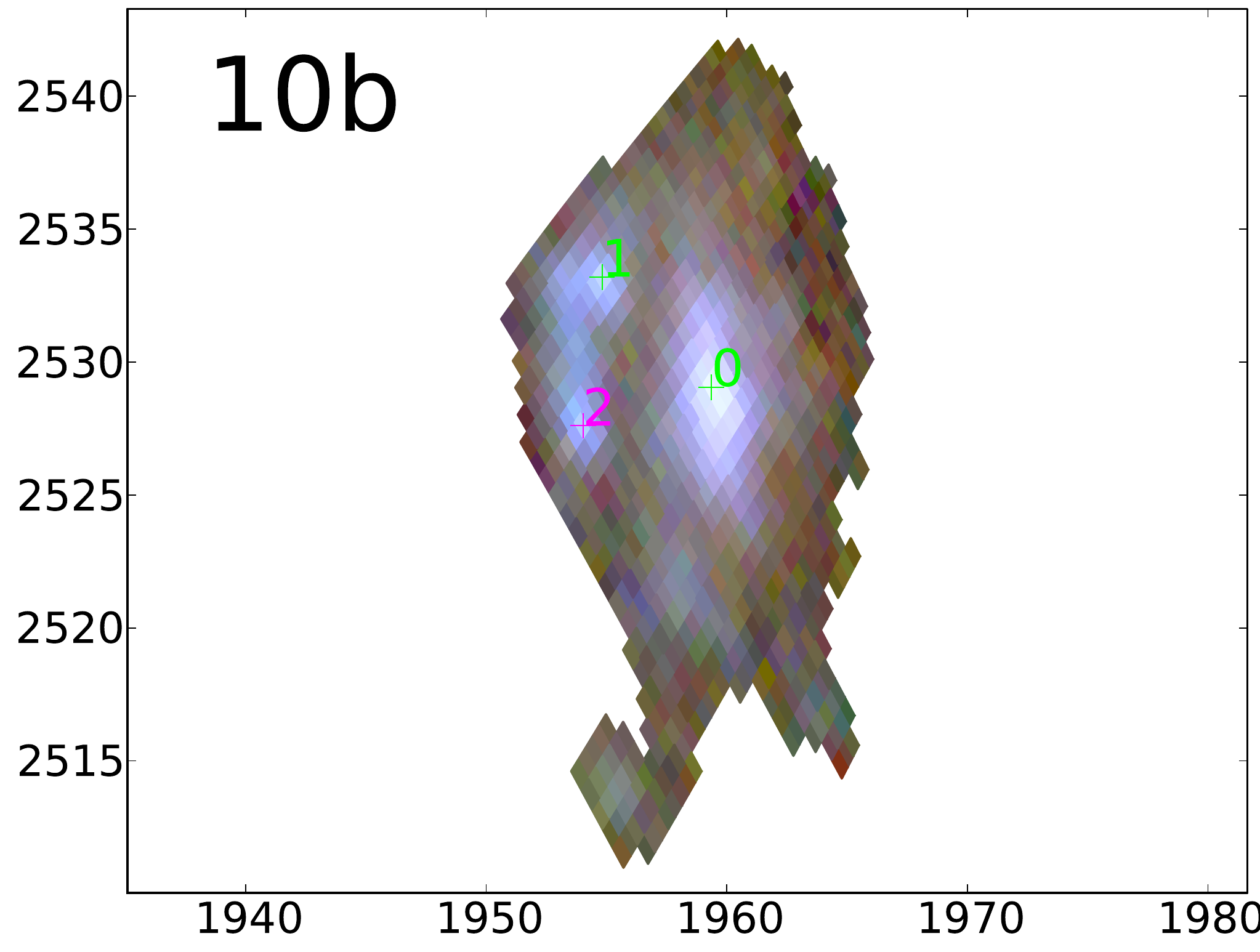}
\includegraphics[width=0.155\hsize]{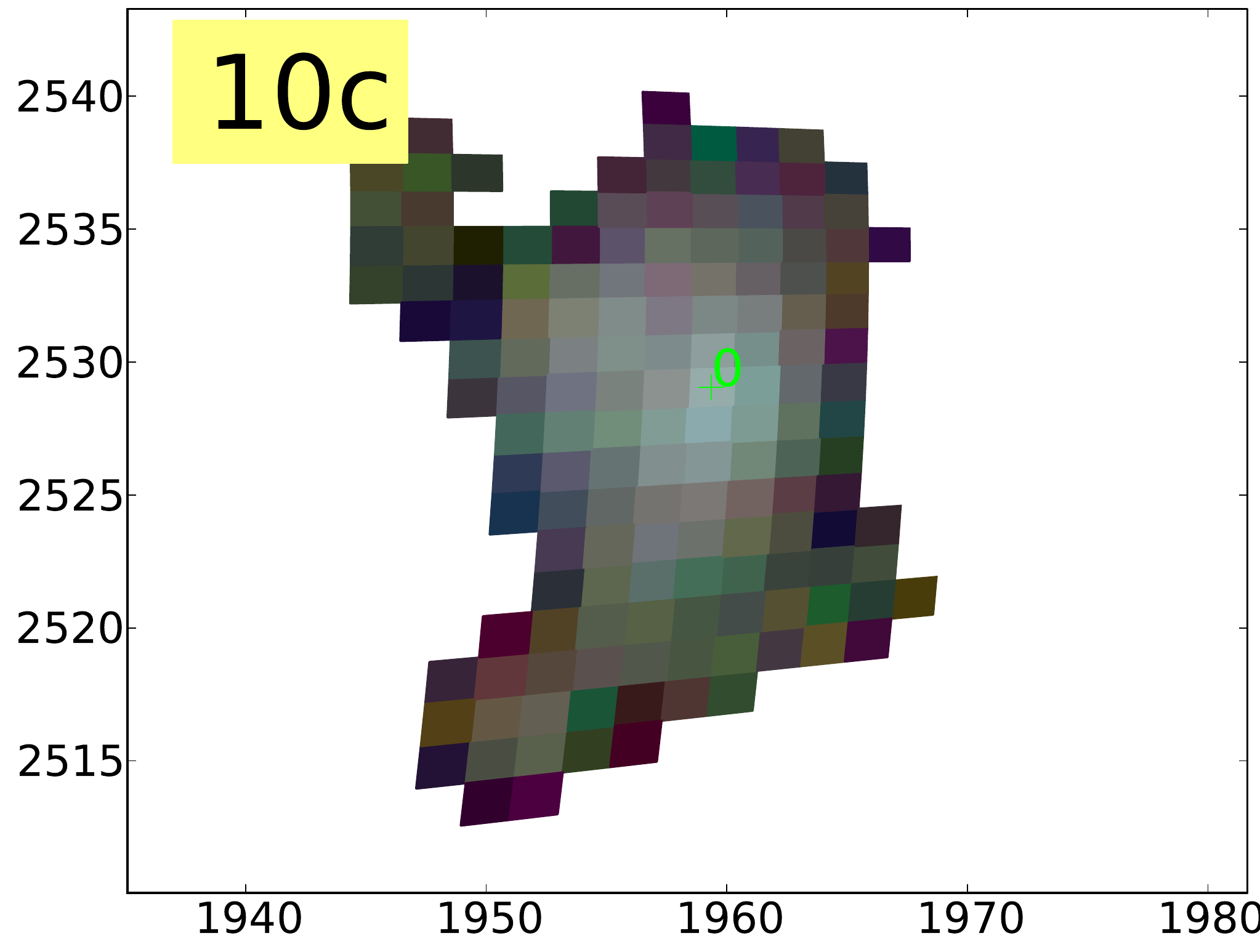}
\hspace{0.03\hsize}
\includegraphics[width=0.155\hsize]{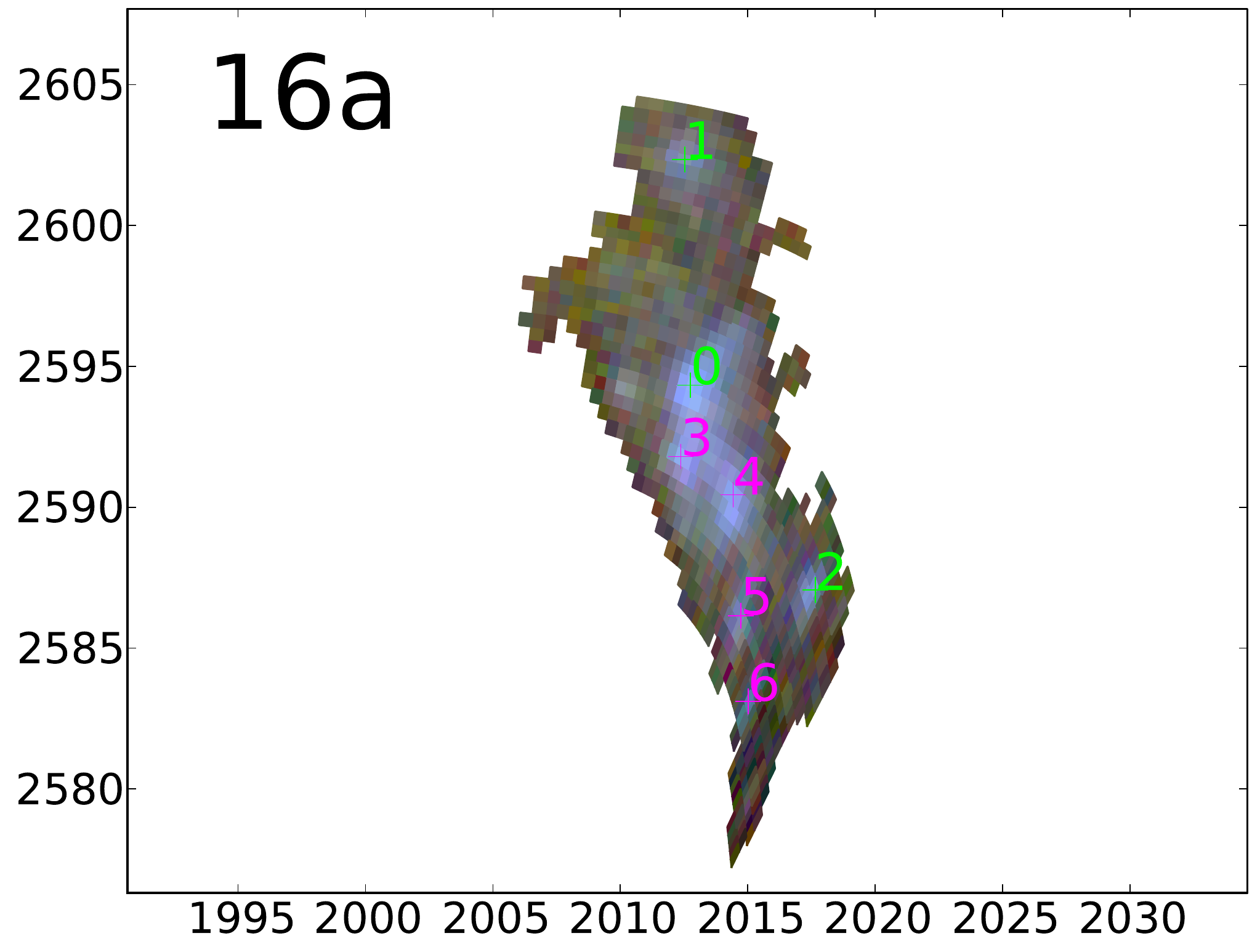}
\includegraphics[width=0.155\hsize]{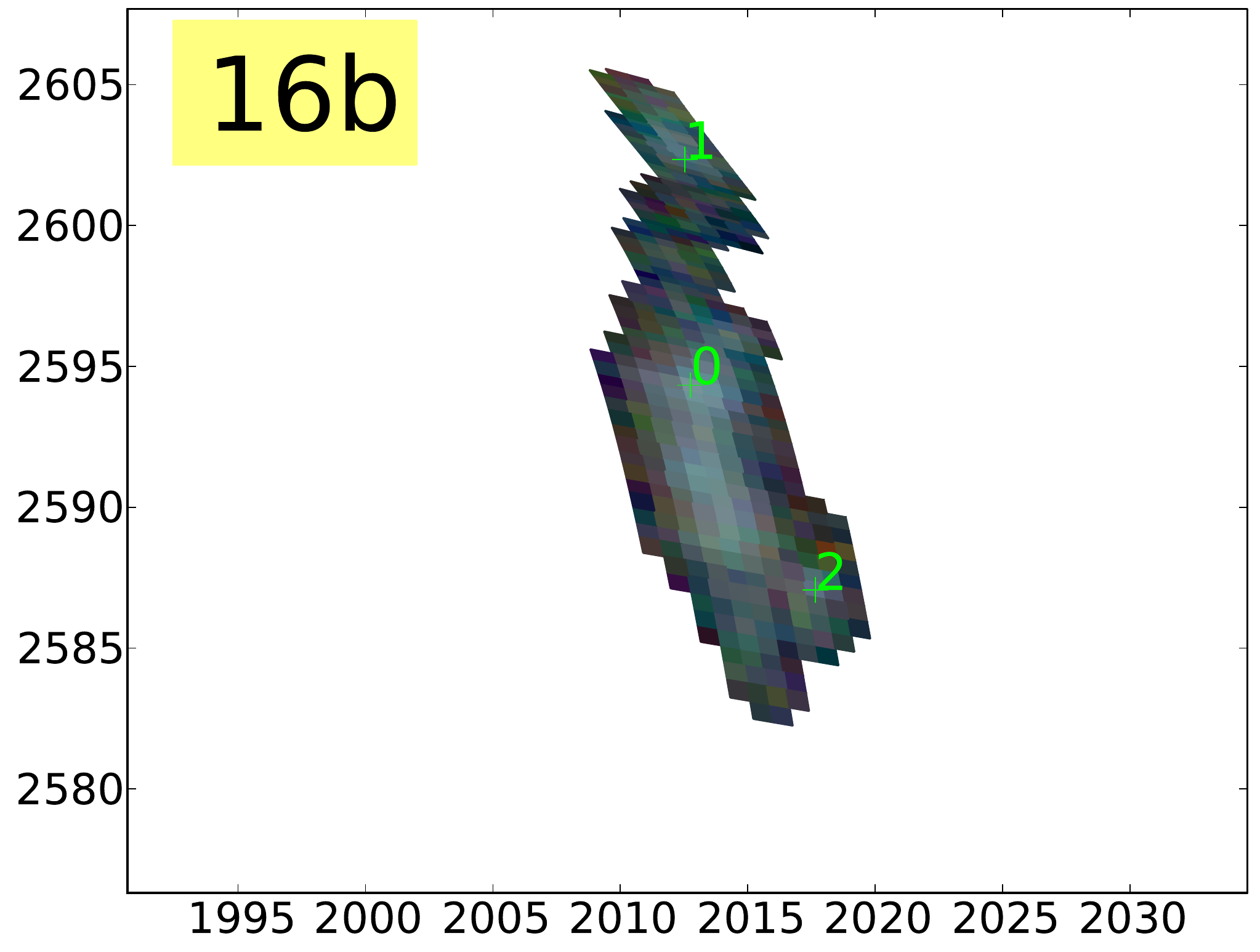}
\includegraphics[width=0.155\hsize]{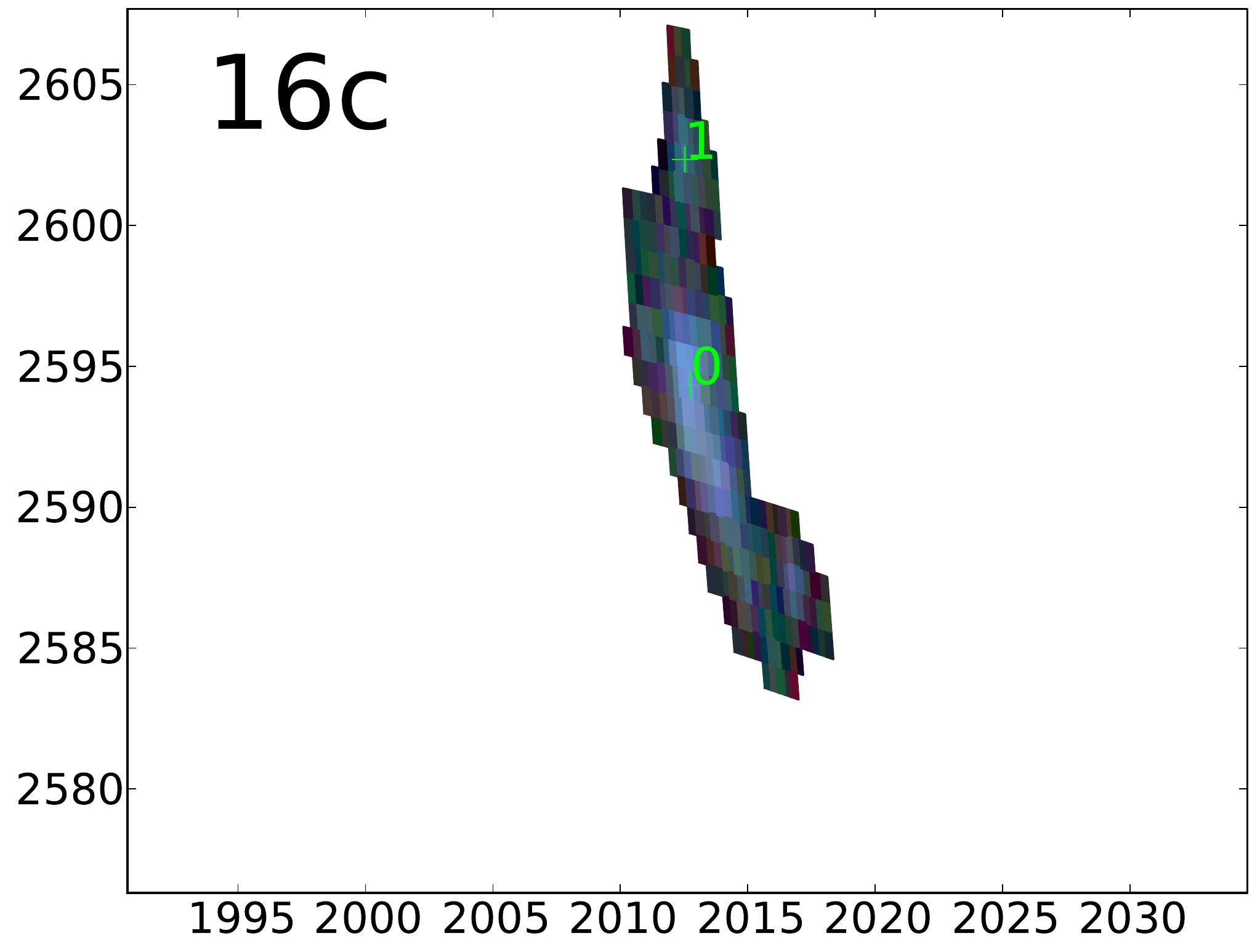}
}
\\
\vspace{0.1in}
\caption[Multiple images delensed to the source plane
]{\label{LPA1689:delensed}Multiple images delensed to the source plane.
Knots are labeled as in Fig.~\ref{LPA1689:images},
but now constrained knots are colored green
and unconstrained knots are colored pink.
Within each system,
all of the constrained knots align exactly among the delensed images.
Unconstrained knots are generally well aligned but not always.
Panels are plotted to the same scale within each system.
Yellow boxes outlining the IDs indicate color image stamps
that were produced from our galaxy-subtracted images.
All other color stamps are taken from the STScI $g\arcmin r\arcmin z\arcmin$ color image.
Coordinates in pixels are given along the axes.
Each pixel measures $0\farcs 05$ across.
}
\end{figure*}

In this section we display color images of the multiple images
used in this work.
Fig.~\ref{LPA1689:images} displays the images as they appear and
Fig.~\ref{LPA1689:delensed} displays the same images delensed,
as they would appear if Abell 1689 did not exist
(according to our best mass model).
The \Nknotss constrained positions are labeled in green
in Fig.~\ref{LPA1689:delensed}.
These delensed positions match perfectly among each set of multiple images.
We have identified other knots, labeled in pink, which we have not constrained.
We have doubts about these identifications
(including a case or two in which it appears that we were obviously mistaken).



Examples of the image stamps are given here.
The remaining images (for all \Ngals image systems) can be found in the online version.


\end{document}